\documentclass{aa}  

\usepackage{graphicx}
\usepackage{txfonts}
\usepackage{xcolor}
\usepackage{graphicx}	
\usepackage{subcaption}
\usepackage{layouts}
\usepackage{tikz}
\begin{document} 

\newcommand{\jmbsize}{40mm}  
\newcommand{\jmtsize}{28mm}
\newcommand{\jmhspace}{-1mm}
\newcommand{\jmvspace}{2mm}

   \title{VALES VII: Molecular and ionized gas properties in pressure balanced interstellar medium of starburst galaxies at $z \sim 0.15$.}
   \titlerunning{Molecular and ionized gas properties in pressure balanced ISM of starburst galaxies}

   \author{Juan Molina\inst{1,2} 
			\and Edo Ibar\inst{3}
			\and Nicol\'as Godoy\inst{3,4}
			\and Andr\'es Escala\inst{2}
			\and Tomonari Michiyama\inst{1}
			\and Cheng Cheng\inst{5,6,3}
			\and Thomas M. Hughes\inst{3,5,7,8}
			\and Maarten Baes\inst{9}
			\and Yongquan Xue\inst{7}
			\and Micha{\l} J. Micha{\l}owski\inst{10}
			\and Paul van der Werf\inst{11}
			\and Xue-Jian Jiang\inst{12}
          }

   \institute{$^1$Kavli Institute for Astronomy and Astrophysics, Peking University, 5 Yiheyuan Road, Haidian District, Beijing 100871, P.R. China\\
			$^2$Departamento de Astronom\'ia (DAS), Universidad de Chile, Casilla 36-D, Santiago, Chile\\
			 \email{jumolina@pku.edu.cn}\\
              $^3$Instituto de F\'isica y Astronom\'ia, Universidad de Valpara\'iso, Avda. Gran Breta\~na 1111, Valpara\'iso, Chile\\
              $^4$N\'ucleo Milenio de Formaci\'on Planetaria -- NPF, Universidad de Valpara\'iso, Av. Gran Breta\~na 1111, Valpara\'iso, Chile\\
              $^5$Chinese Academy of Sciences South America Center for Astronomy, National Astronomical Observatories, CAS, Beijing 100101, China.\\
              $^6$CAS Key Laboratory of Optical Astronomy, National Astronomical Observatories, CAS, Beijing 100101, China.\\
              $^7$CAS Key Laboratory for Research in Galaxies and Cosmology, Department of Astronomy, University of Science and Technology of China, Hefei 230026, China\\
			$^8$School of Astronomy and Space Science, University of Science and Technology of China, Hefei 230026, China\\
			$^9$Sterrenkundig Observatorium, Universiteit Gent, Krijgslaan 281 S9, B-9000 Gent, Belgium\\
			$^{10}$Astronomical Observatory Institute, Faculty of Physics, Adam Mickiewicz University, ul. S{\l}oneczna 36, 60-286 Pozna\'n , Poland\\
			$^{11}$Leiden Observatory, Leiden University, P.O. Box 9513, NL-2300 RA Leiden, The Netherlands\\
			$^{12}$East Asian Observatory, 660 North A'ohoku Place, Hilo, Hawaii 96720, USA
             }

   \date{
   }

 
  \abstract
   {Spatially resolved observations of the ionized and molecular gas are critical for understanding the physical processes that govern the interstellar medium (ISM) in galaxies. The observation of starburst systems is also important as these present extreme gas conditions that may help to test different ISM models. However, matched resolution imaging at $\sim$kpc scales for both ISM gas phases are usually scarce and the ISM properties of starbursts still remain poorly understood.}
   {We aim to study the morpho-kinematic properties of the ionized and molecular gas in three dusty starburst galaxies at $z = 0.12-0.17$ to explore the relation between molecular ISM gas phase dynamics and the star-formation activity.}
   {We employ two-dimensional dynamical modelling to analyse Atacama Large Millimeter/submillimiter Array (ALMA) CO(1--0) and seeing limited Spectrograph for INtegral Field Observations in the Near Infrared (SINFONI) Paschen-$\alpha$ (Pa$\alpha$) observations tracing the molecular and ionized gas morpho-kinematics at $\sim$\,kpc-scales. We use a dynamical mass model, which accounts for beam-smearing effects, to constrain the CO-to-H$_2$ conversion factor and estimate the molecular gas mass content.}
   {One starburst galaxy shows irregular morphology which may indicate a major merger, while the other two systems show disc-like morpho-kinematics. The two disc-like starbursts show molecular gas velocity dispersion values comparable with that seen in local Luminous and Ultra Luminous Infrared Galaxies, but in an ISM with molecular gas fraction and surface density values in the range of the estimates reported for local star-forming galaxies. We find that these molecular gas velocity dispersion values can be explained by assuming vertical pressure equilibrium. We also find that the star-formation activity, traced by the Pa$\alpha$ emission line, is well correlated with the molecular gas content suggesting an enhanced star formation efficiency and depletion times of the order of $\sim 0.1-1$\,Gyr. We find that the star formation rate surface density ($\Sigma_{\rm SFR}$) correlates with the ISM pressure set by self-gravity ($P_{\rm grav}$) following a power law with an exponent close to 0.8.}
   {In dusty disc-like starburst galaxies, our data support the scenario in which the molecular gas velocity dispersion values are driven by the ISM pressure set by self-gravity, responsible to maintain the vertical pressure balance. The correlation between $\Sigma_{\rm SFR}$ and $P_{\rm grav}$ suggests that, in these dusty starbursts galaxies, the star formation activity arises as a consequence of the ISM pressure balance.} 

   \maketitle
%

\section{Introduction} 
Understanding how galaxies build up their stellar mass content within dark matter haloes is a key goal in modern extragalactic astrophysics. One of the best constraints comes from studying the evolution of the star formation rate density (SFRD) across cosmic time \citep{Madau1996,Madau2014}. The overall decline of the SFRD in the last $\sim$10\,Gyr coincides with the decrease of the average fraction of molecular gas mass in galaxies \citep{Tacconi2010,Geach2012,Carilli2013}. A straightforward interpretation is that the molecular gas is the fuel that maintains the star formation activity \citep{Bigiel2008,Leroy2008}. If the gas supply into galaxies is continuously smooth, then the formation of stars may be driven by internal dynamical processes within the interstellar medium (ISM; \citealt{Keres2005,Bournaud2007,Dekel2009b,Spring2017}). It is therefore essential to identify the physical processes that govern the ISM properties to tackle galaxy evolution.

A complete characterization of the ISM involves the understanding of many complex processes that are driven and evolve on different spatial and time scales. ISM models often assume a dynamic equilibrium (e.g. \citealt{Thompson2005,Ostriker2010,FQH2013,Krumholz2018}). In this `quasi steady-state', the ISM gas pressure is set to maintain the vertical pull from galaxy self-gravity. The star formation activity, parametrized by the Kennicutt-Schmidt law \citep{Kennicutt1998a}, arises as a result of the pressure balance (e.g. \citealt{Ostriker2011,Hayward2017}).

It is still unclear which mechanism is the main responsible for setting the pressure support to stabilize the ISM gas against self-gravity. One possibility is stellar feedback (e.g. \citealt{Ostriker2011,Kim2011}). Another possibility comes from the energy released by gravitational instabilities and mass transport within galactic discs \citep{Krumholz2016,Krumholz2018}. Local galaxy spatially-resolved observations show trends in favour of the stellar-feedback regulated model \citep{Sun2020}. Unresolved observations for starbursts also agree with this model \citep{Fisher2019}. However, there is also evidence that additional sources of energy beyond stellar feedback may help support system self-gravity \citep{Zhou2017,Molina2019a}, especially for systems with high star formation rates (e.g. \citealt{Varibel2020}). Luminous and Ultra Luminous Infrared Galaxies (LIRG/ULIRGs) seem also to be in vertical pressure equilibrium set by the release of gravitational energy \citep{Wilson2019}. In any case, to test the pressure balance-based ISM models, galaxy spatially-resolved observations that trace the ISM gas phases, star-formation activity and the stellar component are needed.

Obtaining such a dataset for large galaxy samples is generally time-consuming. While integral field unit (IFU) observations targeting the star formation activity in galaxies are common (e.g. \citealt{Sanchez2012,Bryant2015}), molecular gas spatially-resolved observations are relatively scarce. Observing the spatial distribution of the molecular gas content in star-forming galaxies is still, relative to the optical/near-IR observations, highly time-consuming. This is true even for the present times of Atacama Large Millimeter/submillimeter Array (ALMA) and the NOrthem Ex-tended Millimetre Array (NOEMA). The hydrogen molecule (H$_2$) is not easily detectable at low temperatures in the range of the few hundreds of Kelvin (e.g. \citealt{Papadopoulos1999,Bothwell2013}), and use of molecular gas tracers, such as the carbon monoxide molecule ($^{12}$C$^{16}$O, hereafter CO) emission of rotational low$-J$ transitions (e.g. $J=1-0$), is strictly necessary to indirectly observe this cold gaseous ISM phase \citep{SV2005,Bolatto2013}.

\begin{figure*}
\center
\includegraphics[width=2.0\columnwidth]{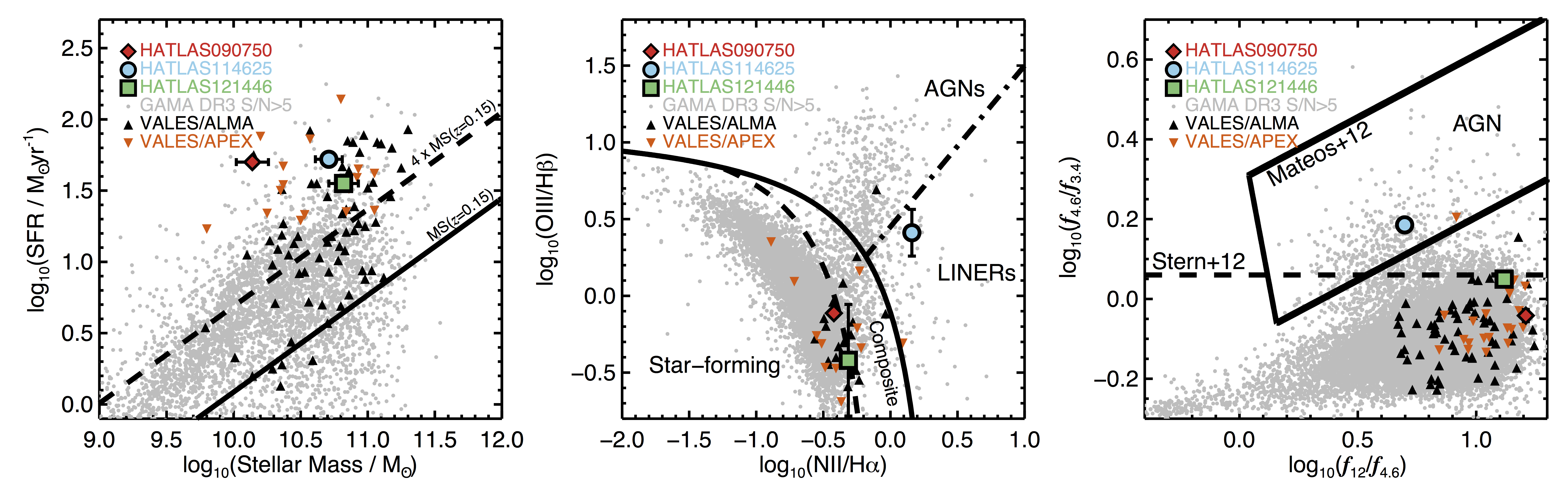}\\
\caption{ \label{fig:fig1}Characterization of the three galaxies presented in this work in terms of stellar mass, SFRs, and AGN activity. \textit{Left:} The SFR-$M_\star$ plane. The solid and dashed lines represent the main-sequence (MS) parametrization suggested by \citet{Whitaker2012} and the $4 \times$\,SFR(MS) starburst threshold, respectively. \textit{Middle:} The BPT-diagram \citep{Baldwin1981}. The dashed curve shows the empirical star-forming threshold \citep{Kauffmann2003}, whereas the solid curve corresponds to the theoretical maximum starburst model \citep{Kewley2001}. These two lines encompass the SFGs-AGN `composite' zone. The dotted-dashed line indicates the division between AGNs and LINERs \citep{Schawinski2007}. \textit{Right:} WISE mid-IR colour-colour diagram. The solid lines delimit the AGN-zone suggested by \citet{Mateos2012}, whereas the dashed line represents the AGN threshold adopted by \citet{Stern2012}. The WISE data 1-$\sigma$ errorbars are smaller than the plotted symbol sizes. The GAMA data are taken from their data-release 3 (GAMA-DR3; \citealt{Baldry2018}) encompassing galaxies at $z<0.35$ (the upper redshift limit for the VALES survey) and with 5-$\sigma$ or higher flux estimates. These three panels indicate that the three galaxies presented in this work can be classified as starbursts, with one target (HATLAS114625$-$014511) likely to be classified as an obscured AGN host galaxy.}
\end{figure*}

In this work, we introduce new detailed $\sim$\,kpc-scale morpho-kinematics observations toward three starburst galaxies taken from the Valpara\'iso ALMA/APEX Emission Line Survey (VALES; \citealt{Villanueva2017,Cheng2018}) at $z\sim0.12-0.18$. The VALES survey is designed to target low-$J$ CO emission line transitions in dusty galaxies extracted from the \textit{Herschel} Astrophysical Terahertz Large Area Survey ($H$-ATLAS; \citealt{Eales2010}). VALES extracts sources from the equatorial Galaxy And Mass Assembly (GAMA) fields \citep{Driver2016}, which present wide broad-band imaging and photometry in multiple bands sampling the galaxy Spectral Energy Distribution (SED) from far-ultraviolet (far-UV) to IR. The VALES survey covers the redshift range of 0.02$<z<$0.35, stellar masses ($M_\star$) from $\approx 6$ to $11\times 10^{10}$\,$M\odot$ and IR-luminosity range of $L_{8-1000\mu {\rm m}} \approx 10^{10-12}$\,$L_\odot$ (see \citealt{Villanueva2017} for more details).

We characterize the molecular gas morpho-kinematics by observing the CO($J=1-0$, $\nu_{\rm{rest}}=115.271$\,GHz) molecule by ALMA. These sub-mm observations are complemented by spatially-resolved seeing-limited ionized gas phase measurements taken by the Spectrograph for INtegral Field Observations in the Near Infrared (SINFONI) IFU located at the European Southern Observatory Very Large Telescope (ESO-VLT). The ionized gas ISM phase is traced by observing the nebular Paschen alpha (Pa$\alpha$) emission line ($\lambda_{\rm{rest}}=1.8751$\,$\mu$m). Our observations are one of the few that use the CO and Pa$\alpha$ emission lines to study the ISM dynamics in dusty starbursts.

We assume a $\Lambda$CDM cosmology with $\Omega_\Lambda=0.73$, $\Omega_{\rm m}=0.27$, and $H_0$=70\,km\,s$^{-1}$\,Mpc$^{-1}$. Thus, at a redshift range of $z=0.1-0.2$, a spatial resolution of 0$\farcs$6 corresponds to a physical scale between $1.0-1.8$\,kpc.

\section{Observations \& Data Reduction}

\subsection{The three targeted galaxies}
\label{sec:data}

We select three galaxies taken from the VALES survey at $z \approx 0.12 - 0.18$. These systems were selected based on their likelihood to be molecular gas-rich systems, i.e., with expected molecular gas fractions $f_{\rm H_2} \equiv M_{\rm H_2}/(M_{\rm H_2} + M_\star) > 0.3$ after assuming a Milky-way like CO-to-H$_2$ conversion factor $\alpha_{\rm CO, MW} = 4.6$\,$M_\odot$\,(K\,km\,s$^{-1}$\,pc$^{2}$)$^{-1}$ \citep{Bolatto2013}. Our `gas-rich' criterion takes into account two observational facts: (1)  the negligible cosmic evolution of $f_{\rm H_2}$ in the redshift range $z = 0 - 0.2$ \citep{Villanueva2017,Tacconi2018}; and (2) local galaxies have average molecular gas fractions of $\sim 0.1$ \citep{Leroy2009,Saintonge2017} with only a few of these presenting $f_{\rm H_2} > 0.3$ ($\approx 1$\,\% based on XCOLD\,GASS survey $M_{\rm H_2}$ measurements re-scaled by assuming $\alpha_{\rm CO, MW}$; \citealt{Saintonge2017}).

In Fig~\ref{fig:fig1}, we present the global properties for these three galaxies compared to full VALES and GAMA surveys. We adopt the star-forming galaxy (SFG) `main-sequence' parametrization suggested by \citet{Whitaker2012}. The main-sequence corresponds to the tight correlation between the galaxy stellar masses and star formation rates (SFRs). Our three targets are representative of the starburst galaxy population.

Using the Baldwin-Phillips-Terlevich (BPT) diagram \citep{Baldwin1981}, we show that two systems lie just below the limit of the pure star-forming region \citep{Kauffmann2003}. The remaining target (HATLAS114625$-$014511) is located in the low ionization nuclear emission line region (LINER). The H$\beta$, [O{\sc iii}], H$\alpha$ and [N{\sc ii}] flux measurements are presented in Appendix~\ref{sec:AppendixA}. By using the \textit{Wide-field Infrared Survey Explorer} (WISE; \citealt{Wright2010}) mid-IR colour diagram (right panel in Fig.~\ref{fig:fig1}; \citealt{Stern2012,Mateos2012}), HATLAS114625$-$014511 would be classified as an AGN host galaxy, while the other two targets are classified as SFGs in agreement with the BPT-diagram analysis.

\begin{table*}
\centering
\setlength\tabcolsep{4pt}
\caption{\label{table_alma} ALMA observational setup for project 2015.1.01012.S.
}
\small
\begin{tabular}{lcccccccc}
\hline
Source List & Observation & Flux & Bandpass & Phase  & P.W.V. & Number of & Time on  & $\theta_{\rm BMAJ}$\\
                   &  Date & Calibrator & Calibrator & Calibrator & (mm) & antennas  & Target (min) & (arcsec.)\\
\hline
HATLASJ114625.0$-$014511 \&  &   9 Aug.\ 2016  & J1229+0203 & J1229+0203 & J1150$-$0023 & 0.80 & 36        & 35 & $0\farcs52$\\
HATLASJ121446.4$-$011155  &  11 Aug.\ 2016  & J1229+0203 & J1229+0203 & J1150$-$0023 & 0.80 & 38        & 35 & $0\farcs50$\\
\hline
HATLASJ090750.0+010141     &  13 Aug.\ 2016  & J0854+2006     & J0854+2006 & J0909+0121 & 0.63  & 36        & 35 & $0\farcs45$\\
\hline
\end{tabular}
\end{table*}

\subsection{ALMA observations}
\label{sec:alma_obs}

In this work, we describe an ALMA follow up campaign (taken from project 2015.1.01012.S; P.I.: E.\ Ibar) for imaging three VALES galaxies for which we obtained previous bright CO(1-0) detections presented in \citet{Villanueva2017}. Observations were taken on Band-3 with the extended 12\,m array to obtain higher spatial and spectral resolution imaging than previous observations. 

The spectral setup was designed to target the redshifted CO(1-0) emission line (between 97\,GHz and 103\,GHz, depending on the source) using a spectral window in Frequency Division Mode to cover 1.875\,GHz of bandwidth at a native 3906.250\,kHz resolution. The other three spectral windows were used in Time Division Mode and were positioned to measure the continuum emission around the redshifted line. Observations were taken under relatively good weather conditions with precipitable water vapour
(P.W.V.) ranging from 0.6\,mm to 0.8\,mm, and using 36 to 38 antennas with a maximum baseline of 1.5\,km. The phase, bandpass and flux calibrations are listed in Table~\ref{table_alma}.

Data reduction was carried out using the Common Astronomy Software Applications (\textsc{CASA}) and using the provided ALMA pipeline up to calibrated $uv$ products. Data taken in different days were concatenated together after running the pipeline and before imaging. After exploring different imaging approaches using task {\sc tclean}, and guided by our scientific objectives, we decided to use a Briggs weighting ({\sc  robust=0.4}) to reach a major axis full width half maximum for the synthesized beam ($\theta_{\rm BMAJ}$) in a range between $0\farcs45$--$0\farcs52$. For each source, we apply a slight convolution (within {\sc tclean}) to obtain a circular beam. The pixel size is set to $0\farcs1$. All three sources are clearly detected at high significance, and the signal was interactively cleaned down to 2--3-$\sigma$ in spectral channels with confident source emission. 

Final images reach r.m.s. noises of $\sim$\,400--500$\mu$Jy\,beam$^{-1}$ at $\approx 12$\,km\,s$^{-1}$ channel width. The channel width is set to minimize spectral resolution effects \citep{Molina2019a}. The continuum emission image, obtained over 6\,GHz bandwidth reaches noise levels of 13$\mu$Jy\,beam$^{-1}$. Two targets are detected as point sources with peak flux densities of $\sim$\,110$\mu$Jy\,beam$^{-1}$, while HATLASJ121446.4$-$011155 remains undetected.

\begin{table*}[!h]
	\centering
    	\caption{\label{tab:table1} Spatially-integrated measurements for the three starbursts. The far-IR luminosities are calculated across the rest-frame 8--1000\,${\mu}$m wavelength range. $E(B-V)_{\rm Neb}$ is the colour excess estimated by using the observed H$\alpha$-to-Pa$\alpha$ flux ratio. SFR$_{\rm{Pa\alpha}}$ and SFR$_{\rm{Pa\alpha, corr}}$ correspond to the observed and attenuation-corrected Pa$\alpha$-based SFR estimates, respectively. $S_{\rm CO} \Delta v$ is the velocity integrated flux density. $L'_{\rm CO}$ is the CO(1-0) line luminosity taken from \citet{Villanueva2017}.}
    	\vspace{1mm}
	\begin{tabular}{lccc} 
		\hline
		\hline
	 	& HATLASJ090750.0+010141 & HATLASJ114625.4$-$014511 & HATLASJ121446.0$-$011155 \\
		 \hline
		RA (J2000) & 09:07:50.07 & 11:46:25.01 & 12:14:46.47 \\ 
		Dec (J2000) & +01:01:41.47 & $-$01:45:12.81 & $-$01:11:55.55 \\ 
		$z_{\rm spec}$ & 0.12834 & 0.16553 & 0.17981 \\
		$M_\star$ ($\times$\,$10^{10}$\,$M_{\odot}$) & 1.4$\pm$0.4 &  5.1$\pm$1.2 & 6.6$\pm$1.7 \\  
		$L_{\rm{IR}}$ ($\times$\,$10^{10}$\,$L_{\odot}$) & 50$\pm$1 & 53$\pm$1 & 35$\pm$1 \\
		SFR$_{\rm{IR}}$ & 50$\pm$1 & 53$\pm$1 & 35$\pm$1 \\
		f$_{\rm{Pa\alpha}}$  ($\times$\,$10^{-17}$\,erg\,s$^{-1}$\,cm$^{-2}$) & -- & 1069$\pm$111 & 644$\pm$96 \\ 
		$E(B-V)_{\rm Neb}$ & -- & 1.35$\pm$0.05 & 0.91$\pm$0.06 \\
		SFR$_{\rm{Pa\alpha}}$ & -- & 32$\pm$4 & 25$\pm$4 \\
		SFR$_{\rm{Pa\alpha, corr}}$ & -- & 67$\pm$8 & 40$\pm$6 \\
		$S_{\rm CO} \Delta v$ (Jy\,km\,s$^{-1}$) & 6.8$\pm$0.6 &  6.6$\pm$0.6 &  4.6$\pm$0.6 \\
		$L'_{\rm{CO}}$ ($\times$\,$10^{9}$\,K\,km\,s$^{-1}$\,pc$^2$) & 5.4$\pm$0.5 & 8.6$\pm$0.8 & 7.3$\pm$0.9 \\      
		\hline                                                                                        
	\end{tabular}
\end{table*}

\subsection{SINFONI observations}
\label{sec:sinfoni_obs}
We observe the Pa$\alpha$ emission line by using the SINFONI IFU \citep{Eisenhauer2003} on the ESO-VLT in its seeing-limited mode (Project 099.B-0479(A); P.I.\,J.Molina). The SINFONI field-of-view (FOV) is $8''\times8''$ with a pixel angular size of $0\farcs 125$. The spectral resolution is $\lambda/\Delta\lambda\sim 3800$, and OH sky-lines have $\sim 5$\,\r{A} full width at half maximum -- FWHM ($\approx 30$km\,s$^{-1}$ at 2.1$\mu$m). The observations were carried out in service mode between 2017 March 15 and 2017 December 11 in seeing and photometric conditions (point spread function -- PSF FWHM $\approx 0\farcs4$--$0\farcs8$ in $K$-band). In addition, two different jittering patterns were used during the observing runs in order to boost the observation signal-to-noise ratio (S/N) in one galaxy.

\subsubsection{`OSSO' Jittering}
To observe the HATLASJ1146251$-$014511 and HATLASJ121446.4$-$011155 galaxies (hereafter, HATLAS114625 and HATLAS121446, respectively), we used the traditional ABBA chop sequences, nodding $16\farcs 0$ across the IFU. That means that the traditional jittering OBJECT-SKY-SKY-OBJECT (`OSSO') pattern was implemented. We used one observing block (OB) per target, implying a total observing time of $\approx$3.2\,ks per source. The raw datasets for these two sources were reduced by using the standard SINFONI \textsc{ESOREX}\footnote{http://www.eso.org/sci/software/pipelines/} data reduction pipeline. 

\subsubsection{`OOOO' Jittering}
We perform an on-source experimental jittering pattern increase the S/N of the Pa$\alpha$ emission line in one galaxy. In this experimental observation, the pointing was kept fixed at the galaxy location. Thus, an OBJECT-OBJECT-OBJECT-OBJECT (`OOOO') jitter sequence was used. Based on previous analyses by Godoy et al. (in prep), this observing approach provides reliable results for emission line with S/N$\gtrsim 15$.

To reduce the data, first, we use the SINFONI \textsc{esoreflex} and \textsc{esorex} pipelines. Then, sky emission lines are subtracted using \textsc{SkyCor} \citep{Noll2014}, while \textsc{Molecfit} \citep{Kausch2015} is implemented to remove telluric absorption bandpass lines (Godoy et al. in prep.). This is necessary as we do not have `sky' observations.

To test this experimental jitter pattern, we choose the brightest galaxy in our small sample, HATLASJ090750.0+010141 (hereafter, HATLAS090750). By using this method, the observed emission line S/N is expected to increase by $\sim \sqrt{2}$ compared to the use of an `OSSO' jitter pattern due to the extra on-source time. For this observation, the exposure time was also set to $\approx 3.2$\,ks. More details about this experimental observation are reported in Appendix~\ref{sec:Jitter-results}.

\subsubsection{Flux calibration} 
The standard star observation is used to perform the flux calibration. First, the galaxy spectrum is corrected in each pixel by atmospheric telluric absorptions and by the SINFONI $K$-band transmission curve. We do this by collapsing the standard star datacube in the spectral axis using a wavelength range free from significant telluric absorptions. A two-dimensional Gaussian function is fitted to this spectrally-collapsed image. Then, we extract the spectrum from the standard star by using an aperture size of $2 \times$\,FWHM in diameter. We use this standard star spectrum to normalize the galaxy spectrum observed in each pixel. We take into account the different total exposure times.

Then, in each pixel, we multiply the normalized spectrum by a representative stellar black-body profile. To obtain this black-body curve, we fit a black-body function to the standard star magnitudes collated in the Visual Observatory SED Analyser (VOSA, \citealt{Bayo2008}). This allows us to estimate the stellar surface temperature -- thus the black-body function shape -- and the normalization constant to construct the representative standard stellar black-body profile as seen in the SINFONI \textit{K}-band. We note that the typical relative uncertainty for the conversion factor is $\sim$5\% (e.g. \citealt{Piqueras2012}).

Even though we can provide reliable flux calibrations for HATLAS114625 and HATLAS121446, the different on-source ('OOOO') observing mode for HATLAS090750 impeded a proper calibration from its standard star observation. The flux calibration for this observation requires us to carefully model the sky for the standard star observation and, hence the stellar spectrum. However, we were unable to obtain an accurate stellar atmospheric model for the standard star (HD\,56006) due to its uncertain stellar parameters. More details about these uncertainties are presented in Appendix~\ref{sec:Jitter-results}.

\subsubsection{Spatial resolution} 
We also use the spectrally-collapsed standard star image to determine the PSF FWHM ($\theta_{\rm PSF}$) for each $K$-band observation. By fitting a two-dimensional Gaussian function, we determine $\theta_{\rm PSF} \approx 0\farcs62$, $0\farcs39$ and $0\farcs81$ for HATLAS090750, HATLAS114625 and HATLAS121446, respectively.

\subsection{Stellar Mass and IR-based SFR estimates}
\label{sec:M_star-SFR_FIR}
The stellar masses for the three galaxies were estimated in \citet{Villanueva2017} by using the photometry provided by the GAMA survey (extending from the far-UV to FIR -- $\sim0.1-500$\,$\mu$m) and by using the Bayesian SED fitting code \textsc{magphys} \citep{Dacunha2008}. We assume a \citet{Chabrier2003} initial mass function (IMF). The $M_\star$ values are presented in Table~\ref{tab:table1}.

The IR-based SFRs (SFR$_{\rm FIR}$) are estimated by using the rest-frame far-IR 8--1000\,$\mu$m luminosity ($L_{\rm{IR}}$) estimates taken from \citet{Ibar2015}. By assuming a \citet{Chabrier2003} IMF, the SFR$_{\rm IR}$ values are calculated following SFR$_{\rm IR}\,(M_\odot$\,yr$^{-1})= $\,$10^{-10}\times$\,$L_{\rm IR}$ ($L_\odot$; \citealt{Kennicutt1998b}) and correspond to the obscured star-formation activity. The IR-based SFRs are consistent with the SFR estimates suggested by \textsc{magphys} but tend to be offset by a factor of $\sim 2$ toward higher values (see \citealt{Villanueva2017} for more details). 

\subsection{CO(1-0) Luminosities}
\label{sec:L_CO}
The total galaxy CO(1-0) velocity-integrated flux densities ($S_{\rm CO(1-0)} \Delta v$) are taken from \citet{Villanueva2017}. Briefly, these were estimated by implementing a two-step procedure. First, the CO(1-0) line is spectrally fitted by a Gaussian profile to determine its FWHM and spectrally-collapse the datacube within $\pm 1 \times$\,FWHM. Then, $S_{\rm CO(1-0)} \Delta v$ values are estimated by fitting a two-dimensional Gaussian function to the spectrally-integrated datacube (moment 0) using the task \textsc{gaussfit} within \textsc{casa}. Finally, the CO(1-0) luminosities ($L'_{\rm CO(1-0)}$) are calculated by following \citet{SV2005};

\begin{equation}
\label{eq:LCO}
L'_{\rm CO(1-0)} = 3.25 \times 10^7\,S_{\rm CO(1-0)} \Delta v\, \nu_{\rm obs}^{-2}\, D^2_{\rm L}\, (1+z)^{-3}\, {\rm [K\,km\,s^{-1}\,pc^2]}, 
\end{equation}

\noindent where $S_{\rm CO(1-0)} \Delta v$ is in Jy\,km\,s$^{-1}$, $\nu_{\rm obs}$ is the observed frequency of the emission line in GHz, $D_{\rm L}$ is the luminosity distance in Mpc, and $z$ is the redshift. Both estimates are presented in Table~\ref{tab:table1}.

\section{ANALYSIS and RESULTS} 

\subsection{Average ISM properties}
\label{sec:ave_ISM}

\begin{figure}
\flushleft
\includegraphics[width=0.95\columnwidth]{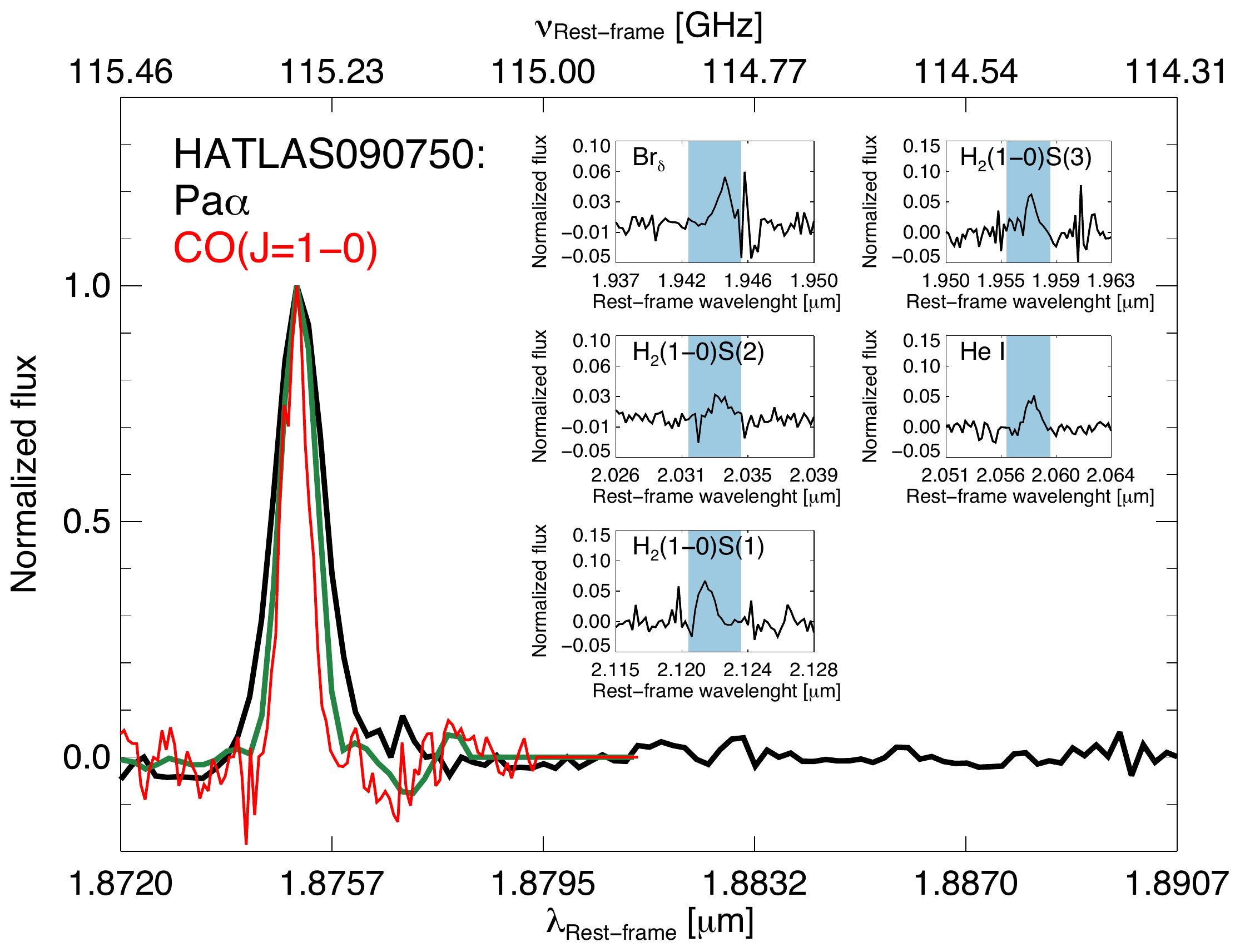}\\
\includegraphics[width=0.95\columnwidth]{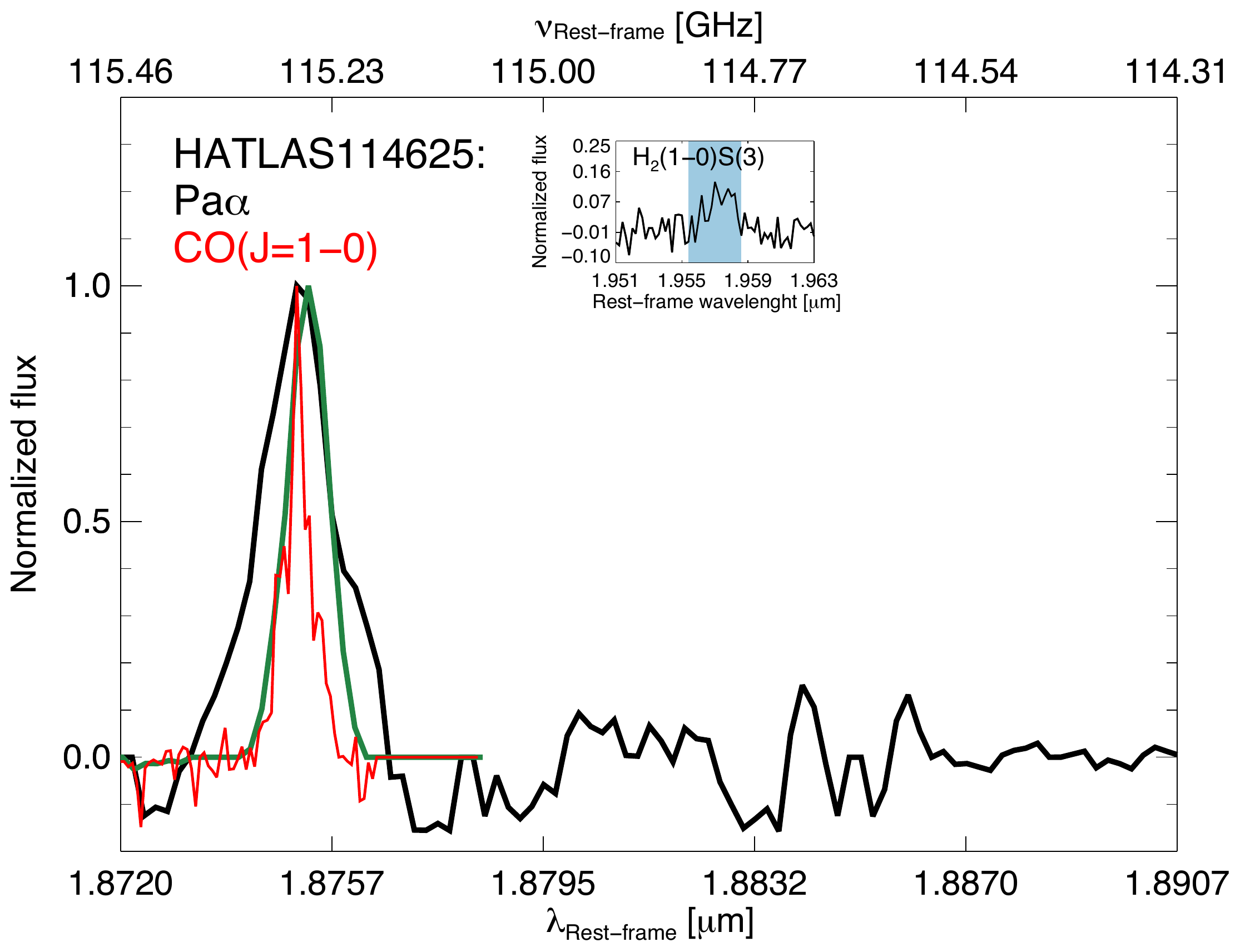}\\ 
\includegraphics[width=0.95\columnwidth]{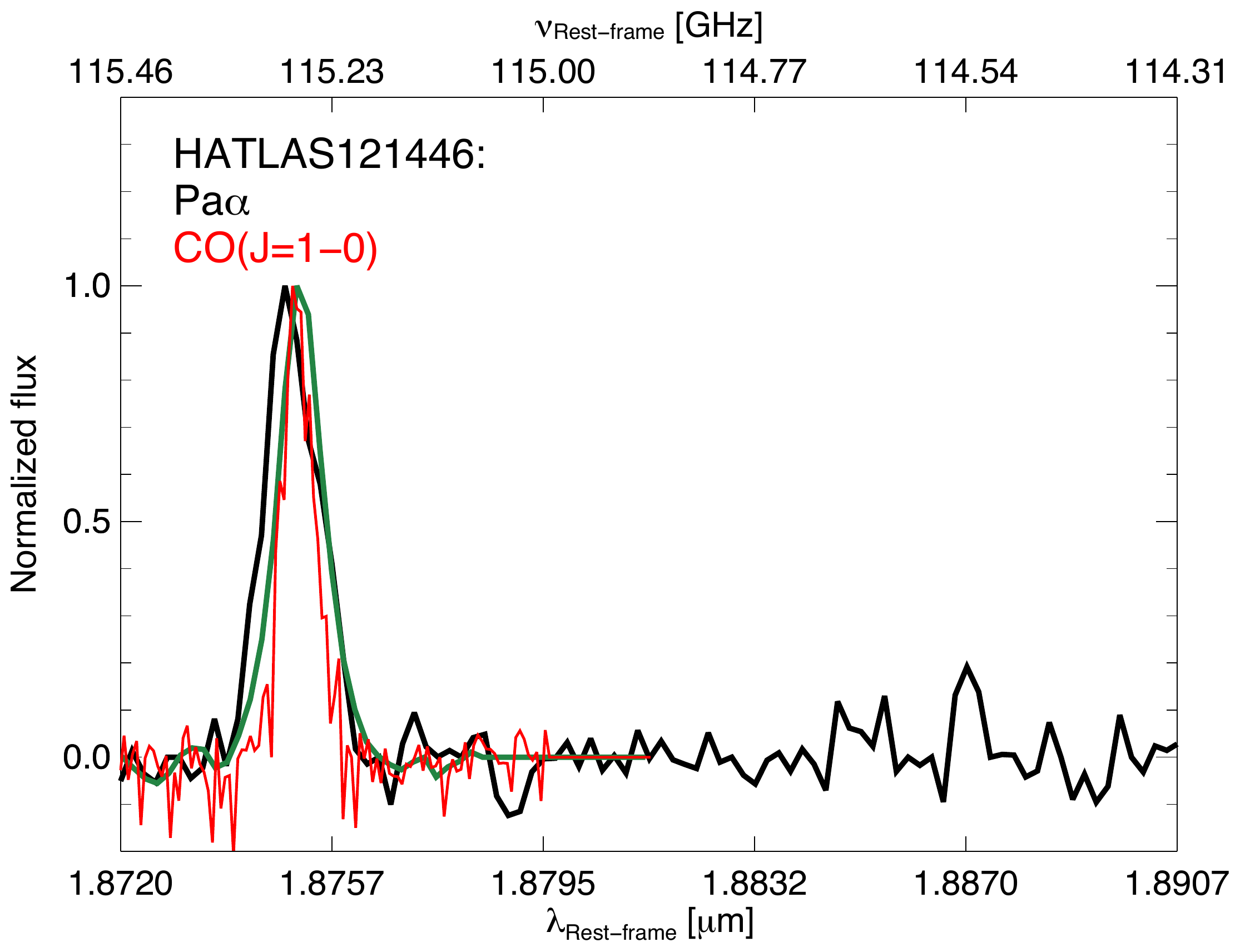}\\
\caption{\label{fig:1d_spectrum} Spatially-integrated rest-frame spectra around the emission lines of interest. The bottom and top x-axes show the rest-frame wavelength and frequency ranges for the Pa$\alpha$ and CO(1-0) emission lines, respectively. For each galaxy, the solid green curve shows the CO(1-0) spectrum convoluted by the SINFONI LSF. From HATLAS090750, we also detect the Br$_\delta$, H$_2$(1-0)S(3), H$_2$(1-0)S(2), H$_2$(1-0)S(1) and He\,{\sc i} near-IR emission lines using as an aperture an encircled zone given by the PSF FWHM and centred at the Pa$\alpha$ luminosity peak. In the case of the HATLAS114625 galaxy observation, we also detect the H$_2$(1-0)S(3) emission. These detections are shown in the sub-plots (blue-shaded area) in each panel (see also Appendix~\ref{sec:NIR-lines}). The CO(1-0) and Pa$\alpha$ emission lines are clearly detected.}
\end{figure} 

To analyse the spatially-integrated emission line fluxes for our three galaxies, first, we collapse the new ALMA and SINFONI datacubes into one-dimensional spectra (Fig~\ref{fig:1d_spectrum}). These spectra were built by stacking the spectrum seen in the individual pixels from which we detected an emission line (see \S~\ref{sec:galaxy_dyn}). Before stacking, we manually shifted the individual emission lines to rest-frame accounting for redshift and the respective pixel line-of-sight (LOS) velocity value (see Fig,~\ref{fig:maps}). Thus, we try to minimize any line broadening produced by rotational motions and we focus on intrinsic individual emission line widths.

In all the three starbursts, the spatially-integrated Pa$\alpha$ emission line seems broader than the CO(1-0) emission line. By convolving the ALMA spatially-integrated spectrum by the SINFONI line spread function (LSF; green curves in Fig.~\ref{fig:1d_spectrum}), we find that the spectral resolution difference is not producing this trend. The difference between the spatially-integrated Pa$\alpha$ and CO(1-0) line widths seems to be caused by broader nuclear Pa$\alpha$ emission lines in the individual pixels in each galaxy (see \S~\ref{sec:kin_pars}). The broad nuclear Pa$\alpha$ emission lines indicate that the ionized gas ISM phase is more affected by turbulent supersonic motions than the molecular gas\footnote{For a typical H{\sc ii} region with a temperature of 10$^4$\,K, we expect a Pa$\alpha$ thermal broadening of $\sim 20$km\,s$^{-1}$. For the molecular gas ISM phase with a temperature of $\lesssim 200$\,K, we expect thermally-broadened CO line widths $\lesssim 0.5$km\,s$^{-1}$.}. We do not detect any broad-line component ($>$\,500\,km\,s$^{-1}$) in the spatially-collapsed SINFONI spectra, suggesting the absence of signatures from a broad line region produced by an active galactic nucleus (AGN).

We use the Pa$\alpha$ emission line fluxes to derive SFR estimates (less affected by attenuation compared to H$\alpha$) using the \citet{Kennicutt1998b}'s conversion for the \citet{Chabrier2003} IMF. By assuming an intrinsic H$\alpha$-to-Pa$\alpha$ ratio equal to 0.116 (Case B recombination, \citealt{Osterbrock2006}), the Pa$\alpha$-based SFRs (SFR$_{\rm Pa \alpha}$) are calculated following SFR$_{\rm Pa \alpha}(M_\odot$\,yr$^{-1}) = 4.0 \times 10^{-41}\times$\,$L_{\rm Pa\alpha}$(erg\,s$^{-1}$). The SFR$_{\rm Pa \alpha}$ values are presented in Table~\ref{tab:table1}. We do not present an SFR$_{\rm Pa \alpha}$ estimate for the HATLAS090750 galaxy as we were unable to obtain a reliable flux calibration for its SINFONI observation.

We compute the nebular $E(B-V)$ colour excess ($E(B-V)_{\rm Neb}$) by using the observed H$\alpha$-to-Pa$\alpha$ flux ratio\footnote{The H$\alpha$ flux estimates are taken from the GAMA survey DR3 (see Table~\ref{tab:appA}).} and assuming a \citet{Calzetti2000} attenuation law. We list the $E(B-V)_{\rm Neb}$ values in Table~\ref{tab:table1}. We note that these $E(B-V)_{\rm Neb}$ values are $\sim 4.7$ and $\sim 2.3$ times higher than the colour excess estimates given by \textsc{magphys} for the stellar component ($E(B-V)_\star \approx$0.29 and $\approx$0.39 for HATLAS114625 and HATLAS121446, respectively). This is expected from local galaxy studies, where the higher $E(B-V)_{\rm Neb}$ values suggest a differential attenuation model in which stars experience attenuation from a diffuse ISM dust component, but the massive young stars experience an additional attenuation as they are embedded in their dusty birth clouds \citep{Calzetti2000}. However, we note that the HATLAS114625's nebular-to-stellar colour excess ratio is twice than the average value found in local galaxies ($\sim 2.3$, \citealt{Calzetti2000}), indicating its highly dusty nature and more in line with the findings of an extreme obscured starburst galaxy population at $z \sim 0.5-0.9$ \citep{Calabro2018}.

By considering the derived $E(B-V)_{\rm Neb}$ values, we estimate attenuation-corrected SFR$_{\rm Pa \alpha}$ (SFR$_{\rm Pa \alpha , corr}$) values of $67\pm8$ and $40\pm6$\,$M_\odot$\,yr$^{-1}$ for HATLAS114625 and HATLAS121446, respectively. These estimates are slightly higher than the SFR$_{\rm FIR}$ values (Table~\ref{tab:table1}), but still consistent with the 2-$\sigma$ uncertainties for both starbursts.

\subsection{Galaxy Dynamics}
\label{sec:galaxy_dyn}

We construct the two-dimensional moment maps by following \citet{Swinbank2012a}. Briefly, the spectrum associated with each pixel corresponds to the average spectrum calculated from the pixels inside a square area that contains the spatial resolution element -- the synthesized beam or PSF. The noise per spectral channel is estimated from a region that does not contain any source emission. We use the \textsc{lmfit} \textsc{Python} package \citep{Newville2014} to fit a Gaussian profile to the emission lines. In the case of the SINFONI observations, we mask the spectrum at the wavelength ranges where OH sky-line features are present and the Pa$\alpha$ line widths are corrected by spectral resolution effects.

We apply an S/N\,$=5$ threshold to determine whether we have detected an emission line or not. If this criterion is not achieved, then we increase the square binned area by one pixel per side and repeat the Gaussian fit again. We iterate up to two more times in order to avoid large binned regions. After the third iteration, if the S/N criterion has not been achieved, we mask that pixel and skip to the next one. 

\begin{figure*}[!h] 
    \center
    \begin{tabular}[t]{cc}
\vspace{\jmvspace}
\begin{subfigure}{\jmbsize}
    \centering
    \tikz[overlay, remember picture] \node[anchor=south, inner sep=-5cm] (090bb) {\includegraphics[width=\jmbsize]{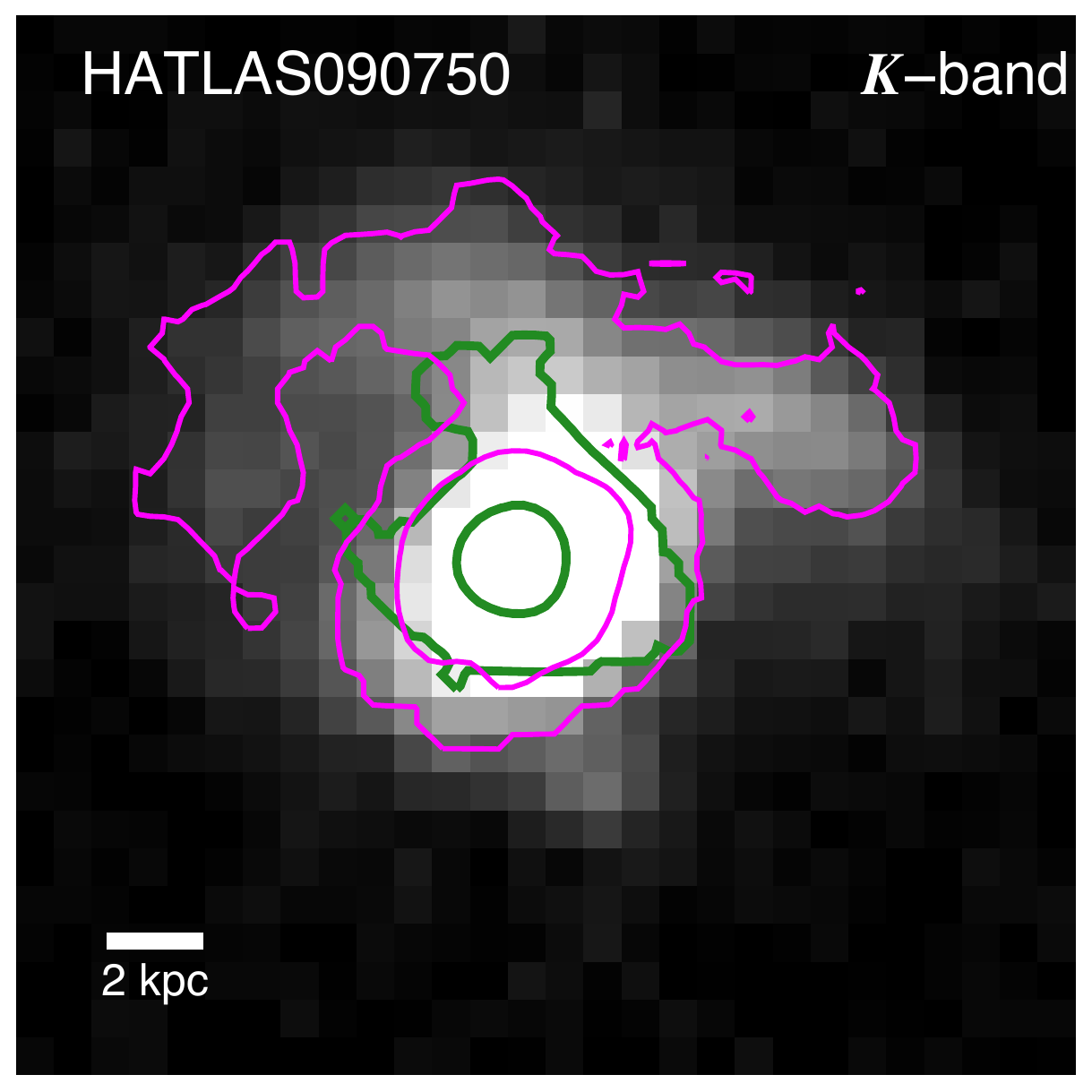}};
\end{subfigure}
    &
        \begin{tabular}{cccc}
        	\hspace{\jmhspace}
            \begin{subfigure}[t]{\jmtsize}
                \centering
                 \tikz[overlay, remember picture] \node[anchor=south, inner sep=0cm] (090pa) {\includegraphics[width=\jmtsize]{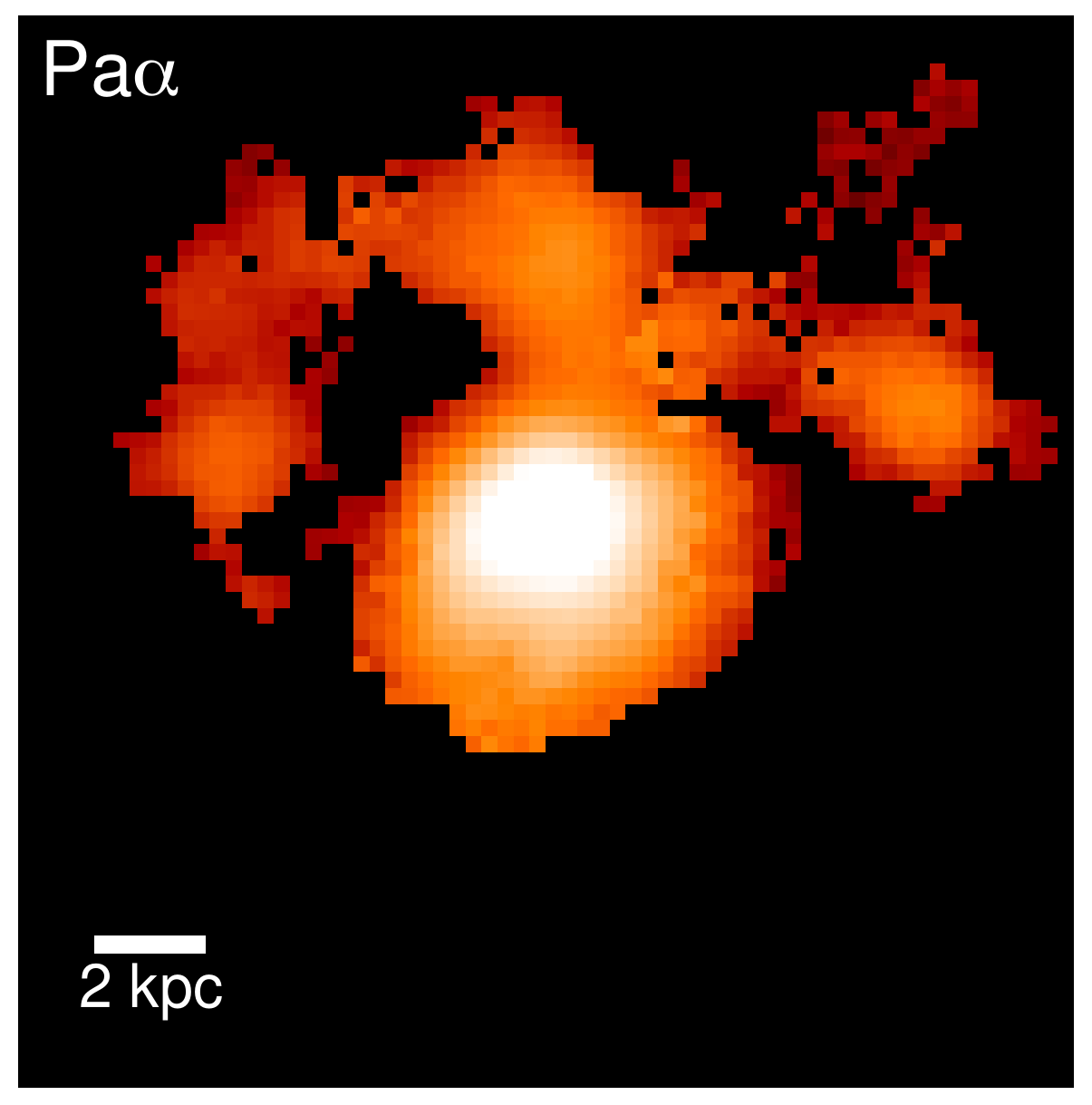}};
            \end{subfigure}
        	\hspace{\jmhspace}
            \begin{subfigure}[t]{\jmtsize}
                \flushleft
                \includegraphics[width=\jmtsize]{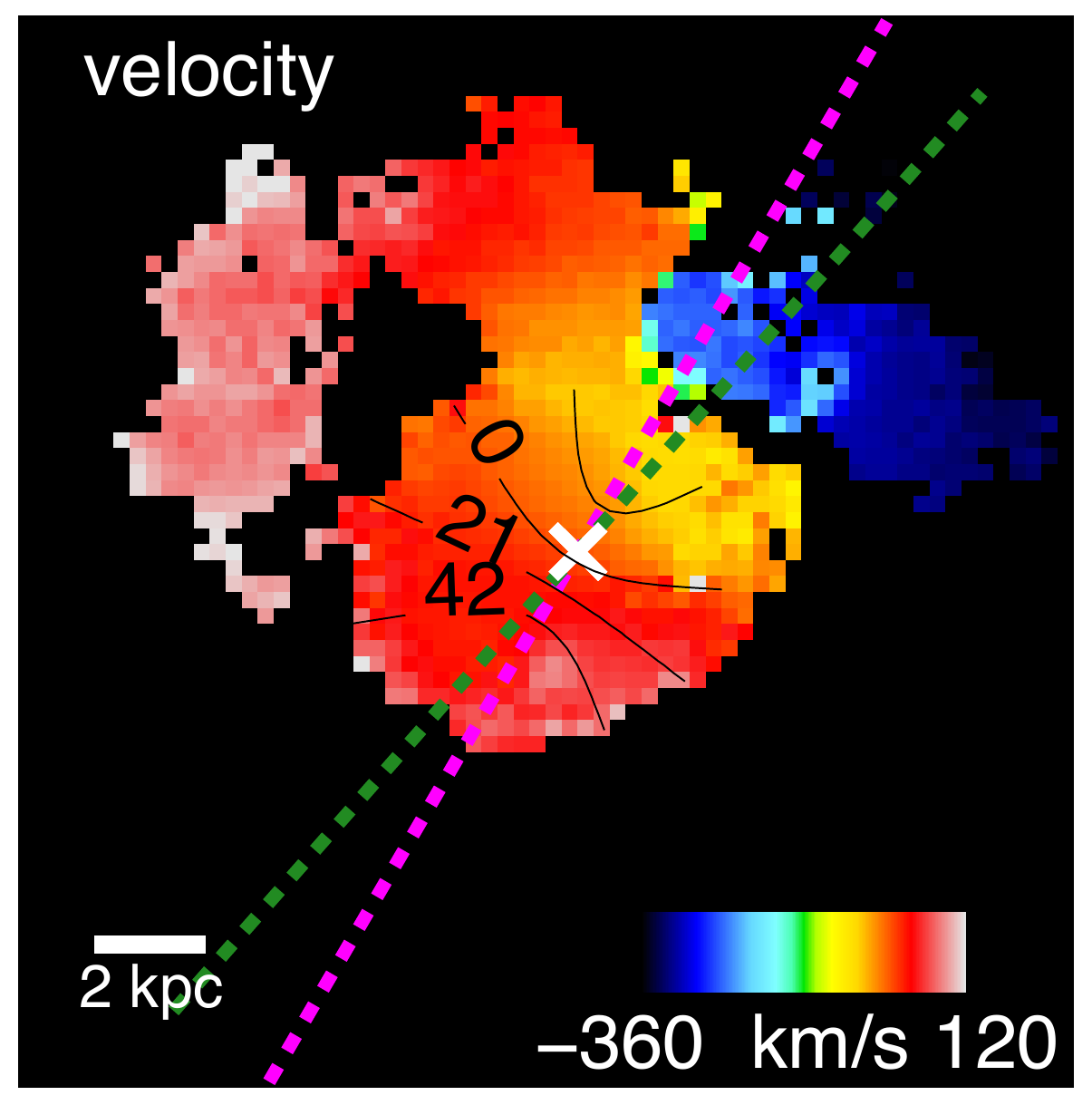}
            \end{subfigure}
        	\hspace{\jmhspace}
            \begin{subfigure}[t]{\jmtsize}
                \flushleft
                \includegraphics[width=\jmtsize]{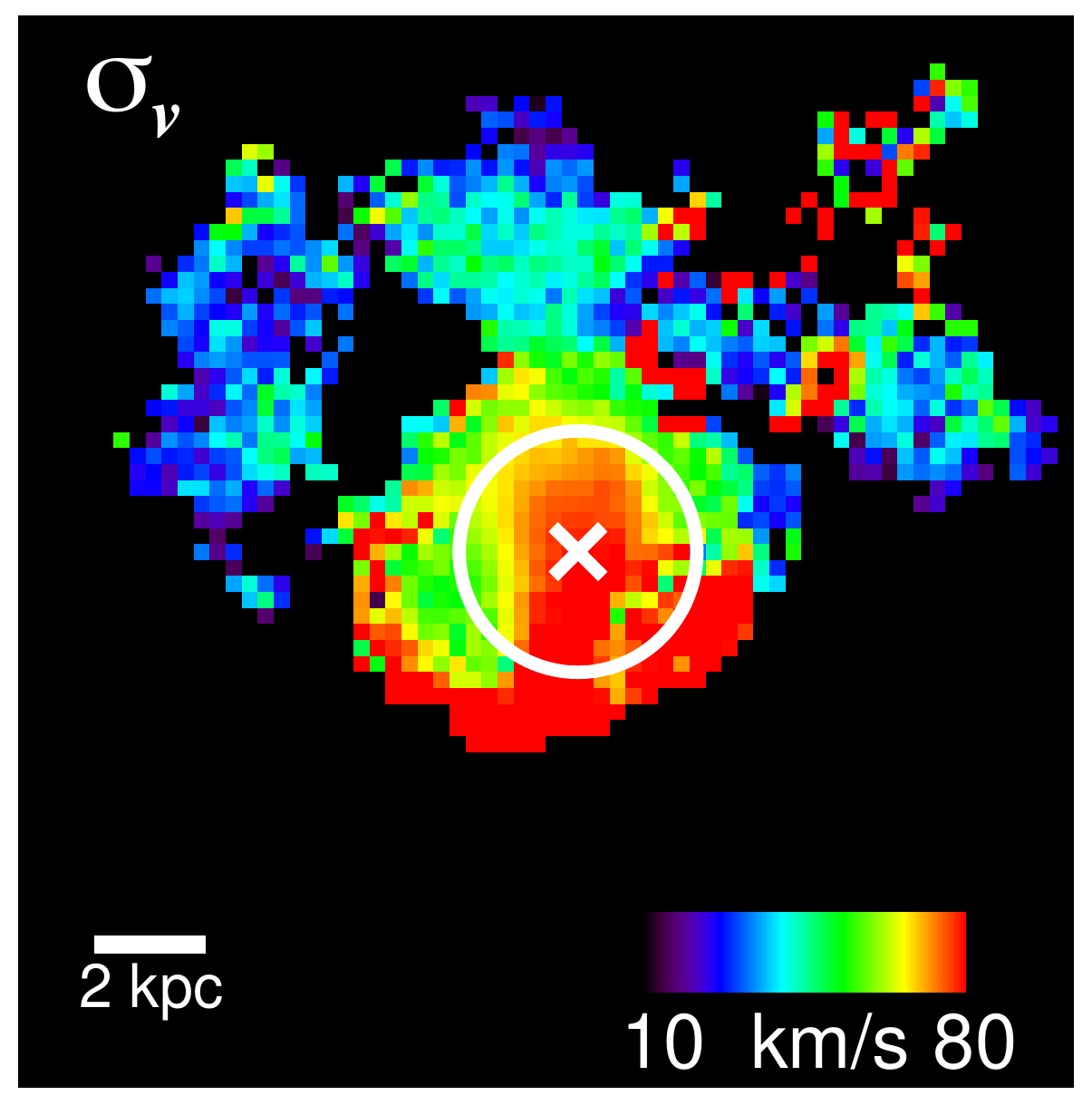}
            \end{subfigure}
        	\hspace{\jmhspace}
            \begin{subfigure}[t]{\jmtsize}
                \flushleft
                \includegraphics[width=\jmtsize]{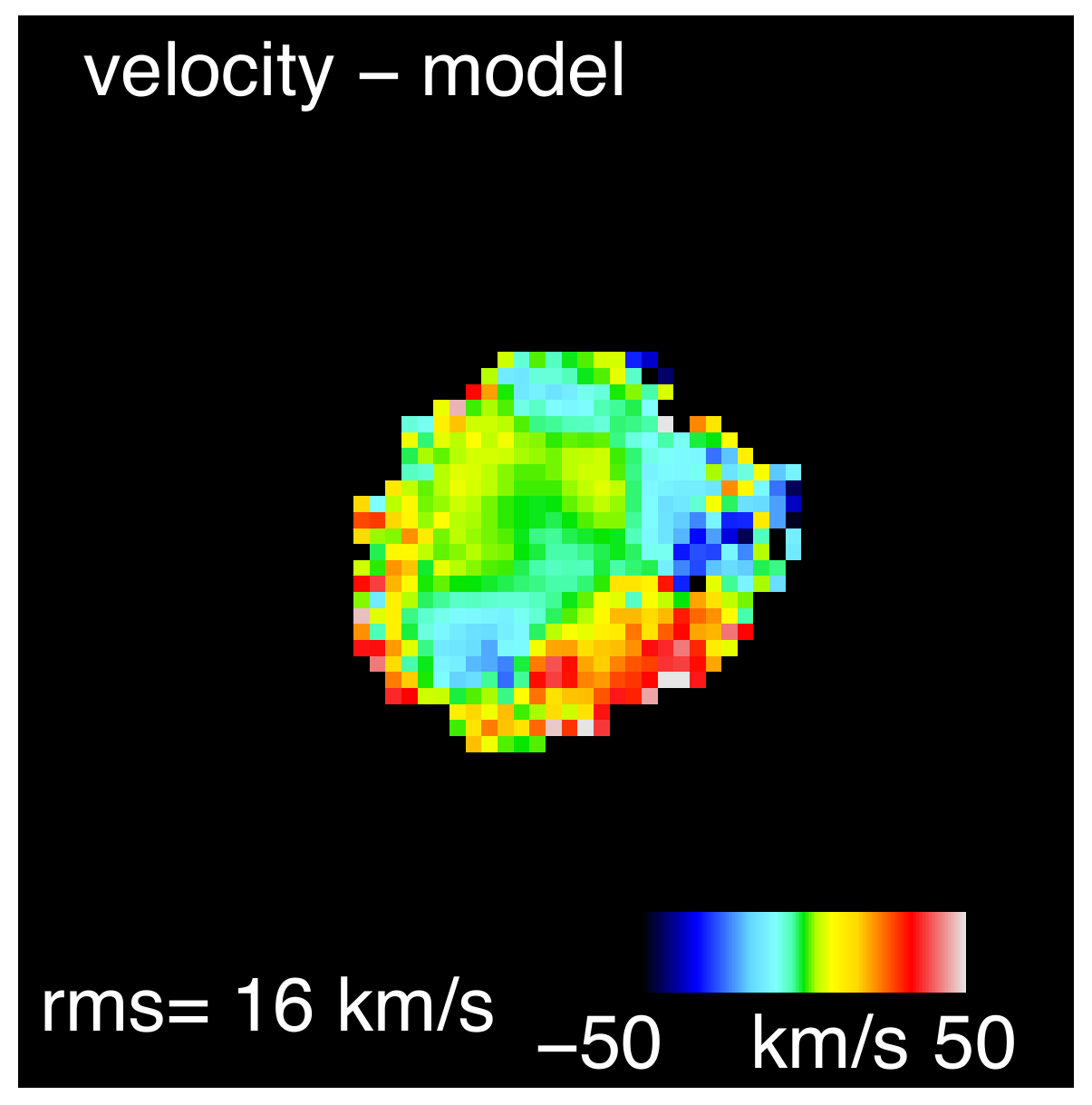}
            \end{subfigure}\\
        	\hspace{\jmhspace}
                \begin{subfigure}[t]{\jmtsize}
                \centering
                \tikz[overlay, remember picture] \node[anchor=south, inner sep=0cm] (090co) {\includegraphics[width=\jmtsize]{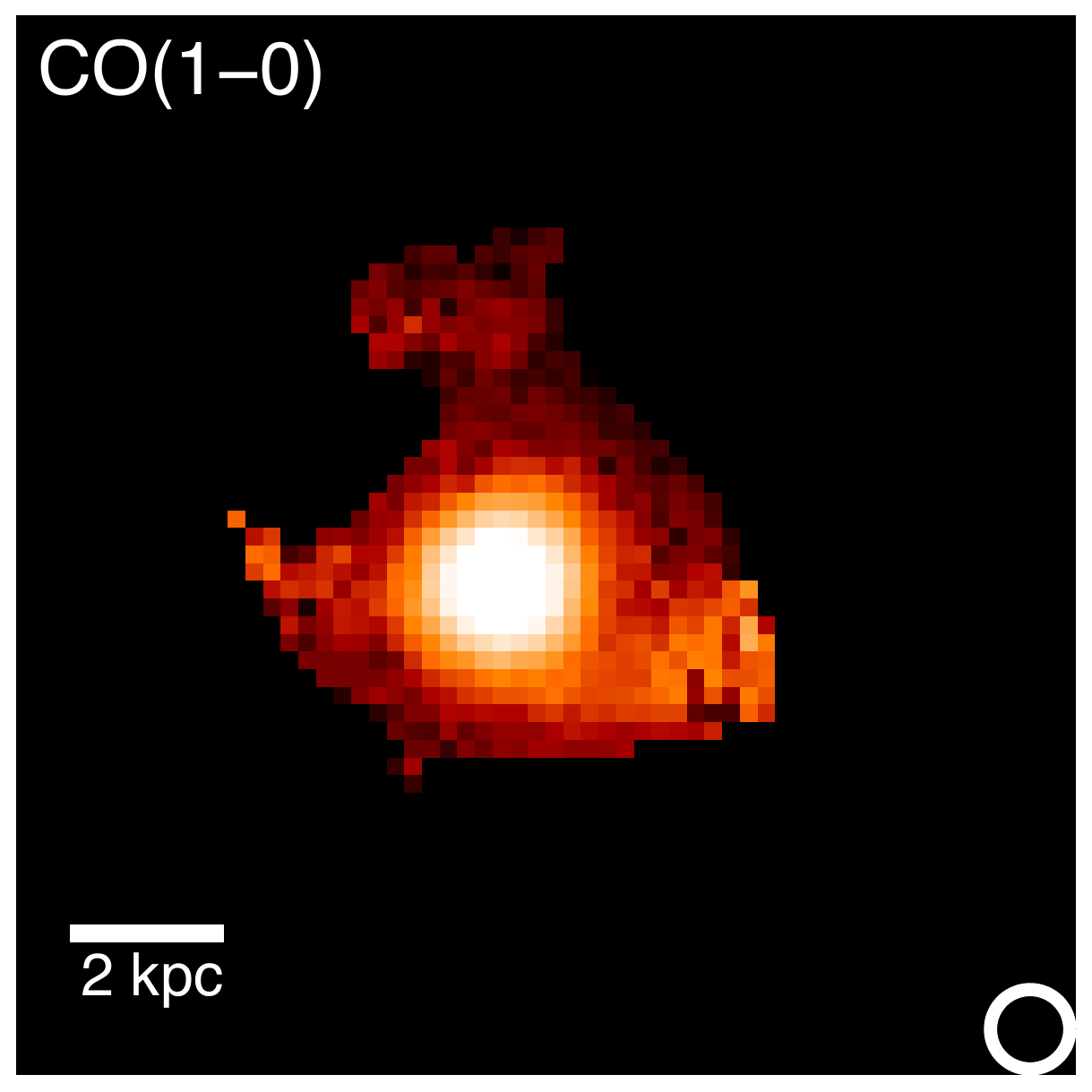}};
            \end{subfigure}
        	\hspace{\jmhspace}
            \begin{subfigure}[t]{\jmtsize}
                \flushleft
                \includegraphics[width=\jmtsize]{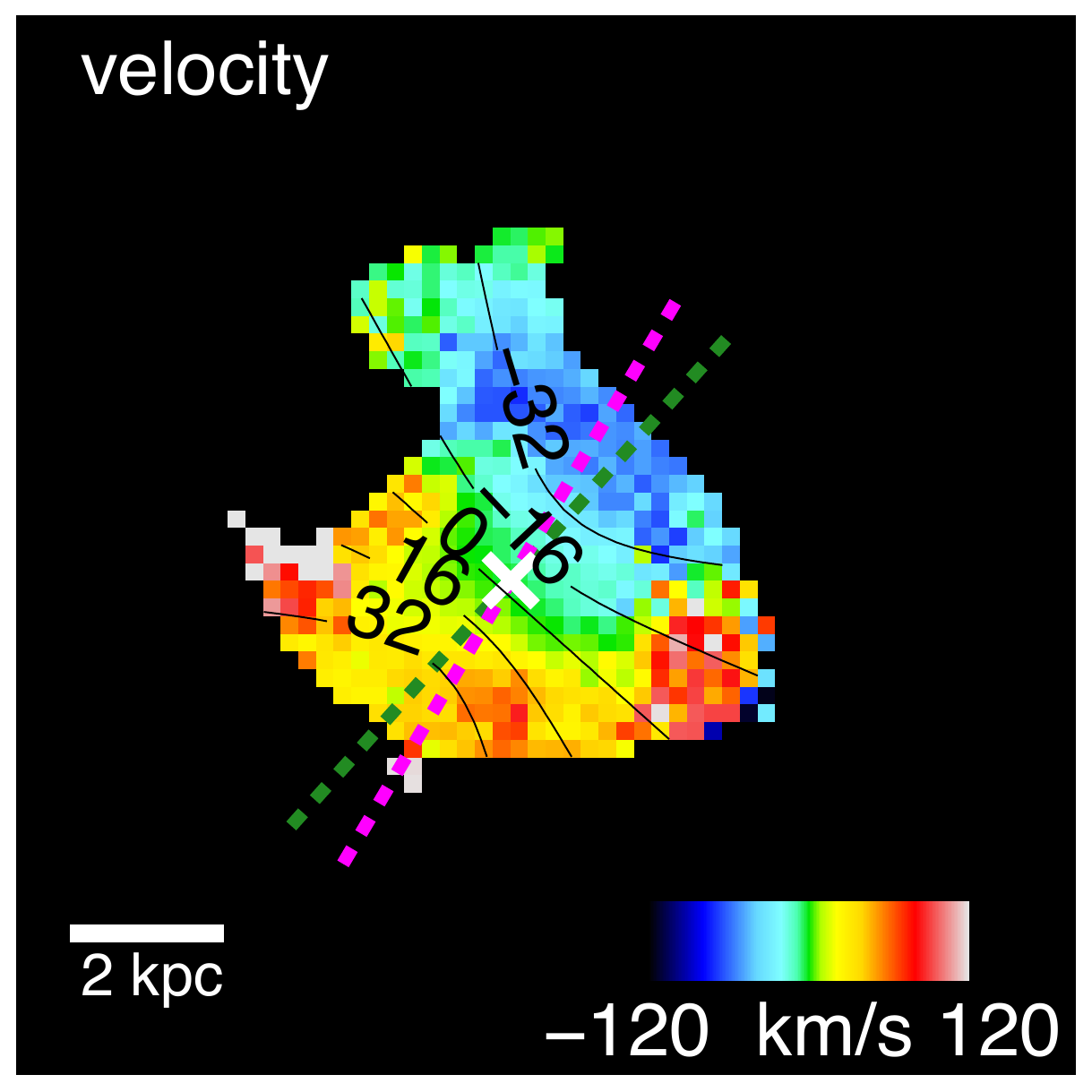}
            \end{subfigure}
        	\hspace{\jmhspace}
            \begin{subfigure}[t]{\jmtsize}
                \flushleft
                \includegraphics[width=\jmtsize]{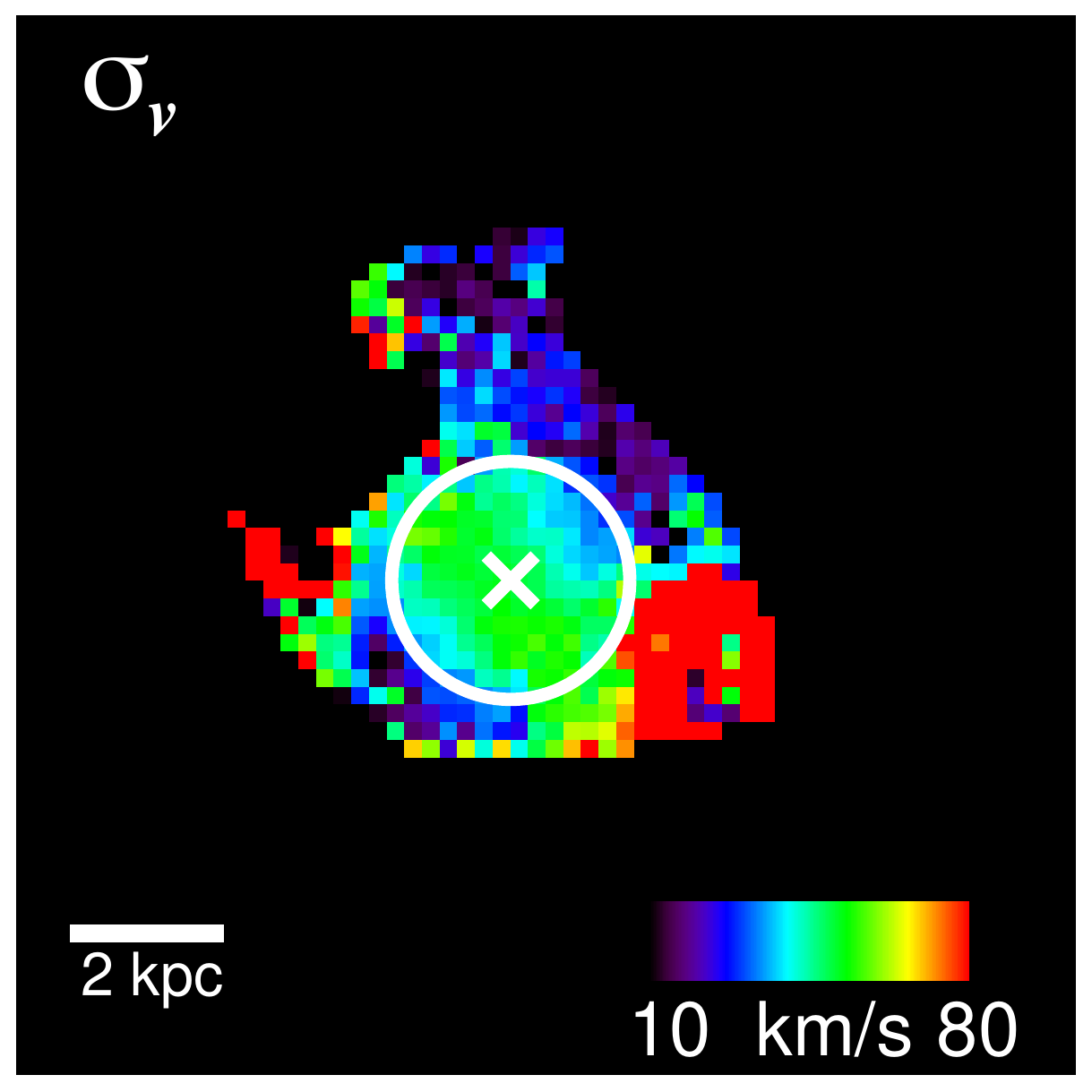}
            \end{subfigure}
        	\hspace{\jmhspace}
            \begin{subfigure}[t]{\jmtsize}
                \flushleft
                \includegraphics[width=\jmtsize]{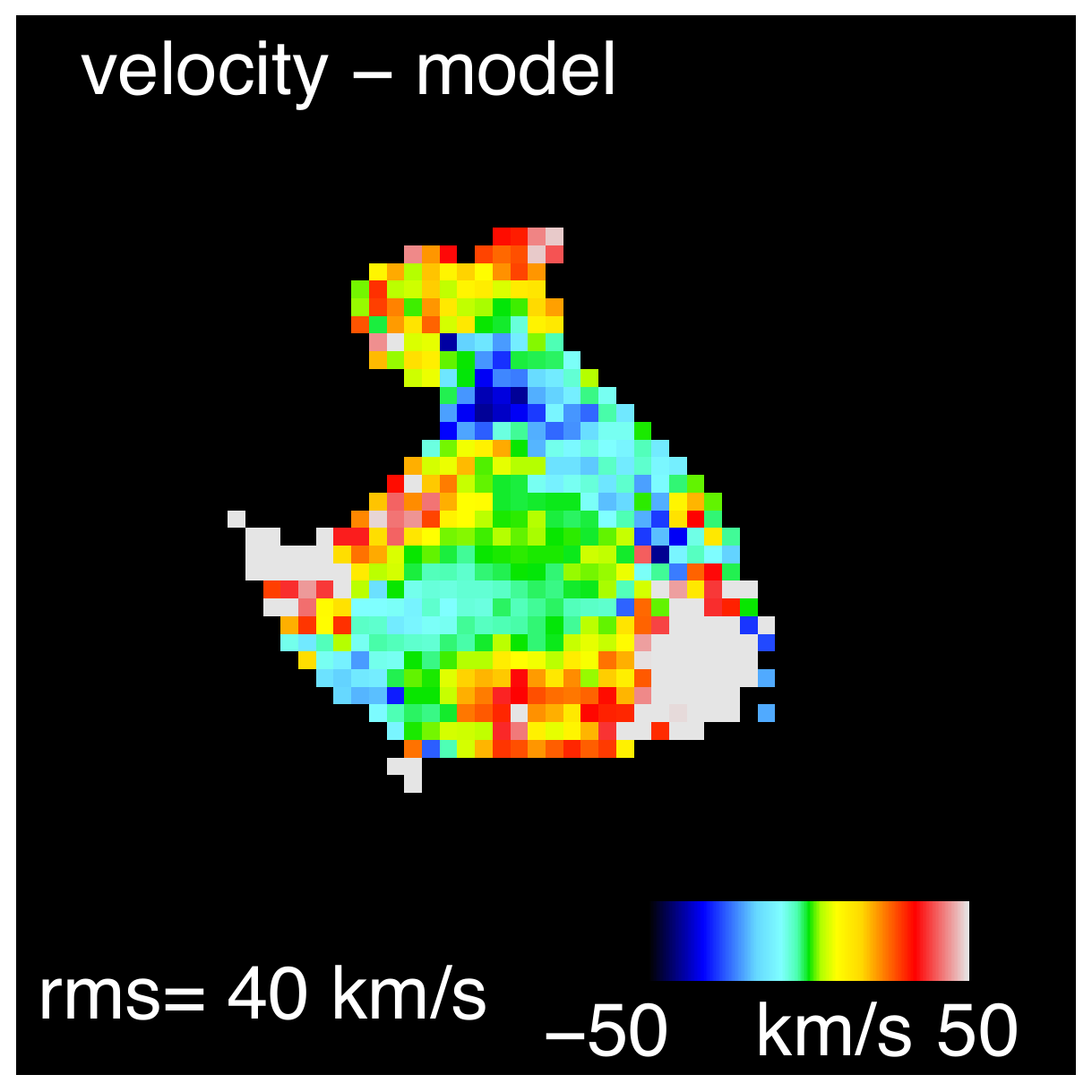}
            \end{subfigure}
            \begin{tikzpicture}[overlay, remember picture]
				\draw[dashed, red!50, ultra thick] ([shift={(19mm, 19mm)}]090bb.east)--([shift={(0.5mm, 13.5mm)}]090pa.west);
				\draw[dashed, red!50, ultra thick] ([shift={(19mm, -19mm)}]090bb.east)--([shift={(0.5mm, -13.5mm)}]090co.west);
			\end{tikzpicture}
        \end{tabular}\\
\vspace{\jmvspace}
\begin{subfigure}{\jmbsize}
    \centering
    \tikz[overlay, remember picture] \node[anchor=south, inner sep=-5cm] (114bb) {\includegraphics[width=\jmbsize]{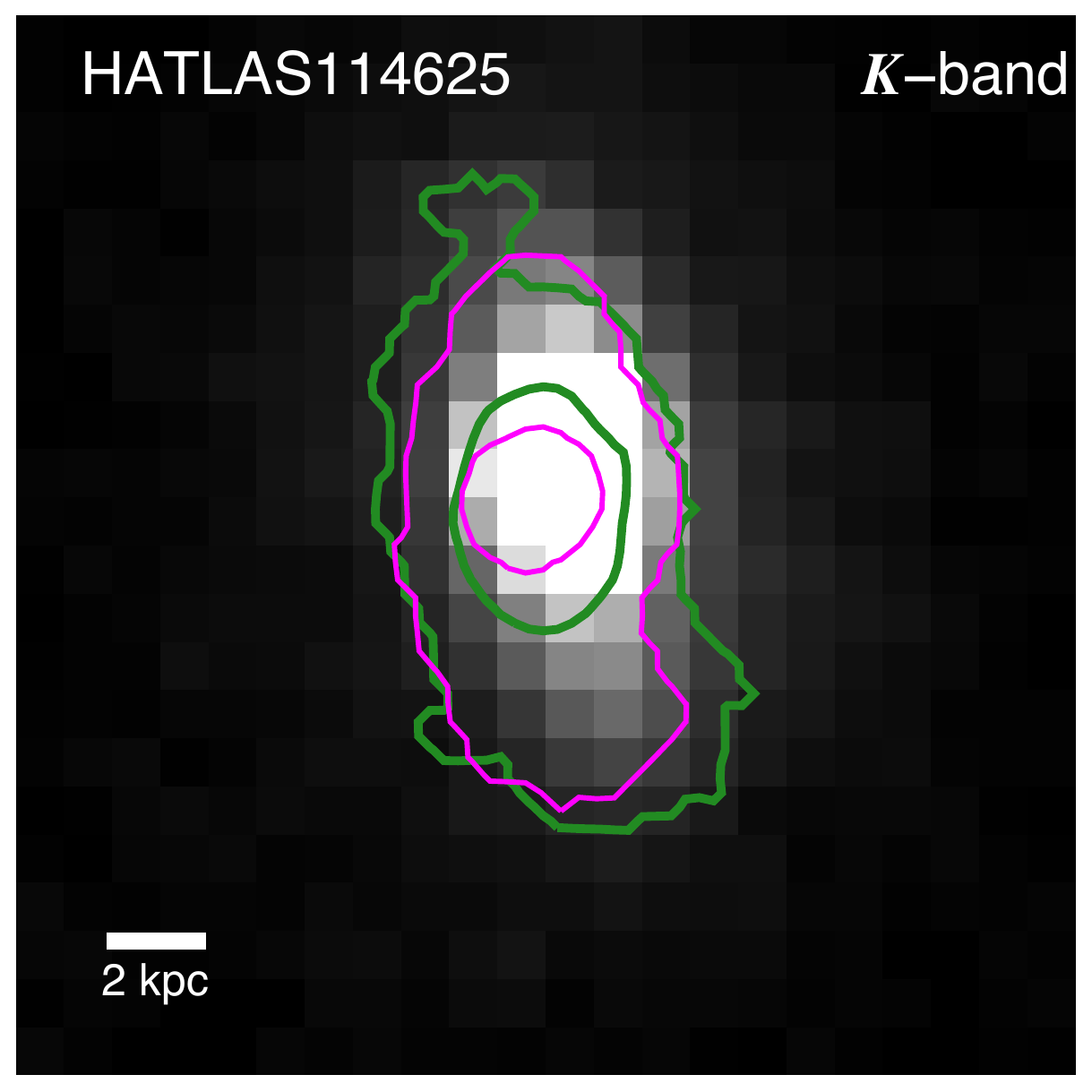}};
\end{subfigure}
    &
        \begin{tabular}{cccc}
        	\hspace{\jmhspace}
            \begin{subfigure}[t]{\jmtsize}
                \centering
                \tikz[overlay, remember picture] \node[anchor=south, inner sep=0cm] (114pa) {\includegraphics[width=\jmtsize]{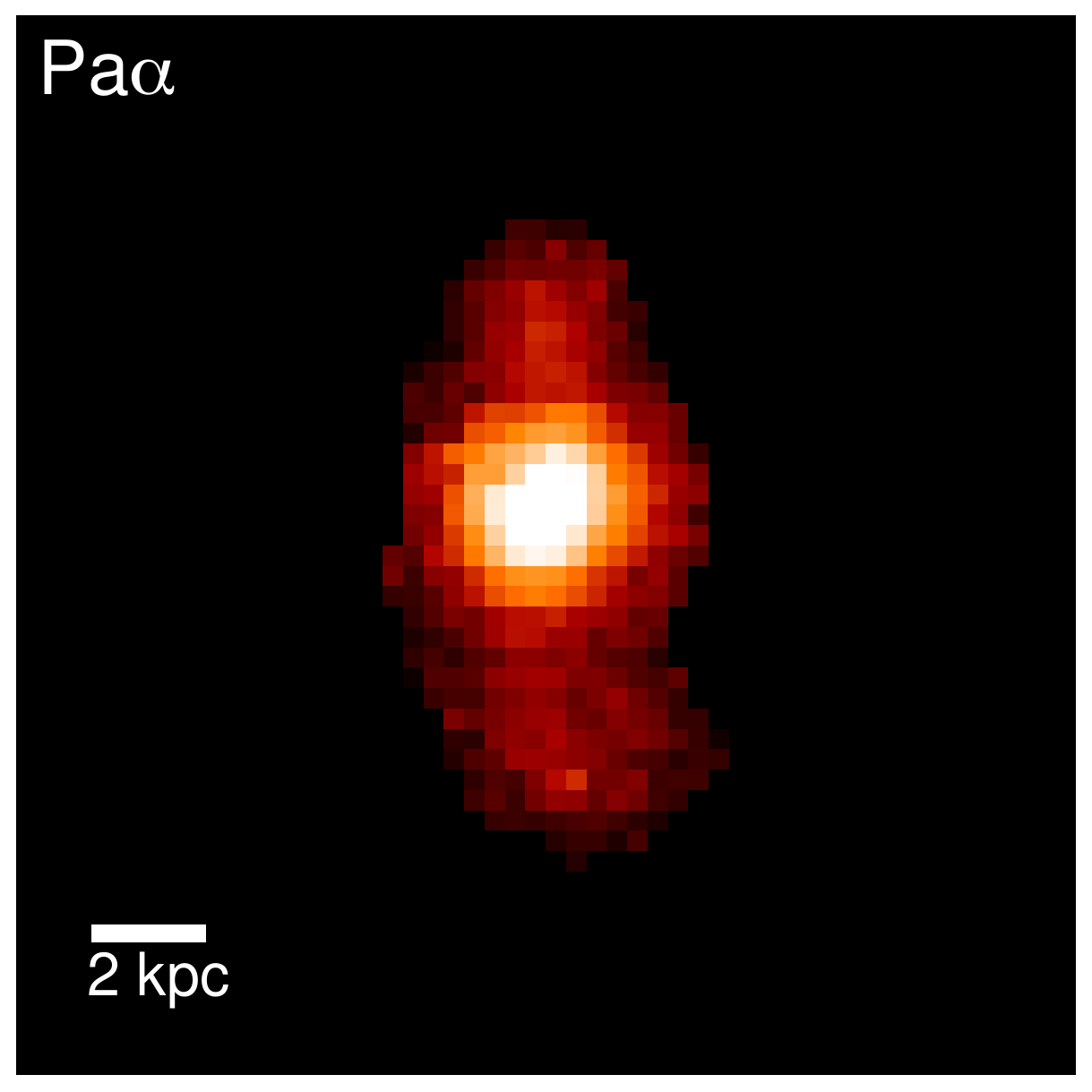}};
            \end{subfigure}
        	\hspace{\jmhspace}
            \begin{subfigure}[t]{\jmtsize}
                \flushleft
                \includegraphics[width=\jmtsize]{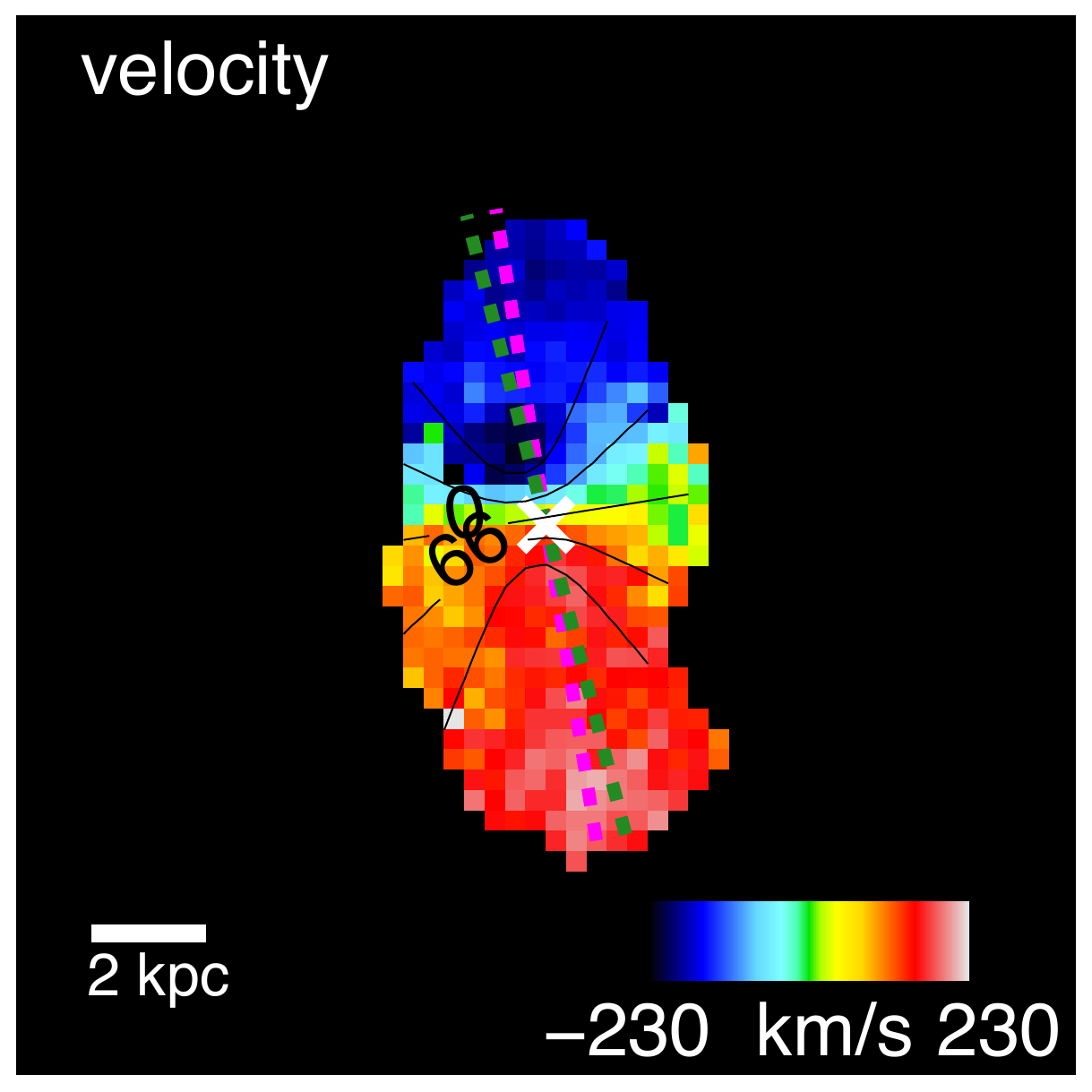}
            \end{subfigure}
        	\hspace{\jmhspace}
            \begin{subfigure}[t]{\jmtsize}
                \flushleft
                \includegraphics[width=\jmtsize]{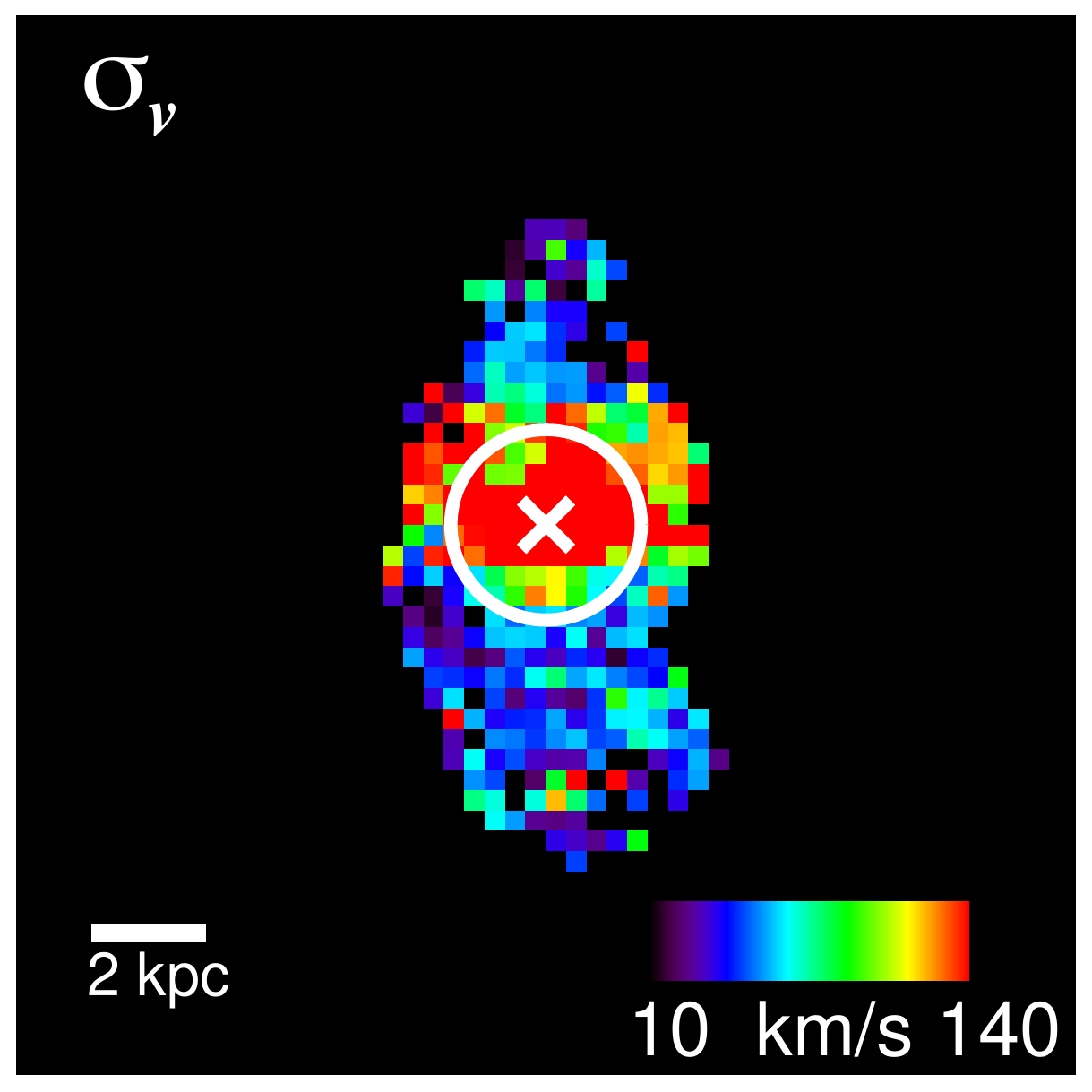}
            \end{subfigure}
        	\hspace{\jmhspace}
            \begin{subfigure}[t]{\jmtsize}
                \flushleft
                \includegraphics[width=\jmtsize]{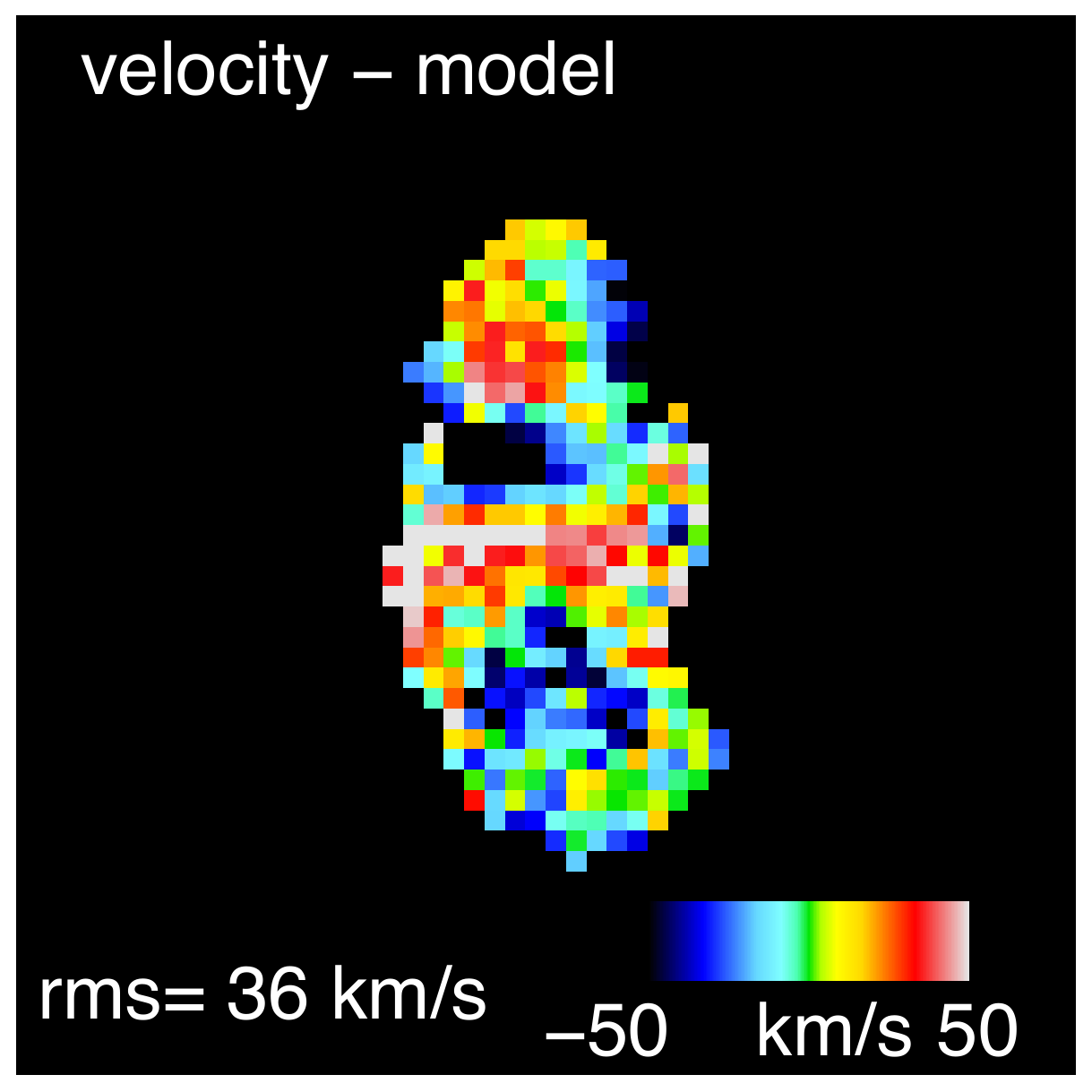}
            \end{subfigure}\\
        	\hspace{\jmhspace}
                \begin{subfigure}[t]{\jmtsize}
                \centering
                \tikz[overlay, remember picture] \node[anchor=south, inner sep=0cm] (114co) {\includegraphics[width=\jmtsize]{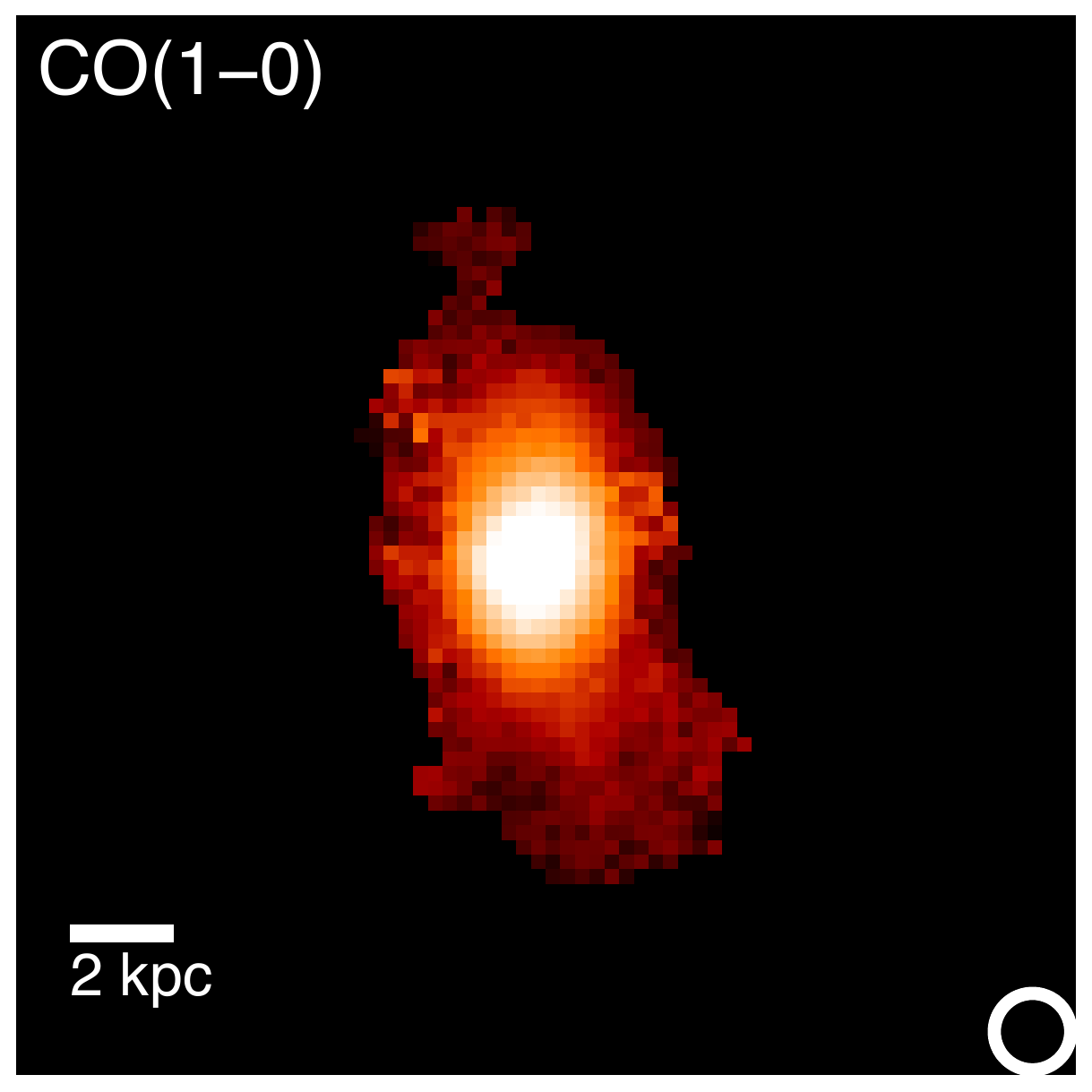}};
            \end{subfigure}
        	\hspace{\jmhspace}
            \begin{subfigure}[t]{\jmtsize}
                \flushleft
                \includegraphics[width=\jmtsize]{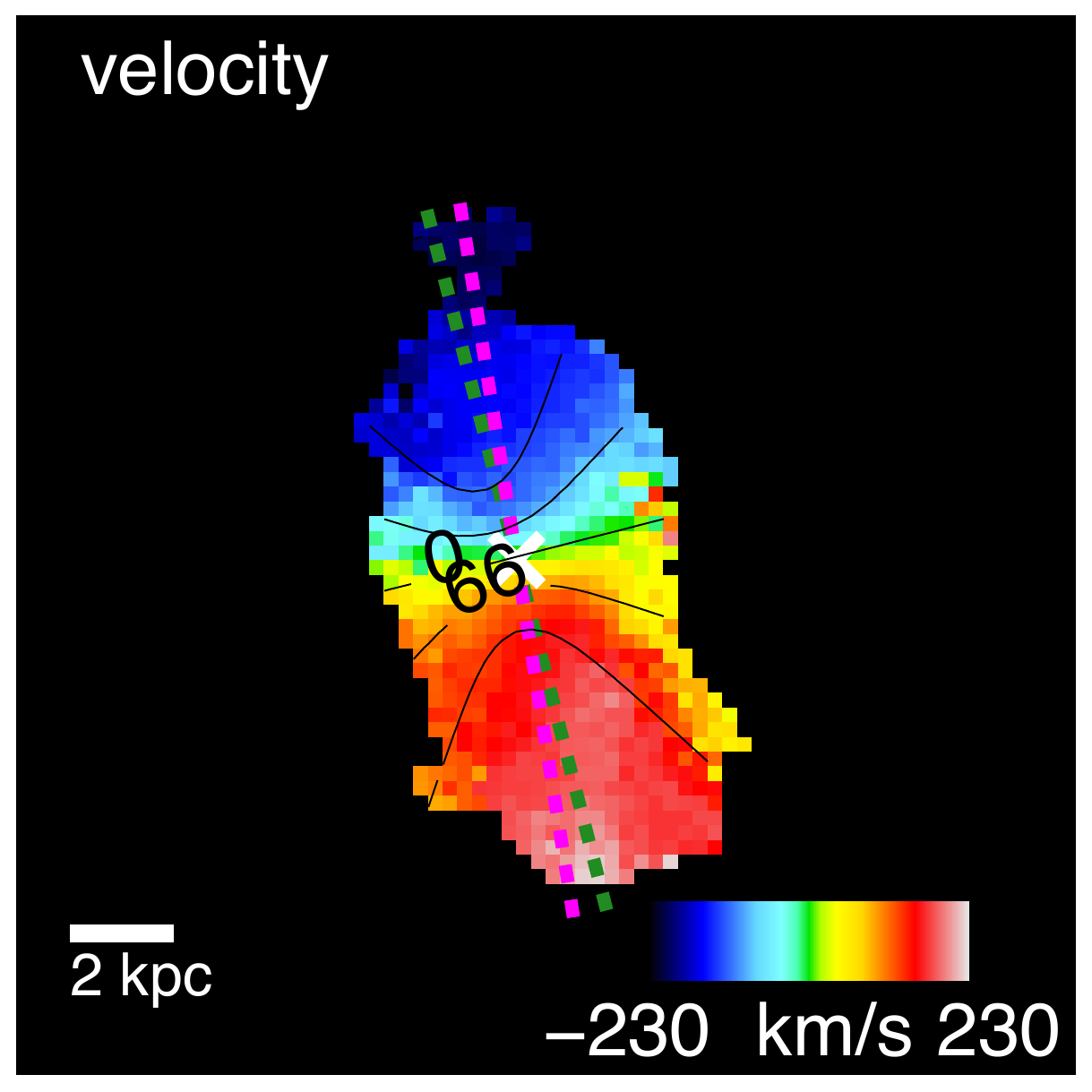}
            \end{subfigure}
        	\hspace{\jmhspace}
            \begin{subfigure}[t]{\jmtsize}
                \flushleft
                \includegraphics[width=\jmtsize]{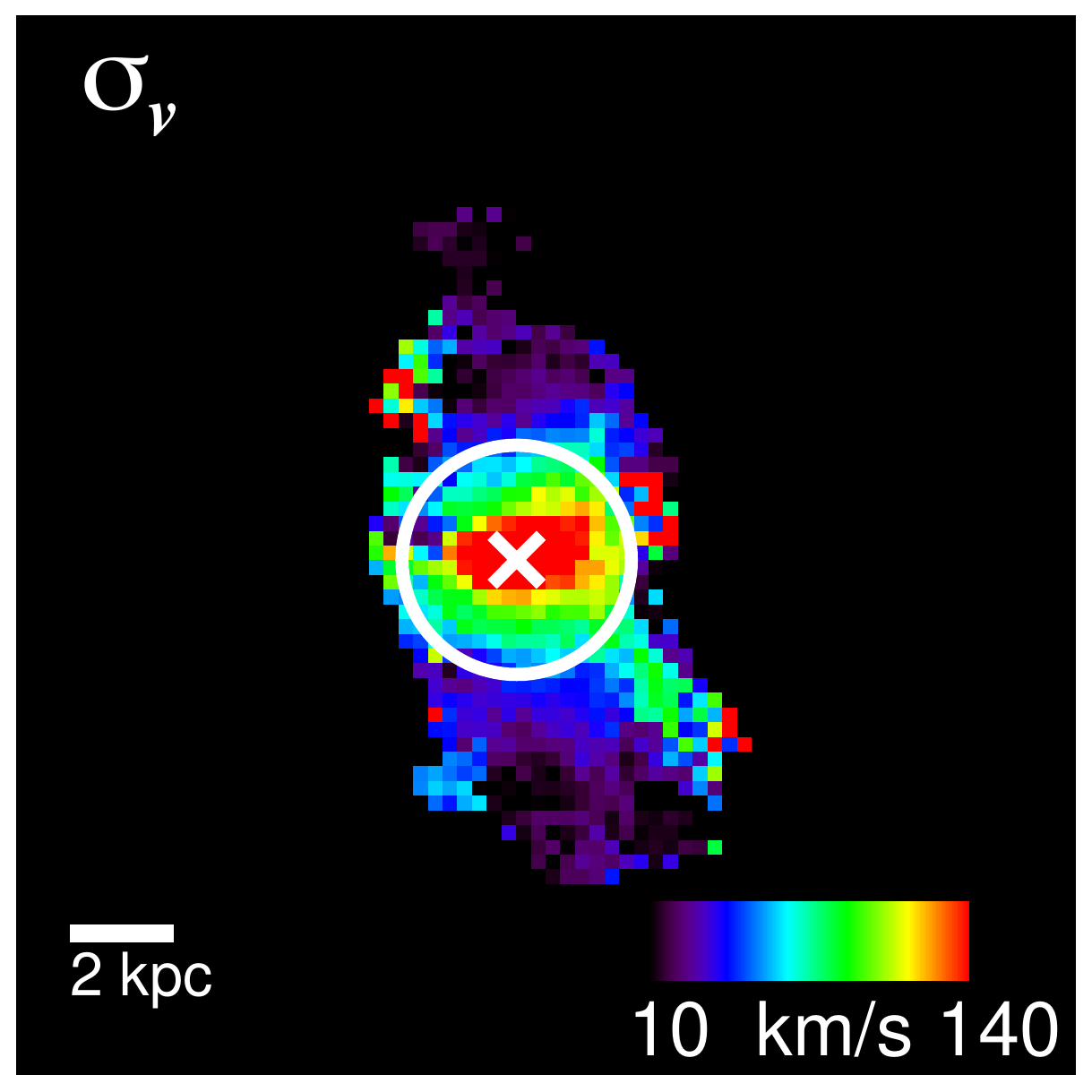}
            \end{subfigure}
        	\hspace{\jmhspace}
            \begin{subfigure}[t]{\jmtsize}
                \flushleft
                \includegraphics[width=\jmtsize]{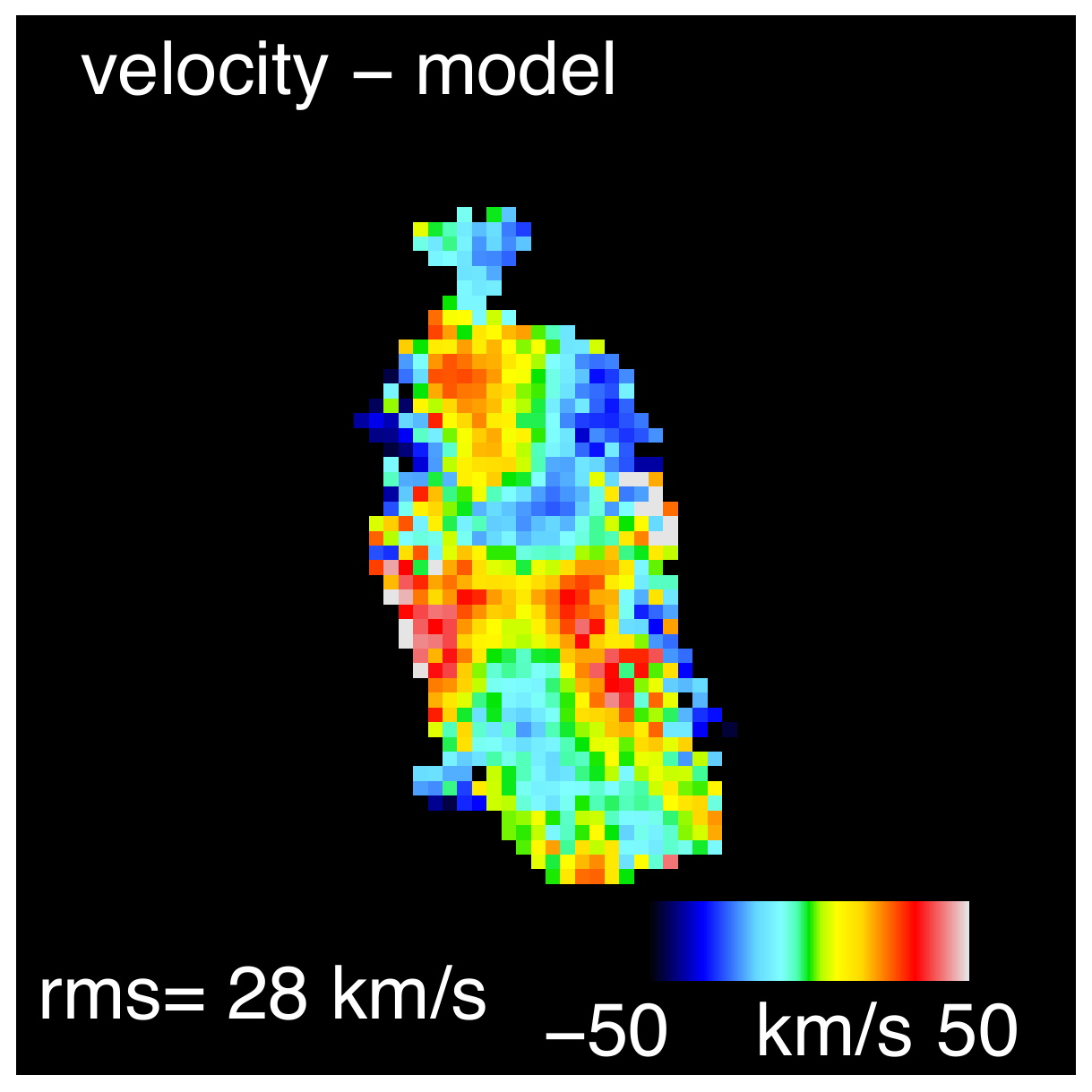}
            \end{subfigure}
             \begin{tikzpicture}[overlay, remember picture]
				\draw[dashed, red!50, ultra thick] ([shift={(19mm, 19mm)}]114bb.east)--([shift={(0.5mm, 13.5mm)}]114pa.west);
				\draw[dashed, red!50, ultra thick] ([shift={(19mm, -19mm)}]114bb.east)--([shift={(0.5mm, -13.5mm)}]114co.west);
			\end{tikzpicture}
        \end{tabular}\\
\vspace{\jmvspace}
\begin{subfigure}{\jmbsize}
    \centering
    \tikz[overlay, remember picture] \node[anchor=south, inner sep=-5cm] (121bb) {\includegraphics[width=\jmbsize]{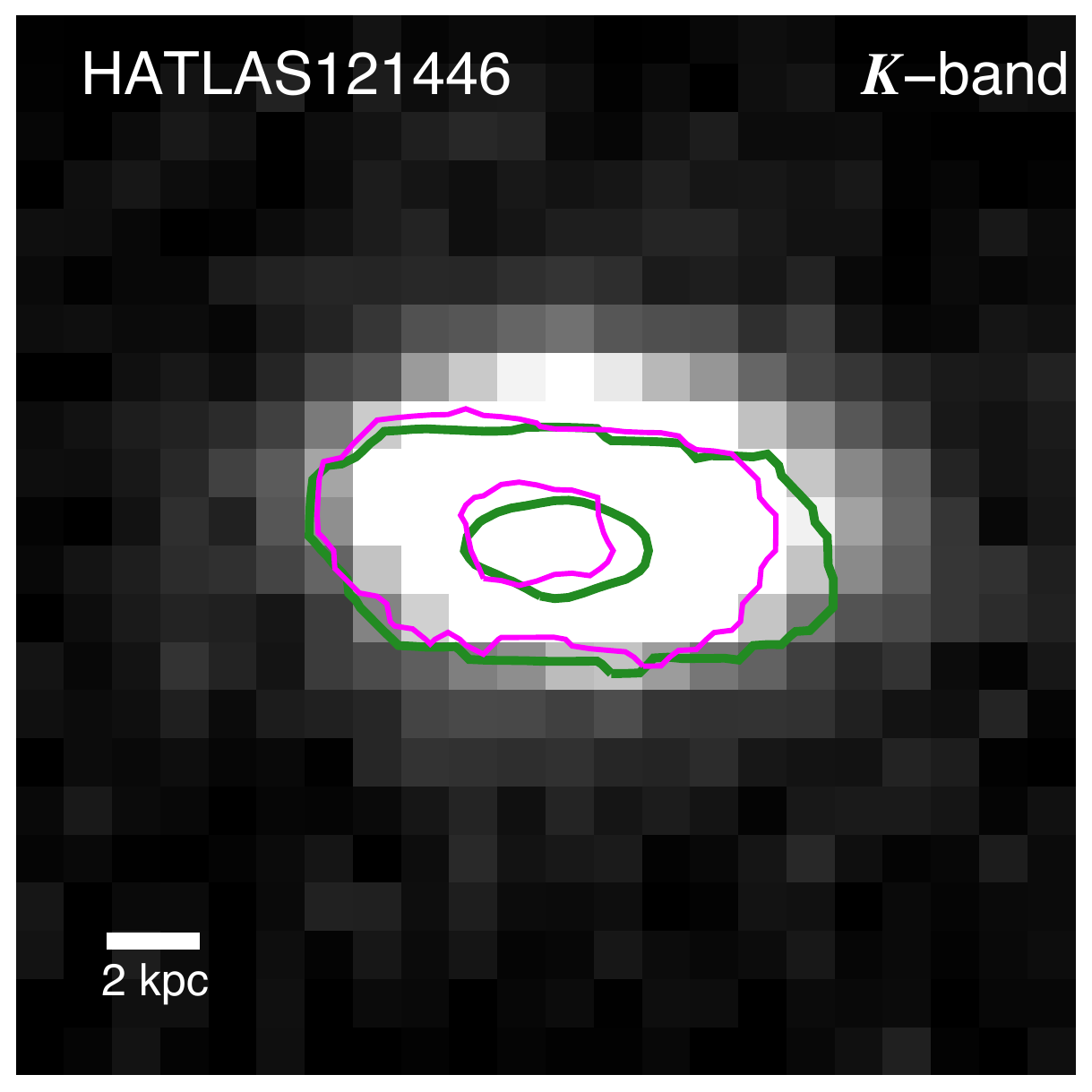}};
\end{subfigure}
    &
        \begin{tabular}{cccc}
        	\hspace{\jmhspace}
            \begin{subfigure}[t]{\jmtsize}
                \centering
                \tikz[overlay, remember picture] \node[anchor=south, inner sep=0cm] (121pa) {\includegraphics[width=\jmtsize]{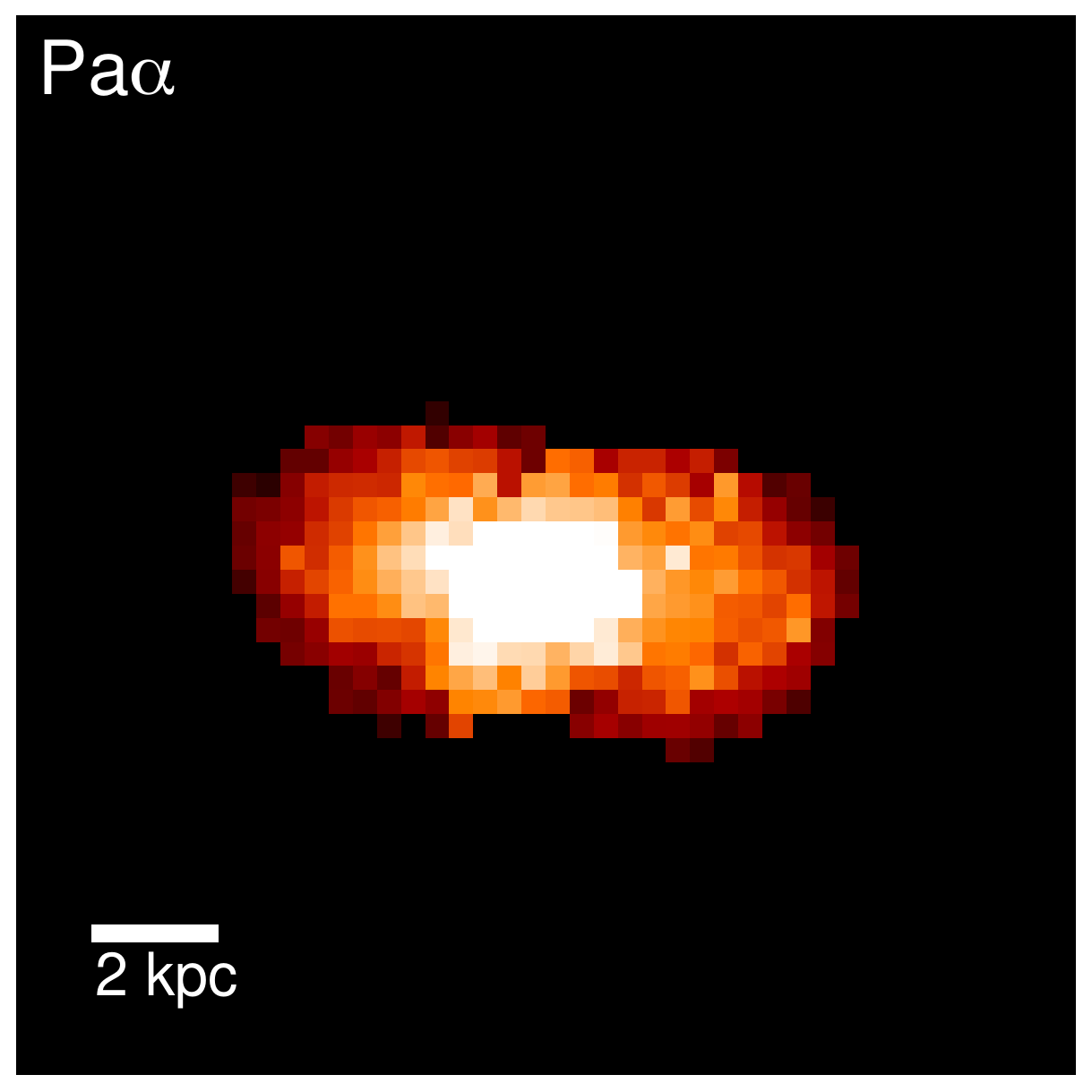}};
            \end{subfigure}
        	\hspace{\jmhspace}
            \begin{subfigure}[t]{\jmtsize}
                \flushleft
                \includegraphics[width=\jmtsize]{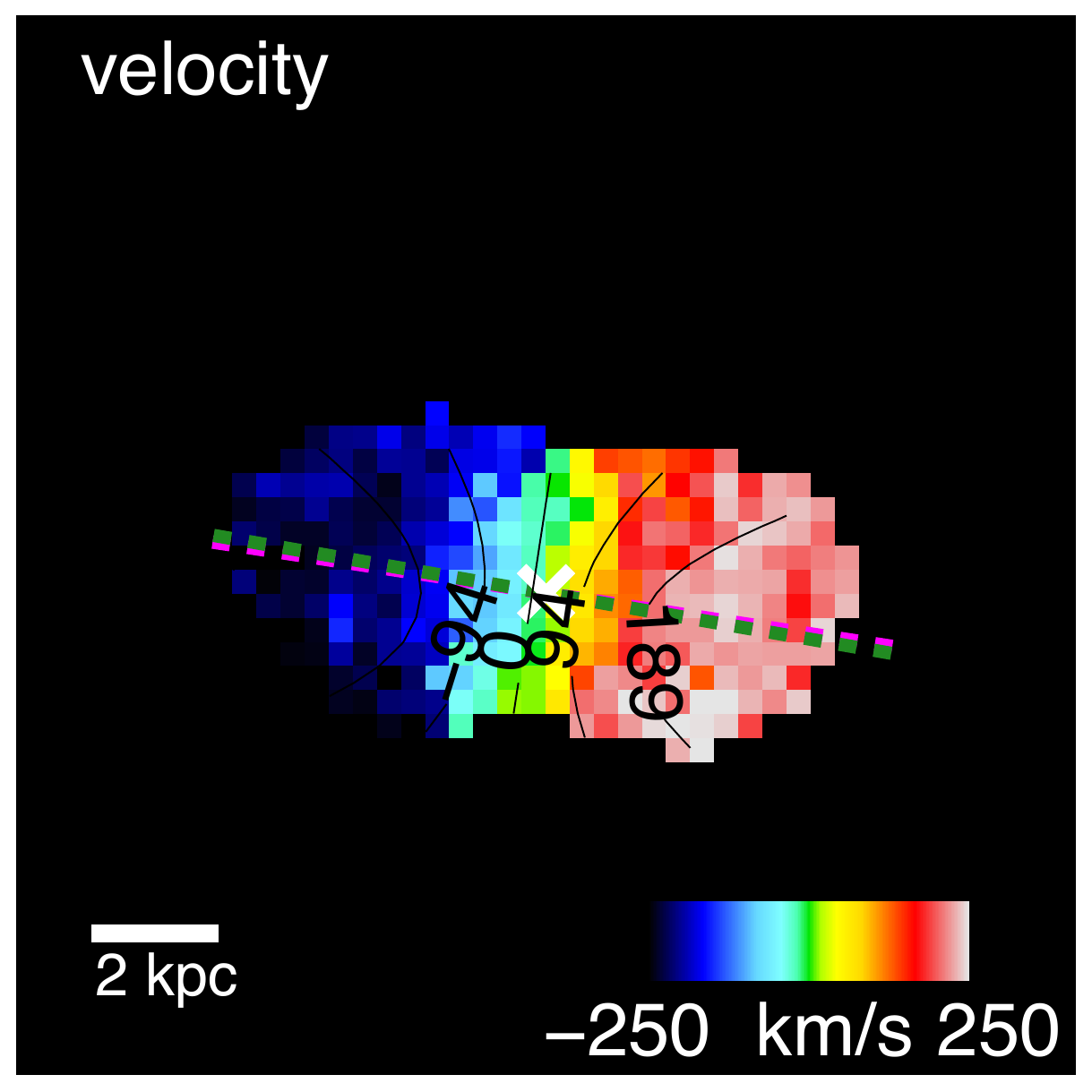}
            \end{subfigure}
        	\hspace{\jmhspace}
            \begin{subfigure}[t]{\jmtsize}
                \flushleft
                \includegraphics[width=\jmtsize]{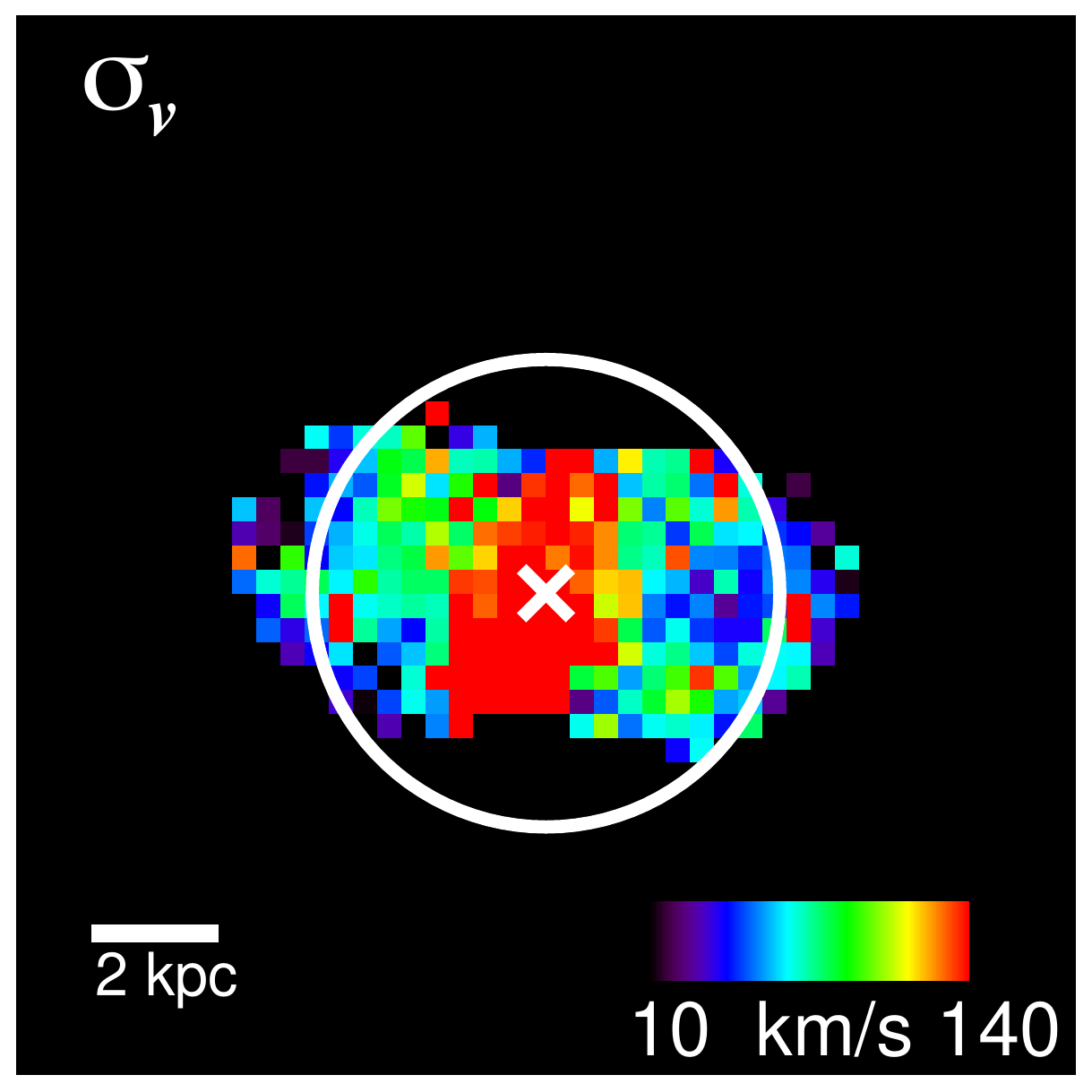}
            \end{subfigure}
        	\hspace{\jmhspace}
            \begin{subfigure}[t]{\jmtsize}
                \flushleft
                \includegraphics[width=\jmtsize]{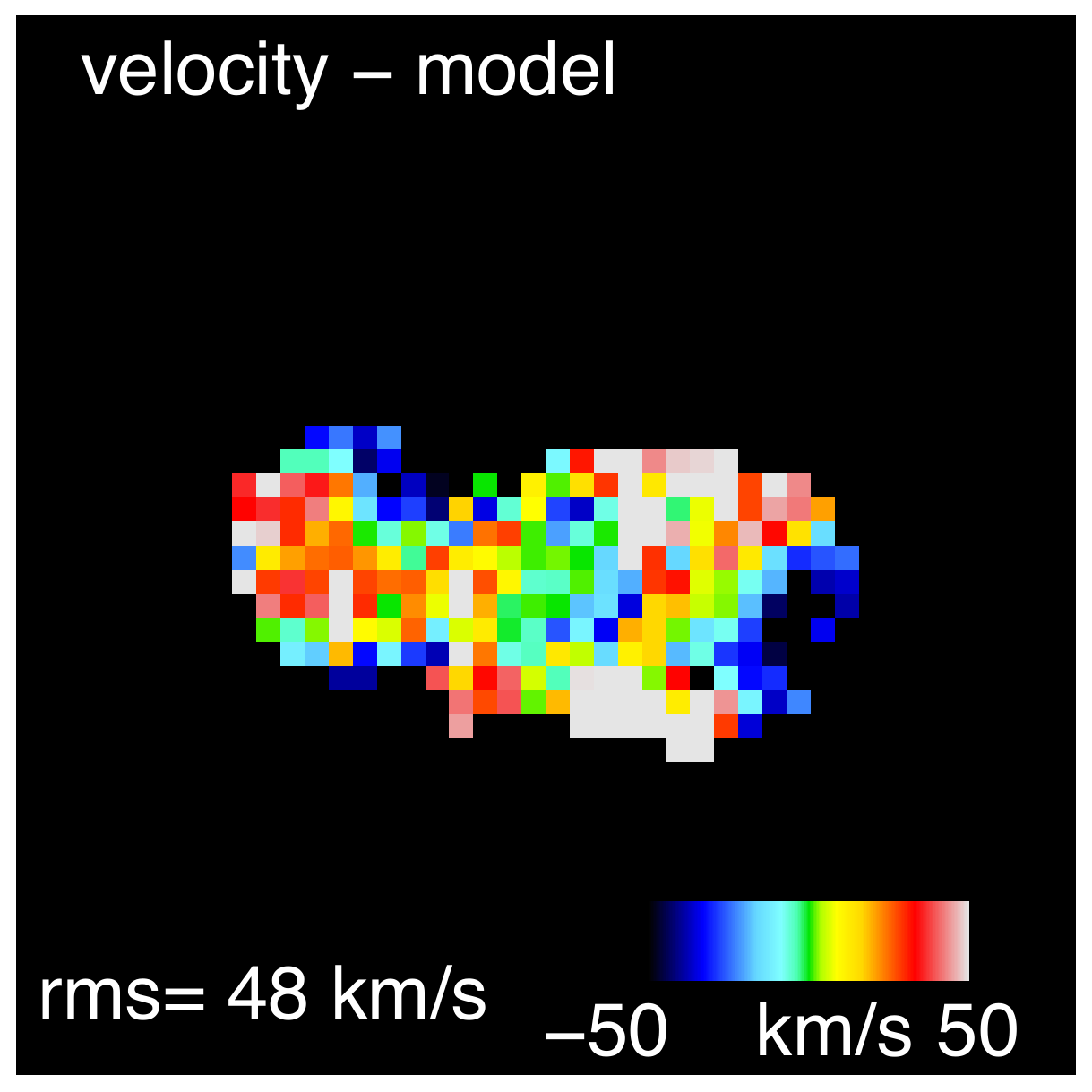}
            \end{subfigure}\\
        	\hspace{\jmhspace}
                \begin{subfigure}[t]{\jmtsize}
                \centering
                \tikz[overlay, remember picture] \node[anchor=south, inner sep=0cm] (121co) {\includegraphics[width=\jmtsize]{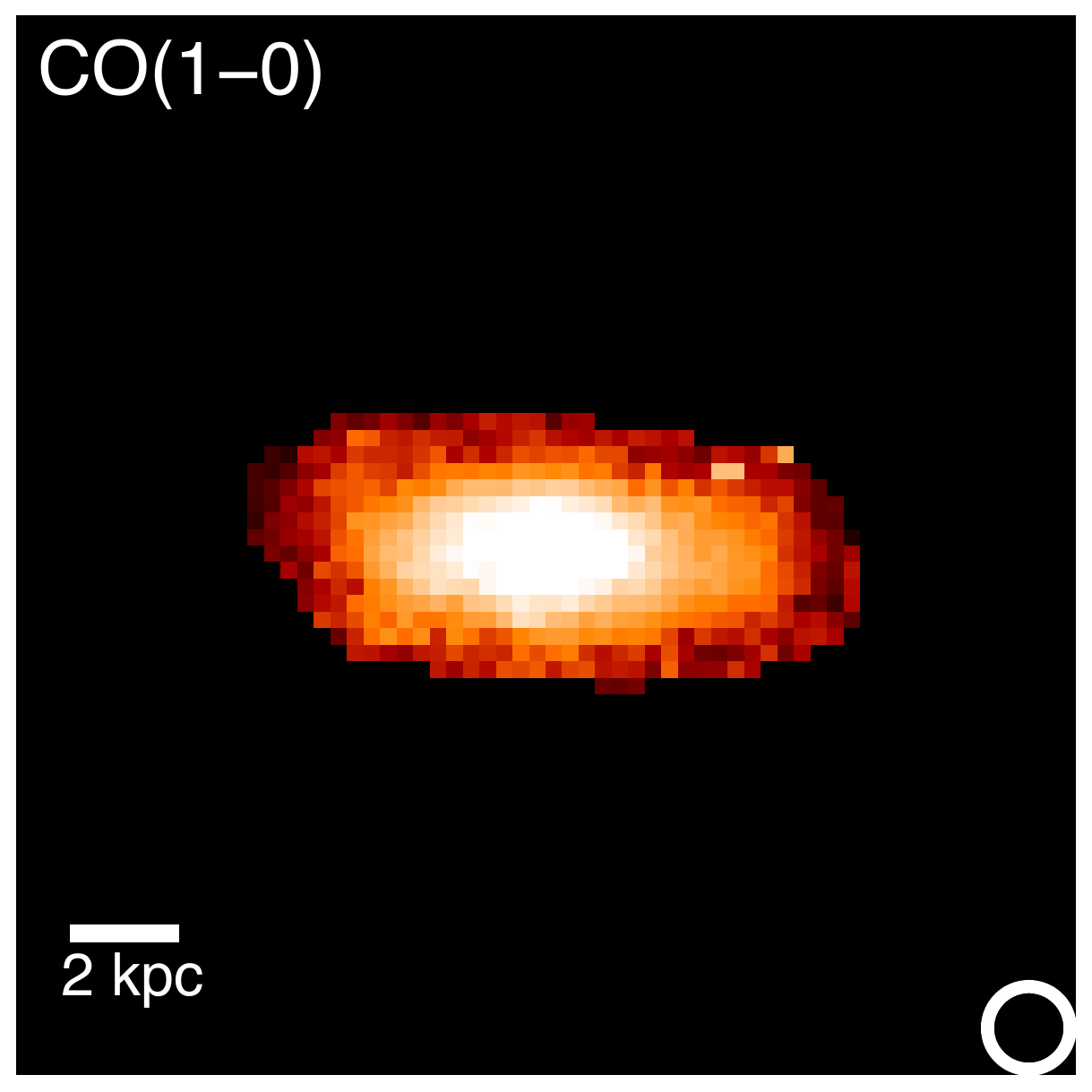}};
            \end{subfigure}
        	\hspace{\jmhspace}
            \begin{subfigure}[t]{\jmtsize}
                \flushleft
                \includegraphics[width=\jmtsize]{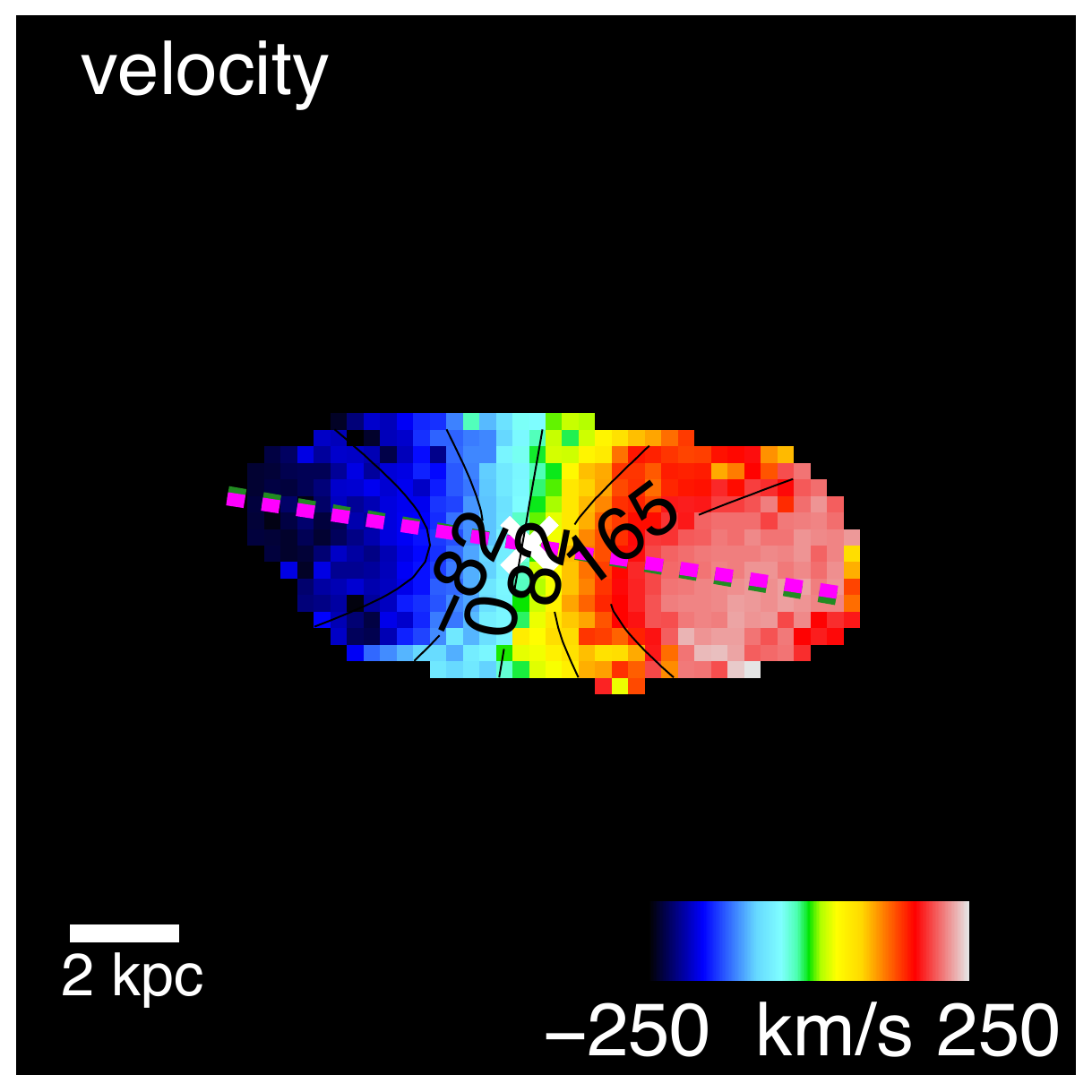}
            \end{subfigure}
        	\hspace{\jmhspace}
            \begin{subfigure}[t]{\jmtsize}
                \flushleft
                \includegraphics[width=\jmtsize]{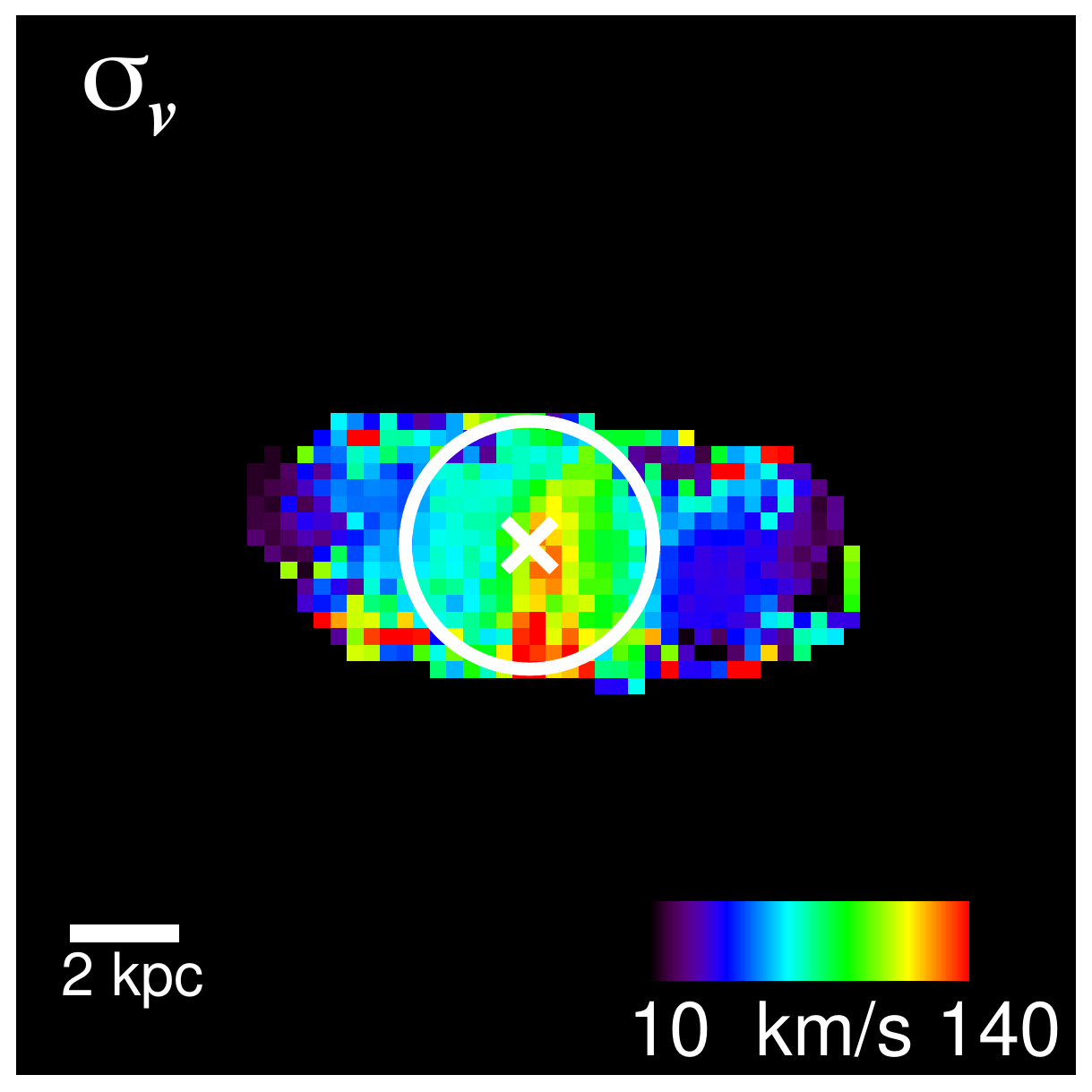}
            \end{subfigure}
        	\hspace{\jmhspace}
            \begin{subfigure}[t]{\jmtsize}
                \flushleft
                \includegraphics[width=\jmtsize]{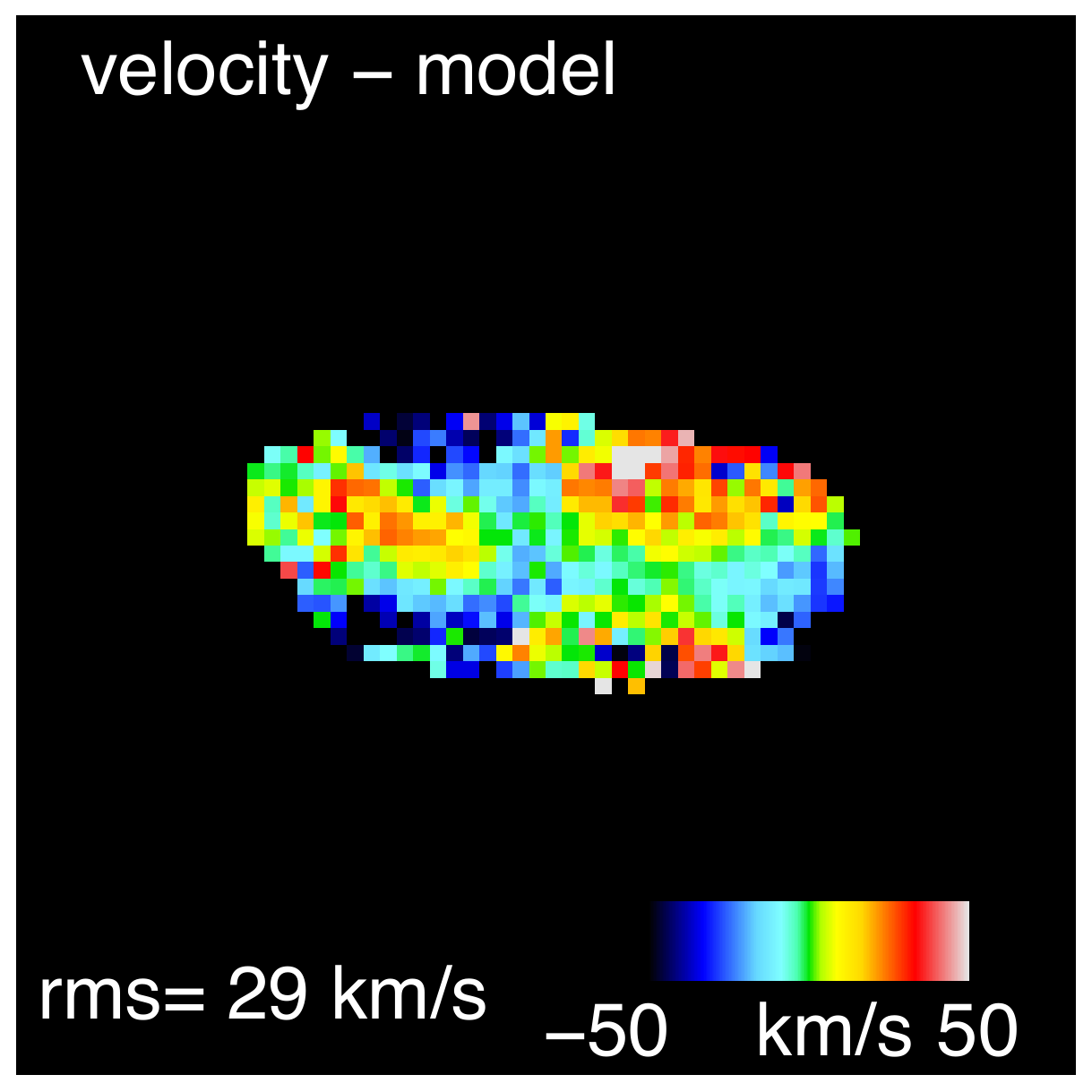}
            \end{subfigure}
             \begin{tikzpicture}[overlay, remember picture]
				\draw[dashed, red!50, ultra thick] ([shift={(19mm, 19mm)}]121bb.east)--([shift={(0.5mm, 13.5mm)}]121pa.west);
				\draw[dashed, red!50, ultra thick] ([shift={(19mm, -19mm)}]121bb.east)--([shift={(0.5mm, -13.5mm)}]121co.west);
			\end{tikzpicture}
        \end{tabular}\\
    \end{tabular}
    \caption{\label{fig:maps} $K$-band, intensity, velocity, velocity dispersion and residual maps (1st to 5th columns) for HATLAS090750 (\textit{Top}), HATLAS114625 (\textit{Middle}) and HATLAS121446 (\textit{Bottom}). For each galaxy, from the 2nd column to the last column, we show the Pa$\alpha$ and CO(1-0) two-dimensional maps, one above the other, respectively. The spatial scale for each observation is shown in each map. The $K$-band map has over-plotted the CO(1-0) and Pa$\alpha$ emissions in green and pink colour contours, respectively. The CO(1-0) intensity map shows the synthesized beam size. In the velocity and velocity dispersion maps, the white cross indicates the location of the best-fitted dynamical centre. The velocity maps have over-plotted the velocity contours from their best-fit disc models, and the green- and pink-dashed lines represent the molecular and ionized gas major kinematic axes, respectively. In each velocity dispersion map, the white circumference represents the boundary of the region masked during the estimation of the global velocity dispersion value. The residual fields are constructed by subtracting the velocity disc models from the velocity maps. The r.m.s. of these residuals are given in each panel. In the case of HATLAS090750 Pa$\alpha$ observation, we only show the modelled central zone in the residual map.}
\end{figure*}
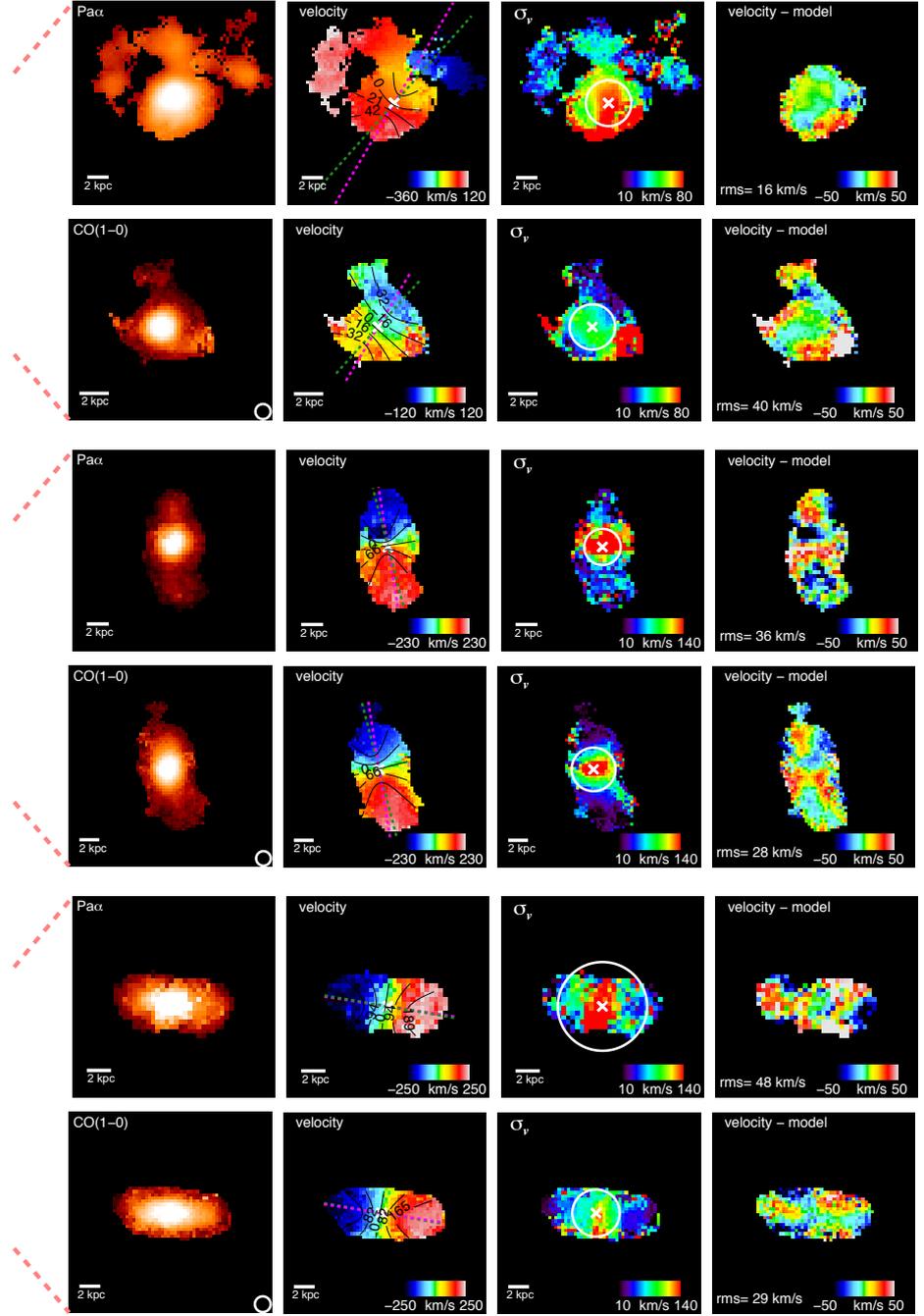

The pixel-by-pixel intensity, velocity and velocity dispersion 1-$\sigma$ uncertainties are estimated by re-sampling via Monte Carlo simulations the flux density uncertainties in the data. The maps from both emission lines are presented in Fig.~\ref{fig:maps}.

The CO(1-0) and Pa$\alpha$ intensity maps present smooth distributions with no clear level of clumpiness, at $\sim$\,kpc-scales, in the three starbursts. These also agree with the stellar morphology seen in $K$-band image. However, we note that OH sky-line features present in the SINFONI observations may add noise to the Pa$\alpha$ two-dimensional maps and this may partly explain the smoother CO(1-0) maps as the ALMA spectra are free from sky-line residuals.

In the particular case of the HATLAS090750 system, the $K$-band and Pa$\alpha$ intensity images show two asymmetric features that may be related to gas inflow/outflow or tidal interaction. These features suggest an on-going merging process. Both features account for $\sim$18\% of the total Pa$\alpha$ flux suggesting on-going star formation activity. One of the asymmetric features has a projected velocity blueshift of $\sim -300$\,km\,s$^{-1}$ compared to the system centre, while the other feature presents a velocity redshift of $\sim 80$\,km\,s$^{-1}$ suggesting that this system has a complex 3D shape. The ALMA observation just traces the CO(1-0) emission coming from the central part of this system, probably due to sensitivity limitations. Interestingly, the central part of this system shows a rotational pattern in the CO(1-0) and Pa$\alpha$ velocity maps, with a peak-to-peak rotational velocity of $V_{\rm max} \sin(i) \sim 90$\,km\,s$^{-1}$.

In contrast, HATLAS114625 and HATLAS121446 show clear disc-like rotational patterns in their CO(1-0) and Pa$\alpha$ velocity maps. The ionized and molecular gas kinematics broadly agree in both starbursts with peak-to-peak rotational velocities of $V_{\rm max} \sin(i) \sim 360 - 460 $\,km\,s$^{-1}$, respectively.

\begin{table} 
	\centering
	\setlength\tabcolsep{2pt}
    	\caption{\label{tab:table2}  $K$-band surface brightness S\'ersic best-fit model parameters taken from the GAMA-DR3 for our sample \citep{Kelvin2012}. $\mu_{0,K}$ is the central surface brightness value. $R_{1/2,K}$ corresponds to the half-light radius. $n_S$ is the S\'ersic photometric index. PA$_K$ indicates the position angle of the photometric major axis. The ellipticity `$e$' is derived from the projected major-to-minor axis ratio on the sky ($e\equiv1-b/a$). The final column denotes the reduced chi-square ($\chi^2_\nu$) value of the best-fit model.}
    \vspace{1mm}
	\begin{tabular}{lcccccc} 
		\hline
		\hline
		Name & $\mu_{0,K}$ & $R_{1/2,K}$ & $n_S$ & PA$_K$ & $e$ & $\chi^2_\nu$ \\
		& (mag/arcsec$^2$) & (kpc) & & (deg) & & \\
		\hline
		HATLAS090750 & 9.36 & 3.66 & 4.92 & 62.1 & 0.26 & 2.19 \\
         HATLAS114625 & 3.78 & 2.76 & 6.80 & $-$82.2 & 0.60 & 1.29 \\
         HATLAS121446 & 15.31 & 2.55 & 1.26 & $-$3.5 & 0.67 & 1.12 \\
		\hline                                                                                        
	\end{tabular}
\end{table}

\subsubsection{Kinematic Modelling}
\label{sec:kin_mod}

We model the ionized and molecular gas ISM kinematics by fitting the two-dimensional LOS velocity fields. The model velocity maps are constructed by assuming an input $\arctan$ rotation curve:

\begin{equation}
V(R)= V_0 + \frac{2}{\pi}V_{\rm asym} \arctan(R/R_{\rm t}),
\end{equation}

\noindent where $R_{\rm t}$ is the radius at which the rotation curve turns over, $V_0$ is the systemic velocity (i.e. redshift) and $V_{\rm asym}$ is the asymptotic rotational velocity \citep{Courteau1997}.

For each observation, the kinematic model considers seven free parameters ($V_0$, $V_{\rm asym}$, $R_{\rm t}$, PA, [$x/y$], and inclination angle). We convolve the velocity model map with the PSF or synthesized beam, and we use the \textsc{emcee} \textsc{Python} package \citep{Foreman2013} to find the best-fit model.

 \begin{table*}
	\centering
	\setlength\tabcolsep{4pt}
    	\caption{\label{tab:table3} Best-fit kinematic parameters for the galaxies in our sample. PA is the kinematic major-axis position angle. $R_{1/2}$ is the half-light radius corrected by beam smearing effects. $\sigma_v$ is the global velocity dispersion value (see \S\ref{sec:kin_pars}). $V_{\rm rot}$ is the rotational velocity measured across the major kinematic axis. $i$ is the inclination angle (for a face-on galaxy, $i = 0$\,deg.). We do not give uncertainty estimates for $i$ as it is constrained by the $K$-band image model. The `CO' and `Pa$\alpha$' sub-indexes indicate the emission line from which the kinematic parameters were estimated.}
    \vspace{1mm}
	\begin{tabular}{lccc} 
        \hline
        \hline
	 & HATLAS090750 & HATLAS114625 & HATLAS121446 \\
     \hline
     $i_{\rm CO}$\,(deg) & 65 & 70 & 80 \\
	PA$_{\rm CO}$\,(deg) & $-132\pm1$ & $-76\pm1$ & $-10\pm1$ \\
	$R_{\rm 1/2,CO}$\,(kpc) & $1.44\pm0.01$ & $2.09\pm0.01$ & $2.74\pm0.01$ \\
	$V_{\rm rot,CO}$\,(km\,s$^{-1}$) & $67\pm2$ & $198\pm22$ & $245\pm5$ \\
	$\sigma_{v,{\rm CO}}$\,(km\,s$^{-1}$) & $35\pm12$ & $26\pm10$ & $34\pm11$ \\
	$\chi^2_{\nu,{\rm CO}}$ &  11.2 & 10.4 & 7.0 \\  
	\hline
	$i_{\rm Pa\alpha}$\,(deg) & 65 & 70 & 80 \\
	PA$_{\rm Pa\alpha}$\,(deg) & $-123\pm1$ & $-81\pm1$ & $-9\pm1$ \\ 
	$R_{\rm 1/2,Pa\alpha}$\,(kpc) & $2.10\pm0.05$ & $1.72\pm0.01$ & $2.52\pm0.03$ \\
	$V_{\rm rot,Pa\alpha}$\,(km\,s$^{-1}$) & $68\pm4$ & $190\pm7$ & $246\pm9$ \\
	$\sigma_{v,{\rm Pa\alpha}}$\,(km\,s$^{-1}$) & $66\pm18$ & $51\pm30$ & $51\pm31$ \\
	$\chi^2_{\nu,{\rm Pa\alpha}}$ & 7.1 & 5.0 & 7.8 \\ 
     \hline
	\end{tabular}
\end{table*}

We use the $K$-band S\'ersic photometric models \citep{Sersic1963} to constrain the inclination angle values. We use the $K$-band best-fit minor-to-major axis ratio ($b/a$; Table~\ref{tab:table2}) as initial guess input to the kinematic modelling and we allow to search the best-fit inclination value within a 3-$\sigma$ range. To better account for the $K$-band model $b/a$ uncertainty, we adopt a $b/a$ ratio 1-$\sigma$ relative error equal to 10\% as suggested by \citet{Epinat2012}. The inclination angle is derived from $b/a$ by considering an oblate spheroid geometry \citep{Holmberg1958}:

\begin{equation}
    \cos^2(i)=\frac{(b/a)^2-{\rm q_0^2}}{1-{\rm q_0^2}},
	\label{eq:inc_eqn}
\end{equation}

\noindent where `\textbf{$i$}' is the galaxy inclination angle and ${\rm q_0}$ is the intrinsic minor-to-major axis ratio (i.e. disc thickness) of the galaxy. For edge-on systems ($i = 90$\,deg), ${\rm q_0} = b/a$. We use ${\rm q_0} = 0.14$ mean value reported for edge-on galaxies at low-redshift ($z<0.05$, \citealt{Mosenkov2015}). 

The model best-fit parameters and $\chi^2_\nu$ values are given in Table~\ref{tab:table3} and the r.m.s. values are shown in each residual map (Fig.~\ref{fig:maps}). The kinematic position angles roughly agree with each other ($\Delta$PA = PA$_{\rm Pa\alpha}$ - PA$_{\rm CO} \lesssim 10$\,deg). For HATLAS114625 and HATLAS121446 galaxies, these also roughly agree with the position angles derived from the $K$-band image modelling.

The best-fit disc model gives a reasonable fit to the inner ionized and molecular gas kinematics of the HATLAS090750 galaxy as suggested by the low reported r.m.s. value. This may indicate a fast relaxation process of the ISM molecular gaseous phase into a disc-like galaxy in the central zone of this system (e.g. \citealt{Kronberger2007}). For the other two galaxies, the r.m.s. values presented in the Pa$\alpha$ velocity residual maps tend to be larger than the values derived from the CO(1-0) observations, suggesting that the ionized gas ISM phase may be a more sensitive tracer of non-circular motions compared to the molecular gas ISM phase. However, these high r.m.s values also are a consequence of the coarser SINFONI spectral resolution compared to the ALMA observations plus additional noise induced by the OH sky-line features present in some pixels at the wavelengths where the Pa$\alpha$ emission line is found.

\subsubsection{Kinematic Parameters}
\label{sec:kin_pars}

\begin{figure}[!h] 
\flushleft
\includegraphics[width=1.0\columnwidth]{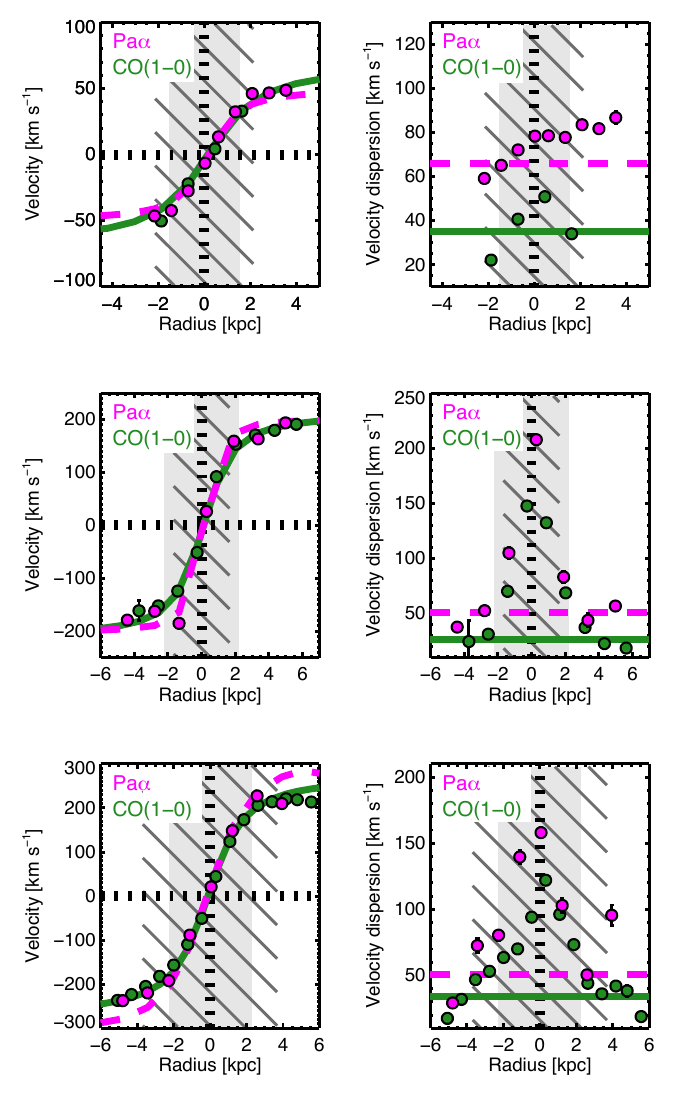}
\caption{\label{fig:1d_profiles} Rotation velocity (\textit{Left}) and velocity dispersion (\textit{Right}) profiles across the major kinematic axis for HATLAS090750 (\textit{Top}), HATLAS114625 (\textit{Middle}) and HATLAS121446 (\textit{Bottom}) galaxies. The error bars show the 1-$\sigma$ uncertainties. The vertical black-dashed line represents the best-fit dynamical centre. The light grey shaded area represents the $3\times$ synthesized beam size region centred at the best-fit molecular gas dynamical centre, whereas the dark grey dashed area represents the $3\times$ PSF FWHM zone centred at the best-fit ionized gas dynamical centre. In the rotation velocity profile panels, the dashed-magenta and solid-green curves show the rotation curves extracted from the beam-smeared Pa$\alpha$ and CO(1-0) two-dimensional best-fit models, respectively. In the velocity dispersion profile panels, the green- and magenta-dashed lines show the median galactic value estimated from the outskirts of the galactic disc (Table~\ref{tab:table3}) for the CO(1-0) and Pa$\alpha$ observations, respectively. We find a good agreement between the rotation curves derived from the ionized and molecular gas ISM phases in the three starbursts.}
\end{figure}

We use the best-fit dynamical models to simulate a slit observation along the major kinematic axis and we extract the one-dimensional rotation velocity and velocity dispersion curves for both ISM phases (Fig.~\ref{fig:1d_profiles}). We consider a slit width equal to the synthesized beam or PSF FWHM. The half-light radii for the ionized and molecular gas ISM phases ($R_{\rm 1/2,Pa\alpha}$, $R_{\rm 1/2,CO}$) are calculated by using a tilted ring approach. From the rotation curve, we define the rotational velocity for the Pa$\alpha$ and CO observations ($V_{\rm rot, Pa\alpha}$, $V_{\rm rot, CO}$) as the inclination corrected values observed at two times the Pa$\alpha$ and CO half-light radii, respectively.

To correct the velocity dispersion values for beam-smearing effects, we apply the correction suggested by \citet{Stott2016}. This corresponds to a linear subtraction of the local velocity gradient $\Delta V / \Delta R$ from the beam-smeared line widths. However, to further consider beam-smearing residual effects from this correction, we define the global velocity dispersion for each gas phase ($\sigma_{v,{\rm CO}}$, $\sigma_{v,{\rm Pa\alpha}}$) as the median value taken from the pixels located beyond three times the synthesized beam or PSF FWHM from the dynamical centre (white circumferences in velocity dispersion maps in Fig.~\ref{fig:maps}).

For HATLAS090750, we find a very compact CO light distribution as suggested by its half-light radius. In its central zone, this system shows a low rotational velocity value ($V_{\rm rot,CO} \sim 70$\,km\,s$^{-1}$) and a high median velocity dispersion $\sigma_{v,{\rm CO}} \sim 35$\,km\,s$^{-1}$, suggesting a molecular gas ISM phase with highly supersonic turbulent motions and a CO-traced kinematic ratio $V_{\rm rot,CO} / \sigma_{v,{\rm CO}} \sim 2 $. The CO- and Pa$\alpha$-based rotation curves clearly agree at the radius at which CO(1-0) is detected (Fig~\ref{fig:1d_profiles}), implying that the $V_{\rm rot,CO}$ and $V_{\rm rot, Pa\alpha}$ values also agree. 

In contrast, the Pa$\alpha$ emission tends to show broader line widths compared to the CO emission line ($\sigma_{v,{\rm Pa\alpha}} \sim 66$\,km\,s$^{-1}$). This is unlikely to be produced by beam-smeared flux coming from the asymmetric features as the broader Pa$\alpha$ line widths are seen across all the major kinematic axis. Assuming that the line widths trace the turbulent kinematic state of the respective ISM gas phase, this result suggests that the molecular gas phase may be able to dissipate the turbulent kinetic energy faster than the ionized gas phase. Another possibility could be an additional energy injection in the ionized gas from stellar feedback such as stellar winds, supernovae feedback and/or Wolf-Rayet episodes (e.g. \citealt{Thornton1998,Crowther2007,Kim2015,Martizzi2015,Kim2017}). The expansion of over-pressured H\textsc{ii} regions is also a possibility \citep{Elmegreen2004}. We remind that we have corrected the SINFONI observations by instrumental line broadening effects.

For HATLAS114625 and HATLAS121446, the molecular and ionized gas ISM phases show similar scale sizes $R_{\rm 1/2, Pa\alpha}/R_{\rm 1/2, CO} \approx 0.85 \pm 0.01$ and $0.96 \pm 0.01$, respectively. For HATLAS121446, these half-light radii estimates also agree with  $R_{1/2,K}$ (see Table~\ref{tab:table2}). However, for HATLAS114625, we find that $R_{1/2,K} / R_{\rm 1/2, CO / Pa \alpha} \approx 1.3-1.6$\,kpc, suggesting that the ionized and molecular gas ISM phases are distributed in a more compact disc-like structure in this galaxy.

For both starbursts, the velocity curves agree and we derive ionized to molecular gas rotation velocity ratios $V_{\rm rot,Pa\alpha} / V_{\rm rot,CO} \approx 1.04 \pm 0.04$ and $0.94 \pm 0.14$, for HATLAS121446 and HATLAS114625, respectively. The consistency between the CO- and H$\alpha$-based velocity curves tend to be found in local galaxies where the ionized gas emission seems to come from recent star formation activity episodes \citep{Levy2018}. 

In both systems, the median $\sigma_{v,{\rm CO}}$ values are lower than the corresponding $\sigma_{v,{\rm Pa\alpha}}$ estimates ($\sigma_{v,{\rm Pa\alpha }} / \sigma_{v,{\rm CO}} \sim 1.5-2$), however, those still agree within 1-$\sigma$ uncertainties. The CO and Pa$\alpha$ velocity dispersion values seen in both galaxies suggest a dominant common nature. The CO and Pa$\alpha$ velocity dispersion profiles (Fig.~\ref{fig:1d_profiles}) suggests even closer $\sigma_{v, {\rm Pa\alpha}}$ and $ \sigma_{v, {\rm CO}}$ values. Nevertheless, we note that our measured $\sigma_{v,{\rm CO}}$ values tend to be higher than the estimates reported from local systems ($\approx 9-19$\,km\,s$^{-1}$, \citealt{Levy2018}). Indeed, these median $\sigma_{v,{\rm CO}}$ values are consistent with the lower end of the velocity dispersion estimates measured from ULIRGs ($\sim 30$--140\,km\,s$^{-1}$, \citealt{Downes1998,Wilson2019}).

We derive an average CO-based rotational velocity to dispersion velocity ratio ($V_{\rm rot, CO} / \sigma_{v, {\rm CO}}$) of $ 8 \pm 3$ and $7 \pm 2$ for HATLAS114625 and HATLAS121446, respectively. If we consider the Pa$\alpha$ observations, we derive $V_{\rm rot, Pa\alpha} / \sigma_{v, {\rm Pa\alpha}} \sim 4 \pm 2$ and $\sim 5 \pm 3$, respectively. Independent of the emission line considered, the $V_{\rm rot} / \sigma_v$ ratios measured for HATLAS114625 and HATLAS121446 suggest that the rotational motions are the main support against self-gravity in both starburst galaxies.

\subsubsection{Comparison with previous VALES works}
\label{sec:VALES_comp}

Using our kpc-scale resolution data ($\sim 0 \farcs 5$), we try to test if the previous kinematic analysis done for the VALES galaxies \citep{Molina2019a} may be biased due to beam-smearing effects. These previous CO(1-0) observations were performed by using a more compact ALMA array configuration, thereby delivering a coarser spatial resolution ($\sim3-4" \approx 5-7$\,kpc). Beam-smearing could hide galaxy morpho-kinematic properties, making it hard to recover unbiased intrinsic parameters when the spatial resolution is of the order of several kpc. 

Even though the galaxies presented in this work, HATLAS114625 and HATLAS121446, were not described in \citet{Molina2019a} as they were not extended enough for a dynamical interpretation, we can still make a brief comparison with the VALES systems that share similar global properties.

We concentrate in VALES sources with similar specific SFR values (sSFR$\equiv$SFR/$M_\star$; $\Delta \log($sSFR$)<0.3$\,dex) than the estimated for HATLAS114625 and HATLAS121446 (sSFR = 10--100\,Gyr, respectively). For these sources, we find that the kinematic maps present marginally resolved rotation ($V_{\rm rot, CO} \approx 40-200$km\,s$^{-1}$) and high velocity dispersion values ($\sigma_{v,{\rm CO}}\approx 40-70$\,km\,s$^{-1}$), implying $V_{\rm rot, CO} / \sigma_{v, {\rm CO}}$ ratios in the range of $1-3$. These values are lower than the ones presented in this work, suggesting that the kinematic parameters presented in \citet{Molina2019a} might be systematically biased due to beam smearing. This comparison is not straightforward as the resolution presented in this work is five to seven times higher than in \citet{Molina2019a}, however, it highlights the importance of high-resolution imaging for extracting more precise dynamical information.

\subsection{The CO-to-H$_2$ conversion factor from dynamics} 
\label{sec:aco_dyn}

A CO-to-H$_2$ conversion factor must be used to estimate molecular gas masses from the CO luminosities ($M_{\rm H_2} = \alpha_{\rm CO} L'_{\rm CO}$, e.g. \citealt{Bolatto2013}). Traditionally, two different $\alpha_{\rm CO}$ values have been considered to calculate $M_{\rm H_2}$ for galaxies as a whole \citep{SV2005}. An $\alpha_{\rm CO,MW} \approx 4.6$\,$M_\odot$\,(K\,km\,s$^{-1}$\,pc$^{2}$)$^{-1}$ value seems to be more appropriate for disc-like galaxies (e.g. \citealt{Solomon1987}), whereas an $\alpha_{\rm CO,ULIRG} \approx 0.8$\,$M_\odot$\,(K\,km\,s$^{-1}$\,pc$^{2}$)$^{-1}$ value has been estimated for ULIRGs and assumed to be representative for merger-like systems (e.g. \citealt{Downes1998}). However, it is unlikely that $\alpha_{\rm CO}$ follows a bi-modal distribution. Models suggest a smooth transition that depends on the ISM physical properties (e.g. \citealt{Narayanan2012}).

\begin{figure*} 
\centering
\includegraphics[width=0.8\columnwidth]{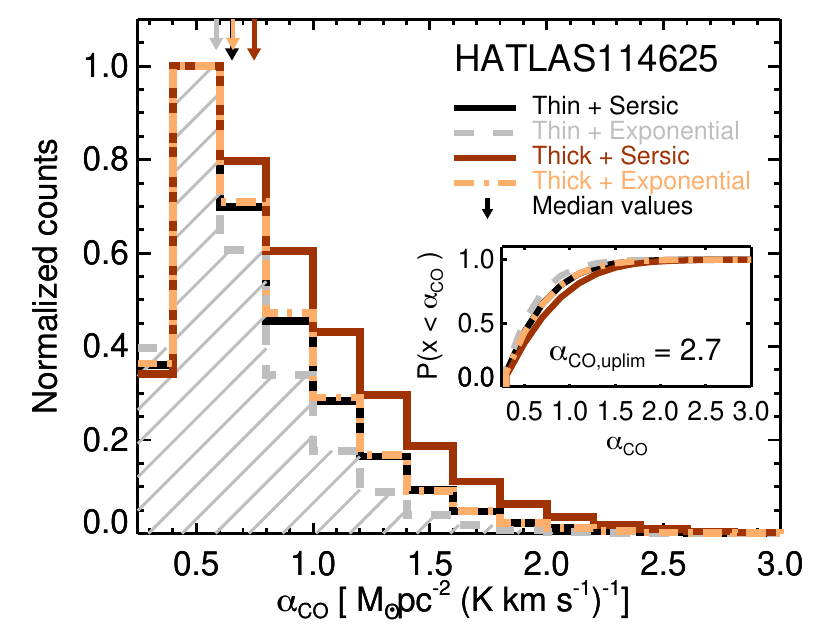}
\includegraphics[width=0.785\columnwidth]{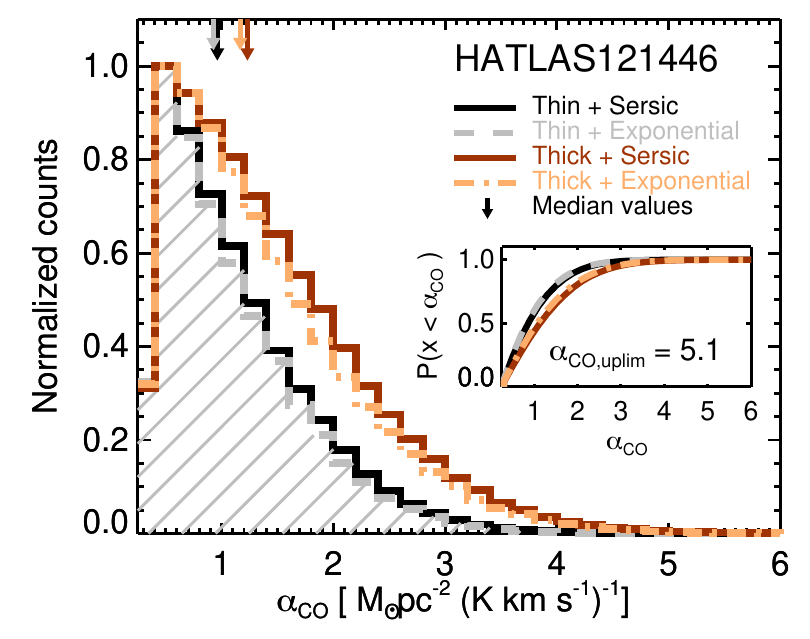}
\caption{\label{fig:aco_pdf} Posterior $\alpha_{\rm CO}$ PDFs for HATLAS114625 (\textit{Left}) and HATLAS121446 (\textit{Right}) starbursts. For each galaxy, we consider four dynamical mass models encompassing different underlying surface density mass distributions and hydrostatic equilibrium approximations. We also show the cumulative probability distribution in each panel with our $\alpha_{\rm CO}$ upper limit defined as $P( x < \alpha_{\rm CO,uplim}) \approx 0.997$ (i.e. 3-$\sigma$) and estimated by using the `thick-disc + S\'ersic' dynamical mass model. The coloured arrows indicate the median $\alpha_{\rm CO}$ value for each PDF. We find median $\alpha_{\rm CO}$ estimates consistent with the ULIRG-like value for both starburst galaxies.}
\end{figure*}

We exploit the dynamical mass estimate [$M_{\rm dyn}(R) = \frac{V_{\rm circ}^2 R}{G}$] to constrain the $\alpha_{\rm CO}$ value. In this procedure, we assume that the dynamical mass estimate corresponds to the sum of the stellar, molecular and dark matter masses (e.g. \citealt{Motta2018,Molina2019b}). This is true when looking at the central regions of galaxies. The H{\sc i} component at larger scales dominates the gas mass, while the ionized gas might have a role as well. Additionally, for the sake of simplicity, we also assume a constant $\alpha_{\rm CO}$ value across each galactic disc. Therefore, by quantifying the dark matter content in terms of the dark matter fraction ($f_{\rm DM}$) at each galactocentric radius, we obtain the following constraint;

\begin{equation}
\label{eq:fdm}
f_{\rm DM}(R) = 1 - \frac{M_\star(R) + \alpha_{\rm CO} L'_{\rm CO}(R)}{M_{\rm dyn}(R)},
\end{equation}

\noindent where the CO luminosities inside each radius are calculated directly from the ALMA observations and the stellar masses are truncated using the $K$-band S\'ersic model profile following \citet{Molina2019b}.

To estimate the dynamical mass values and use Eq.~{\ref{eq:fdm}, first we need to calculate the circular velocity $V_{\rm circ}$ at each galactocentric radius. To do this, we consider two cases, the thin- and thick-disc hydrostatic equilibrium approximations. In the first case, the galaxy support against self-gravity is assumed to be purely rotational and $V_{\rm circ}$ corresponds to the observed rotational velocity ($V_{\rm circ} = V_{\rm rot,CO}$, \citealt{Genzel2015}). In the second case, the galaxy scale height can not be neglected, and the self-gravity is balanced by the joint support between the rotational motions and the pressure gradient across the galactic disc \citep{Burkert2010}. 

In this `thick-disc' approximation, an analytic expression for $V_{\rm circ}$ can be derived by parametrizing the pressure gradients in terms of $\sigma_v$ (which is assumed to be constant across the galactic height and radius) and the mass distribution, which we assume to follow the best-fit S\'ersic model of the $K$-band surface brightness distribution;

\begin{equation}
\label{eq:vcirc_sersic}
V_{\rm circ}^2(R) = V_{\rm rot,CO}^2(R) + 2 \frac{\sigma_v^2 b_{n_S}}{n_S} \left( \frac{R}{R_{1/2,K}} \right) ^{(1/n_s)},
\end{equation}

\noindent where, $V_{\rm rot,CO}(R)$ is the rotation velocity profile (Fig.~\ref{fig:1d_profiles}), $b_{n_S}$ is the S\'ersic coefficient that sets $R_{1/2,K}$ as the $K$-band half-light radius (e.g. \citealt{Burkert2016,Lang2017,Molina2019a}).

The disc radial coordinates are determined by the best-fit two-dimensional model. Additionally, to minimize beam-smearing effects, we only consider the $V_{\rm rot,CO}$ values extracted from a zone beyond three times the synthesized beam FWHM from the dynamical centre (see `$\sigma_v$' panels in Fig.~\ref{fig:maps}). However, as we still expect some residual beam-smearing effect at these radii, we also apply a correction factor ($\lesssim 10$\,\%) to the rotation velocity values based on the ratio between the intrinsic-to-smoothed best-fit $\arctan$ velocity models across the galaxy major kinematic axis (Appendix~\ref{sec:AppendixB}).

We note that this method suffers from a degeneracy between the $\alpha_{\rm CO}$ and $f_{\rm DM}(R)$ parameters, along with it there is a strong dependence on the accuracy of the $M_{\rm dyn}$ and $M_\star$ values. To try to overcome these issues, we use a Markov Chain Monte Carlo (MCMC) technique \citep{Calistro2018,Molina2019a} implemented in \textsc{emcee} \citep{Foreman2013}. We estimate the posterior probability density function (PDF) for the CO-to-H$_2$ conversion factor and the dark matter fraction parameters by sampling the $\alpha_{\rm CO}$--$f_{\rm DM}(R)$ phase-space defined in Eq.~\ref{eq:fdm} and by considering the likelihood of the estimated $L'_{\rm CO}$, $M_{\rm dyn}$ and $M_\star$ values.

Additional to the thin- and thick-disc dynamical model assumptions, we explore the effect of the chosen underlying mass distribution by assuming that the galaxies follow an exponential total-mass surface density distribution \citep{Freeman1970}. We note that this assumption produces a variation in our thick-disc $M_{\rm dyn}$ and truncated $M_\star$ estimates. Thus, we employ a total of four different dynamical models per galaxy.
 
We do not derive an $\alpha_{\rm CO}$ value for the HATLAS090750 system as this on-going merger may not fulfil the virial assumption necessary to obtain a dynamical mass estimate.

\begin{table} 
	\centering
	\setlength\tabcolsep{2pt}
	\renewcommand{\arraystretch}{1.5}
    	\caption{\label{tab:table5} CO-to-H$_2$ conversion factor, molecular gas masses and gas fractions for HATLAS114625 and HATLAS121446 starbursts}
    \vspace{1mm}
	\begin{tabular}{lcc} 
		\hline
		\hline
	 	& HATLAS114625 & HATLAS121446 \\
     	\hline
     	$\alpha_{\rm CO}$\,$M_\odot$\,(K\,km\,s$^{-1}$\,pc$^{2}$)$^{-1}$ & 0.7$^{+0.5}_{-0.3}$ & 1.2$^{+1.0}_{-0.6}$ \\
		$M_{\rm H_2}$\,($\times$\,$10^{9}$\,$M_\odot$) & 6.0$^{+4.3}_{-2.6}$ & 10.3$^{+8.7}_{-5.3}$ \\
     	$f_{\rm H_2} $ & 0.11$^{+0.07}_{-0.05}$ & 0.14$^{+0.10}_{-0.07}$ \\
		\hline                                                                                        
	\end{tabular}
\end{table}

In Fig~\ref{fig:aco_pdf} we show the $\alpha_{\rm CO}$ posterior PDFs for HATLAS114750 and HATLAS121664 starbursts. We note that the thick-disc S\'ersic mass-profile model suggests slightly higher $\alpha_{\rm CO}$ values than the other three models for both galaxies. This is produced by two effects; (1) the additional pressure gradient support against self-gravity, which is low for our galaxies as suggested by the $V_{\rm rot, CO}/\sigma_{v,{\rm CO}} \sim 7$ ratios; and (2) surface density profiles steeper than the ones derived from an exponential model profile as indicated by the S\'ersic indexes $n_{S} \gtrsim 1$. This tends to increase the $M_{\rm dyn}$ values by a larger amount compared to the truncated $M_\star$ values at smaller galactocentric radii. 

We note that a possible systematic overestimation of $M_\star$ by \textsc{magphys} may bias the $\alpha_{\rm CO}$ estimates toward lower values than the reported ones. This scenario is unlikely as we have input a large wavelength SED coverage ($\sim0.1-500$\,$\mu$m) to obtain accurate $M_\star$ values (see also \citealt{Michalowski2014}). However, to be conservative, we assume $\alpha_{\rm CO}$ upper limit values ($\alpha_{\rm CO, uplim}$) given by the PDFs 3-$\sigma$ range. We obtain $\alpha_{\rm CO, uplim} = 2.7$ and 5.1\,$M_\odot$\,(K\,km\,s$^{-1}$\,pc$^{2}$)$^{-1}$ for HATLAS114625 and HATLAS121446, respectively.

In the remaining of this work, we estimate the molecular gas masses by adopting the median CO-to-H$_2$ conversion factor value derived from the thick-disc S\'ersic mass-profile dynamical model, i.e., by considering the model that suggests the higher median $\alpha_{\rm CO}$ value. This election does not affect our results as the differences between the four dynamical models are marginal compared to the uncertainties behind the galaxy estimates as seen by the broad $\alpha_{\rm CO}$ PDFs. Our analysis suggests low $\alpha_{\rm CO}$ values that are consistent with the ULIRG-like value for both starbursts. We present this value along with the molecular gas estimates in Table~\ref{tab:table5}.

A direct result of adopting that $\alpha_{\rm CO}$ value is that our early expectation about observing `gas-rich' systems was wrong. Indeed, the measured $f_{\rm H_2}$ values are consistent with the average estimate for local star-forming galaxies ($f_{\rm H_2} \sim 0.1$, \citealt{Leroy2009,Saintonge2017}). This suggests that these starburst galaxies may not be `gas-rich' as originally expected, implying that without a robust molecular gas estimate, it is not straightforward to catalogue these systems as possible analogues of the high-$z$ SFG population.

\section{Discussion}

\subsection{What sets the molecular gas velocity dispersions?}
\label{sec:vel_disp}
HATLAS114625 and HATLAS121446 present $\sigma_{v,{\rm CO}}$ values that are comparable with the lower end estimates observed in ULIRGs ($\sigma_{v,{\rm CO}} \approx $30--140\,km\,s$^{-1}$, \citealt{Downes1998,Wilson2019}). However, both galaxies show regular disc-like kinematics with little evidence of interactions that may enhance the internal $\sigma_{v,{\rm CO}}$ values, suggesting that the high molecular gas velocity dispersion values may be produced by internal secular processes. 

\citet{Wilson2019} found that, in ULIRGs, the $\sigma_{v,{\rm CO}}$ values roughly increase with the molecular surface density ($\Sigma_{\rm H_2}$), following a power-law relationship with a tentative exponent of $\sim 0.5$. \citet{Wilson2019} suggested that this correlation can be explained if ULIRGs are in vertical pressure balance. In this section, we explore if their model is able to explain the $\sigma_{v,{\rm CO}}$ values measured for the HATLAS114625 and HATLAS121446 galaxies. We choose \citet{Wilson2019}'s ISM pressure balance model because we lack of stellar velocity dispersion measurement for our sources. This quantity is required to calculate the pressure set by self-gravity in other ISM models such as, for example, in the traditional \citet{Elmegreen1989}'s ISM model.

In \citet{Wilson2019}'s model, a downward pressure on the molecular gas (modelled as a gas layer) is produced by the disc self-gravity, plus an additional contribution from the dark matter halo. This pressure can be calculated as;

\begin{equation}
\label{eq:Pgrav}
P_{\rm grav,W+19} = 0.5 \pi G \Sigma_{\rm H_2} \Sigma_{\rm Tot} (1 + \gamma),
\end{equation}

\noindent where $\Sigma_{\rm Tot}$ is the disc total mass surface density and $\gamma$ is a factor that accounts for the vertical pull toward the galaxy mid-plane produced by dark matter. This factor depends inversely on $V_{\rm rot}/\sigma_v$ squared, thus, it contributes a small correction for both galaxies ($\gamma \sim 0.05$). 

The upward pressure is parametrized as a function of the average mid-plane density $\rho_{\rm mid}$ and the thermal plus turbulent velocity dispersions;

\begin{equation}
\label{eq:PISM}
P_{\rm ISM} = \rho_{\rm mid} \sigma_{v,{\rm H_2}}^2 (1 + \psi) = \frac{\Sigma_{\rm H_2} \sigma_{v,{\rm H_2}}^2 (1 + \psi)}{2 h_{\rm H_2}},
\end{equation}

\noindent where $\rho_{\rm mid} = \Sigma_{\rm H_2} / 2 h_{\rm H_2}$, $h_{\rm H_2}$ is the molecular gas disc scale height, $\sigma_{v,{\rm H_2}}$ is the molecular gas velocity dispersion (hereafter we assume $\sigma_{v,{\rm H_2}} \approx \sigma_{v,{\rm CO}}$) and $\psi$ is a factor which accounts principally for the magnetic-to-thermal support ratio ($\sim 0.3$, \citealt{Kim2015}) as the cosmic ray to turbulent support ratio is negligible (see \citealt{Wilson2019}, for more details).

The vertical equilibrium condition requires $P_{\rm grav,W+19} = P_{\rm ISM}$ and allows us to write the \citet{Wilson2019}'s Eq.~6 in a more compact form;

\begin{equation}
\label{eq:h}
h_{\rm H_2} = \frac{\sigma_{v, {\rm CO}}^2}{\pi G \Sigma_{\rm Tot}} \times \left( \frac{1 + \psi}{1 + \gamma} \right).
\end{equation}

\noindent From this equation, if galaxies have similar $\Sigma_{\rm H_2} / \Sigma_{\rm Tot}$ ratio, then $\sigma_{v,{\rm CO}} \propto \Sigma_{\rm H_2}^{0.5}$, and $h_{\rm H_2}$ is constant across the galactic radius, i.e., the correlation found by \citet{Wilson2019} for local ULIRGs.

\begin{figure*}
\includegraphics{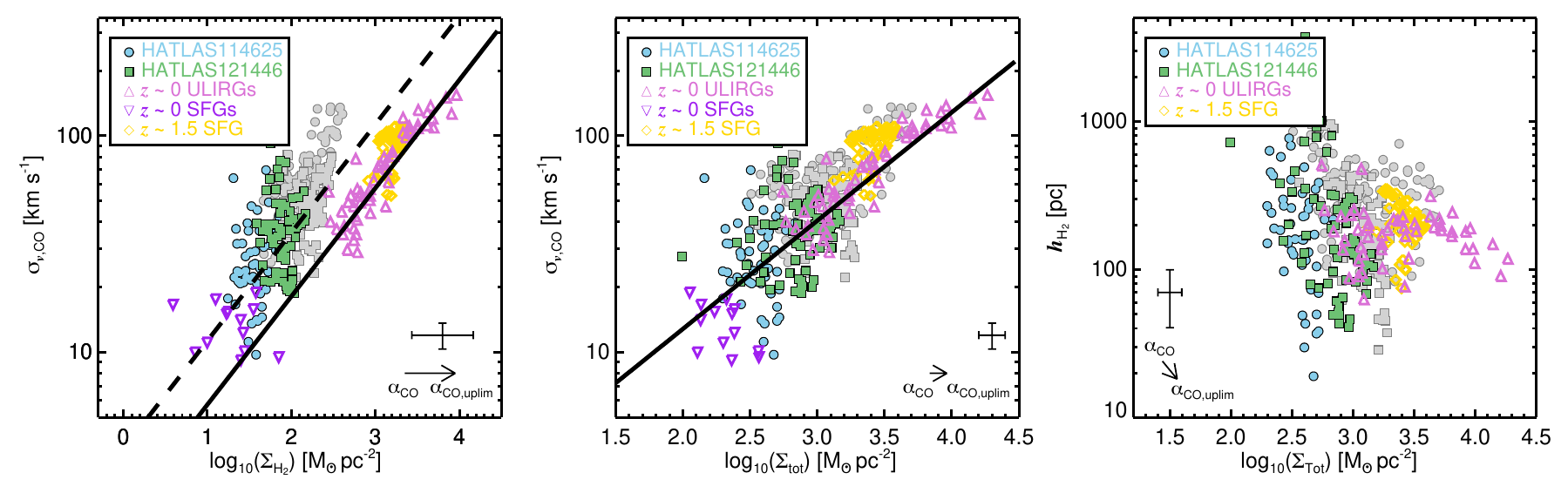}
\caption{\label{fig:Pressure_plot} \textit{Left:} Pixel-by-pixel molecular gas velocity dispersion estimates as a function of molecular gas surface density. For each starburst galaxy, in grey colour, we show the $\sigma_{v,{\rm CO}}$ values that may be overestimated due to beam-smearing residual effects. The error bar in the lower-right corner indicates the typical 1-$\sigma$ uncertainty, whereas the arrow represents the systematic uncertainty given by the use of our $\alpha_{\rm CO}$ upper limit instead of the adopted value. The `$z\sim0$ ULIRG' sample is taken from \citet{Wilson2019}. The `$z\sim0$ SFGs' sample estimates are galactic average values measured for a sub-sample of galaxies taken from the CARMA-EDGE survey \citep{Bolatto2017,Levy2018}. The `$z\sim1.5$ SFG' data correspond to the $\sim$kpc-scale measurements for a main-sequence galaxy presented in \citet{Molina2019b}. The solid line represents the empirical relationship suggested by \citet{Wilson2019}. The dashed line shows the empirical relationship corrected by the average $\Sigma_{\rm H_2} / \Sigma_{\rm Tot}$ and $V_{\rm rot,CO}/ \sigma_{v,{\rm CO}}$ ratios measured for both systems. \textit{Middle:} Pixel-by-pixel $\sigma_{v,{\rm CO}}$ values as a function of the total surface density. \textit{Right:} Molecular gas scale height as a function of the total gas surface density. The last two panels are colour-coded in the way as the \textit{left} panel. The vertical pressure equilibrium model gives a reasonable representation of our data.} 
\end{figure*}

In the left panel of Fig.~\ref{fig:Pressure_plot}, we plot the pixel-by-pixel $\sigma_{v,{\rm CO}}$ values as a function of $\Sigma_{\rm H_2}$ (corrected by projection effects) for HATLAS114625 and HATLAS121446. All the values associated with the pixels that reside inside the central galactic region\footnote{It is defined as the region within three times the synthesized beam size from the best-fit galactic dynamical centre (see Fig.~\ref{fig:maps}).} are shown in grey colour, highlighting that these $\sigma_{v,{\rm CO}}$ values are likely to be overestimated due to beam-smearing residual effects. 

We also show the values presented by \citet{Wilson2019} for the local ULIRG sample (measured at $450-650$\,pc scales) and the average galactic values for a sub-sample of 17 SFGs taken from the EDGE-CALIFA survey \citep{Bolatto2017}. For these SFGs, the $\sigma_{v,{\rm CO}}$ values are taken from \citet{Levy2018}, whereas the average $\Sigma_{\rm H_2}$ values are calculated by using the $M_{\rm H_2}$ and $R_{1/2, \rm CO}$ estimates presented in \citet{Bolatto2017} and assuming a radial spatial extension of $2 \times R_{1/2, \rm CO}$. The `$z\sim1.5$' SFG data correspond to $\sim$kpc-scale measurements for a main-sequence galaxy presented in \citet{Molina2019b}. 

Despite of comparing with data observed at different spatial resolutions, both starbursts exhibit $\sigma_{v, \rm CO}$ values mainly in the range between the local SFGs and ULIRGs, but their $\sim$kpc-scale $\Sigma_{\rm H_2}$ values are comparable to the average estimates measured for the local SFGs and much lower than the estimates reported for the ULIRG sample. However, similar to the ULIRGs, the starburst data seem to follow a roughly $\sigma_{v,{\rm CO}} \propto \Sigma_{\rm H_2}^{0.5}$ power-law relationship. This is shown by the dashed line which represents the pressure balance model suggested by \citet{Wilson2019}, but scaled to the average $\Sigma_{\rm H_2}/ \Sigma_{\rm Tot}$ and $V_{\rm rot,CO}/ \sigma_{v,{\rm CO}}$ ratios measured for both systems. Additionally, the solid line shows \citet{Wilson2019}'s model for the local ULIRGs.

We now consider the total surface density $\Sigma_{\rm Tot}$ (Fig.~\ref{fig:Pressure_plot}). We approximate $\Sigma_{\rm Tot}$ by the sum of $\Sigma_{\rm H_2}$ and the stellar surface density $\Sigma_\star$ ($\Sigma_{\rm Tot} \equiv \Sigma_{\rm H_2} + \Sigma_\star $). For each starburst, the pixel-by-pixel $\Sigma_\star$ values are calculated by scaling the SINFONI $K$-band continuum image surface brightness distribution (Fig.~\ref{fig:SINFOcont}) to the global $M_\star$ value derived by \textsc{magphys}.

HATLAS114625 and HATLAS121446 starbursts tend to be located in the lower $\Sigma_{\rm Tot}$ limit covered by the ULIRG sample. Despite of the large scatter ($\approx$0.22\,dex, for non-masked values), the vertical pressure balance model (solid line) gives a reasonable representation of the data. We note that the systematic uncertainty added by the adopted $\alpha_{\rm CO}$ conversion factor is low as the $\Sigma_{\rm Tot}$ values are mainly dictated by $\Sigma_\star$ in both starbursts (sources have low integrated molecular gas fractions; see Table~\ref{tab:table5}). The scatter is probably increased by the use of a constant mass-to-light ratio to estimate $\Sigma_\star$.

By using Eq.~\ref{eq:h} we estimate roughly $h_{\rm H_2}$ for both starbursts. We plot the $h_{\rm H_2}$ pixel-by-pixel distribution in the right panel of Fig.~\ref{fig:Pressure_plot}. From the non-masked pixels, we obtain $h_{\rm H_2} \sim 200^{+250}_{-130} $ and $160^{+570}_{-80}$\,pc median values for HATLAS114625 and HATLAS121446, respectively. Those values are consistent with the average estimate reported for the ULIRG systems ($\sim 150$\,pc; \citealt{Wilson2019}).

Our data support the scenario in which the molecular gas velocity dispersion on large scales ($\sim$kpc-scales) is set by the local gravitational potential of the galaxy through the reaching of the vertical pressure balance as suggested by \citet{Wilson2019}. 

We note that, the main difference between the two starbursts analysed in this work and the ULIRGs presented by \citet{Wilson2019} is that, in the former, the vertical gravitational pressure is mainly dictated by the stellar component ($f_{\rm H_2} \sim 0.1$) and not by a nearly equal gravitational contribution from stars and gas. Indeed, if in Eq~\ref{eq:Pgrav} we take the approximation $\Sigma_{\rm Tot} \sim \Sigma_\star$ and we assume vertical pressure equilibrium, then we obtain $\sigma_{v,{\rm CO}} \propto \Sigma_\star^{0.5}$, suggesting that, even in starburst systems, the molecular gas dynamical properties can be set by the stellar gravity.

\begin{figure*}
\includegraphics[width=1.0\columnwidth]{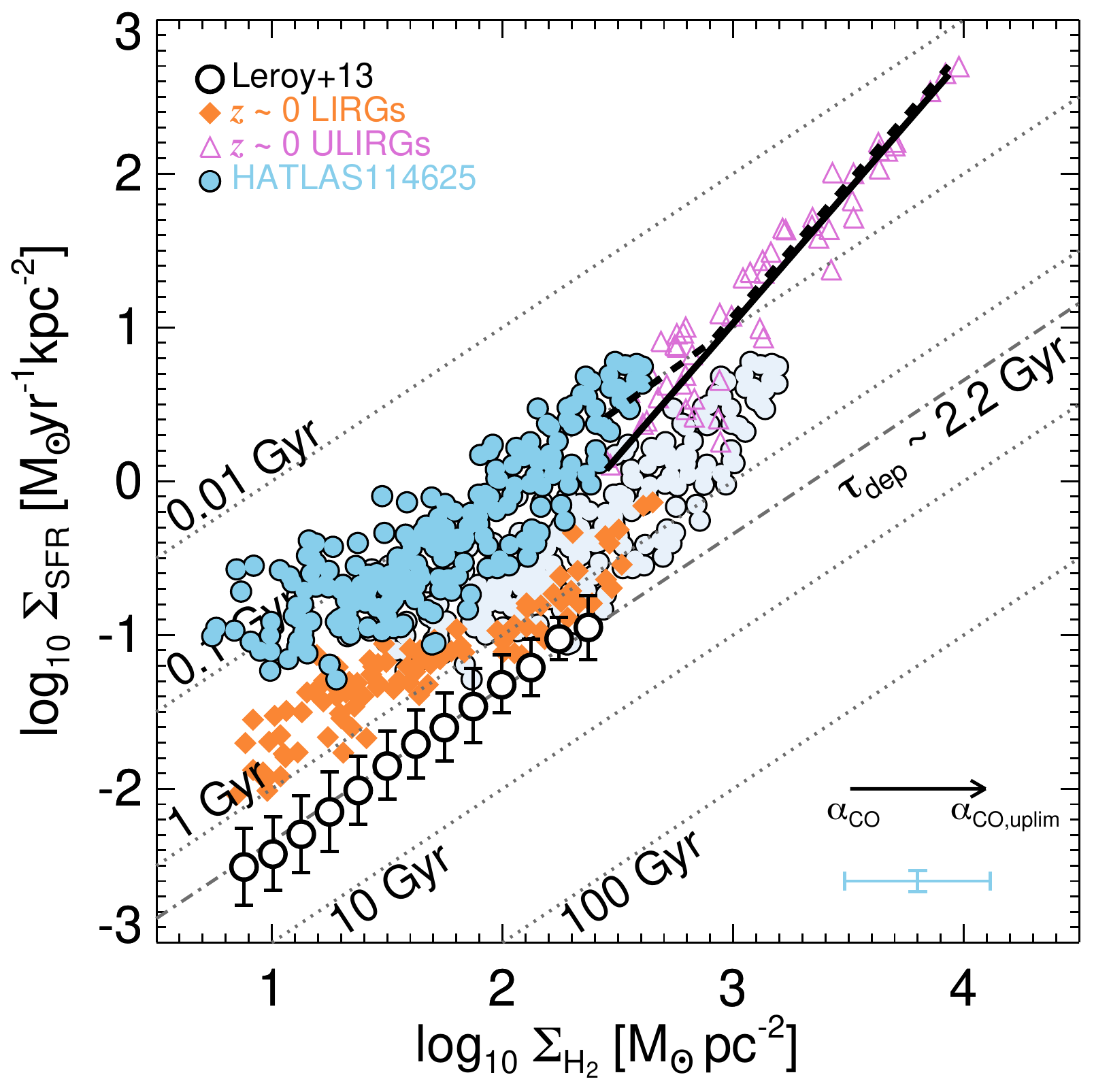}
\includegraphics[width=1.0\columnwidth]{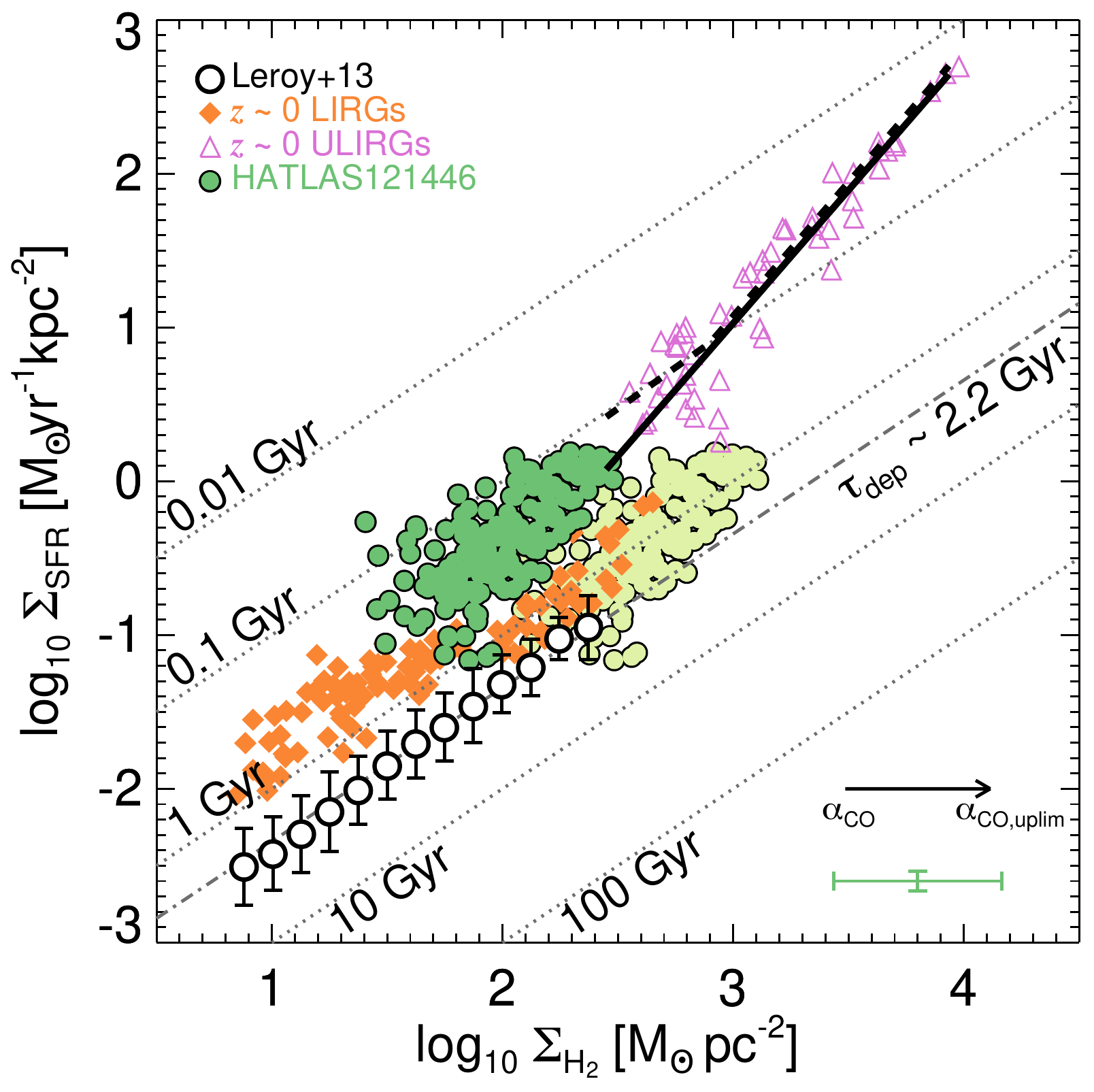}
\caption{\label{fig:SF_plot} $\Sigma_{\rm SFR}$ against $\Sigma_{\rm H_2}$ for HATLAS114625 (\textit{Left}) and HATLAS121446 (\textit{Right}) starbursts. In each panel, the error bar in the bottom-right corner represents the typical 1-$\sigma$ uncertainty. The arrow indicates the horizontal shift of the data produced if we assume the $\alpha_{\rm CO,uplim}$ value instead of the median $\alpha_{\rm CO}$ estimate to calculate $\Sigma_{\rm H_2}$. This is also highlighted by the lightly-coloured data showed in the background. The dotted lines indicate fixed $\tau_{\rm dep}$ values. We show $\sim$kpc-scale spatially-resolved observations of two $z \sim 0$ LIRGs (orange diamonds, \citealt{Espada2018}) and the median trend observed for the HERACLES nearby galaxy survey (open circles, \citealt{Leroy2013}). The dot-dashed line represents the best-fit for the HERACLES $\sim$kpc-scale median values. We also present spatially-resolved estimates for local ULIRGs measured at $\sim$350-650\,pc scales \citep{Wilson2019}. In solid and dashed lines we show the double and single power-law best fits reported by \citet{Wilson2019} for the ULIRG data, respectively. Independent of the $\alpha_{\rm CO}$ value assumed, we find lower $\tau_{\rm dep}$ values than that measured from local normal star-forming galaxies.} 
\end{figure*}

Momentum injected by stellar feedback may be insufficient to produce the observed $\sigma_{v,{\rm CO}} - \Sigma_{\rm Tot}$ trend. Hydrodynamical simulations suggest that stellar feedback can just account for $\sigma_v$ values up to $\sim$6--10\,km\,s$^{-1}$ for the diffuse gas component and with a moderate increase with gas surface density \citep{Ostriker2011,Shetty2012}. However, our data sample the $\sigma_{v, {\rm CO}} \gtrsim 15$\,km\,s$^{-1}$ range and additional pressure sources, such as stellar feedback, may still set the $\sigma_{v, {\rm CO}}$ values below this limit.

Resolution effects should be present as our $\sim$kpc-scale measurements may underestimate the ambient pressure at smaller scales. This effect has been recently measured by high-resolution ($\sim 60$\,pc) molecular gas observations in nearby galaxies \citep{Sun2020}. Indeed, \citet{Sun2020} suggest a correction for the $\sim$kpc-scale pressure estimates. However, their observations cover a considerable lower galactic pressure range ($\sim 10^{4-6}/k_B$\,K\,cm$^{-3}$, see Fig~\ref{fig:SFR_pressure}) and, thus, extrapolating such a correction and applying it to our measurements is uncertain. Nevertheless, these high-resolution observations also suggest that the molecular gas is in pressure balance with its weight and the local ISM self-gravity (see also \citealt{Schruba2019}).

Another major caveat in our analysis comes from the assumption behind Eq.~\ref{eq:Pgrav}. This equation corresponds to a corrected form of the \citet{Spitzer1942} formula for an isothermal layer embedded in a spherical mass component. It does not consider a multi-component composition of the ISM and may not be appropriate to describe the vertical pressure produced by a gaseous plus stellar ISM. For example, in the traditional \citet{Elmegreen1989}'s ISM pressure formula, the $\Sigma_\star$ term is weighted by the ratio between the molecular-to-stellar velocity dispersions ($s \equiv \sigma_{v,{\rm CO}}/ \sigma_{v,\star}$)\footnote{Compared to Eq.~\ref{eq:Pgrav}, $\Sigma_{\rm Tot}$ is replaced by [$\Sigma_{\rm H_2} + s\, \Sigma_\star$] and $\gamma = 0$.}. In this case, the additional vertical pressure set by $\Sigma_\star$ can be neglected in the limit $s<<1$. Only if $s \sim 1$, then Eq.~\ref{eq:Pgrav} is recovered. Thus, Eq.~\ref{eq:Pgrav} should be considered as an upper limit case of the \citet{Elmegreen1989}'s formula. 

\subsection{The star-formation activity traced at $\sim$kpc-scales}
\label{sec:SF_law}

Our CO(1-0) and Pa$\alpha$ observations are ideal for studying the star formation activity in dusty starburst galaxies. The CO(1-0) emission provides a direct estimate to the molecular gas mass (albeit an $\alpha_{\rm CO}$), and Pa$\alpha$ does not suffer from significant extinction (compared to H$\alpha$), facilitating a direct view to the star formation activity in dustier environments. 

The star formation activity can be described as a power-law relationship between the SFR surface density ($\Sigma_{\rm SFR}$) and total gas surface density ($\Sigma_{\rm gas}$) or $\Sigma_{\rm H_2}$, the well-known Kennicutt-Schmidt relationship \citep{Kennicutt1998a}. For typical local star-forming galaxies, when $\Sigma_{\rm H_2}$ is used, it is well-characterized by a linear relation with an observed average molecular gas depletion time $\tau_{\rm dep} \equiv \Sigma_{\rm H_2} / \Sigma_{\rm SFR} = 2.2 \pm 0.3$\,Gyr \citep{Leroy2013}. However, this linear trend seems not to be followed by galaxies with enhanced SFRs as those tend to exhibit shorter molecular gas depletion times or higher star formation efficiencies (SFE$ \equiv \tau_{\rm dep}^{-1}$, e.g. \citealt{Daddi2010}).

In Fig~\ref{fig:SF_plot}, we show the pixel-by-pixel distribution in the $\Sigma_{\rm SFR}$--$\Sigma_{\rm H_2}$ plane for HATLAS114625 and HATLAS121446. The $\Sigma_{\rm SFR}$ and $\Sigma_{\rm H_2}$ quantities are directly estimated from the spatially-resolved SINFONI and ALMA observations assuming the median $\alpha_{\rm CO}$ value (Table~\ref{tab:table5}) and employing the dynamical modelling to correct by projection effects. 

We use the \textsc{hastrom} task written in the Interactive Data Language (\textsc{IDL}) to register the images on the same pixel scales and orientation. While implementing this routine, we consider that the total flux is conserved in each map. We prefer not to include the HATLAS090750 system in our analysis due to its complex geometry and uncertain $\alpha_{\rm CO}$ value.

We compare our $\Sigma_{\rm SFR}$--$\Sigma_{\rm H_2}$ estimates with $\sim$kpc-scale local galaxy measurements and the $\sim$sub-kpc data from local ULIRGs. Briefly, the $\sim$kpc-scale data are represented by the median trend reported from the HERA CO-Line Extragalactic Survey (HERACLES, \citealt{Leroy2008}) for normal star-forming systems and measurements from two LIRGs (NGC3110 and NGC232; \citealt{Espada2018}). The local ULIRG data correspond to CO(1-0)-based $\sim$350-650\,pc-scale estimates presented in \citet{Wilson2019}.

The two galaxies presented in this work exhibit $\tau_{\rm dep}$ values in the range of $\sim$0.1--1\,Gyr, i.e., $\tau_{\rm dep}$ values comparable to that derived for ULIRGs ($\tau_{\rm dep} \lesssim 0.1$\,Gyr), but in a low $\Sigma_{\rm H_2}$ environment. $\tau_{\rm dep}$ internal galactic trends are not clear due to the considerable data scatter. 

If we assume our $\alpha_{\rm CO,uplim}$ estimates, we obtain $\tau_{\rm dep}$ median values of $\sim$0.5 and 0.6\,Gyr for HATLAS114625 and HATLAS121446, respectively. In this case, both starbursts mainly present $\tau_{\rm dep}$ values within $0.2--2.2$\,Gyr, a range similar to that reported for the two local LIRGs (0.2--1.6\,Gyr; \citealt{Espada2018}). In any case, the depletion times estimated for both galaxies are lower than the $\tau_{\rm dep}$ median values reported for local SFGs with similar $\Sigma_{\rm H_2}$ and $f_{\rm H_2}$ values \citep{Leroy2013}. It suggests that the enhancement of the SFE seen in starburst galaxies may not be only related to high $\Sigma_{\rm H_2}$ estimates.

\subsection{Pressure regulated star-formation activity}
\label{sec:P-SFR}

If the star-formation activity also depends on the dynamics of the molecular gas, then $\Sigma_{\rm SFR}$ should also correlate with the physical variables that regulate these properties. Thus, if the vertical pressure equilibrium sets the molecular gas properties on larger scales ($\sim$\,kpc-scales, \S~\ref{sec:vel_disp}), then $\Sigma_{\rm SFR}$ should correlate with the ISM pressure set by self-gravity $P_{\rm grav}$. The correlation between $\Sigma_{\rm SFR}$ and ISM pressure has been suggested and observed previously by many authors (e.g. \citealt{Genzel2010,Ostriker2010,Ostriker2011,Bolatto2017,Herrera2017,Fisher2019}) and also used to explain the so-called `extended' Kennicutt-Schmidt relation \citep{Shi2011}. Indeed, the mid-plane hydrostatic pressure seems to predict the star formation efficiency better than gas surface density in atomic-dominated regimes \citep{Leroy2008}.

\begin{figure}
\includegraphics[width=1.0\columnwidth]{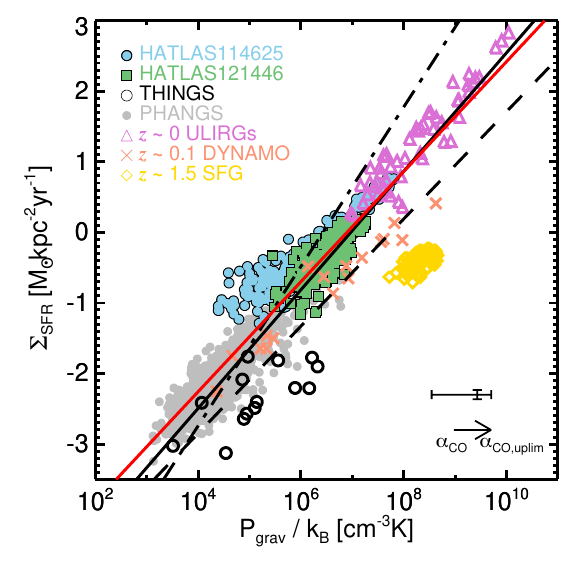}
\caption{\label{fig:SFR_pressure} $\Sigma_{\rm SFR}$ against the ISM pressure set by self-gravity. The PHANGS data correspond to the manually convolved kpc-scale measurements reported by \citet{Sun2020} for 28 nearby-galaxies. The ULIRG data correspond to the values measured at $\sim$350-650\,pc scales by \citet{Wilson2019}. The DYNAMO data are based on galactic ionized gas velocity dispersion measurements along with spatially-unresolved molecular gas observations \citep{Fisher2019}. We also present the $\sim$kpc-scale values measured for a typical star-forming galaxy at $z \sim 1.5$ (SHiZELS-19; \citealt{Molina2019b}). The dashed line corresponds to the best-fit presented by \citet{Fisher2019} for the DYNAMO and THINGS data, while the solid black line corresponds to the best-fit given by \citet{Sun2020} to the PHANGS data. The dot-dashed line corresponds to the parametrization given by \citet{Kim2013} for their set of hydrodynamical simulations. The solid red line represents our best-fit. The starburst and ULIRG data are consistent with the trend reported from the nearby galaxies. The $z \sim 1.5$ star-forming galaxy data are in clear offset.}
\end{figure}

In Fig.~\ref{fig:SFR_pressure}, we show $\Sigma_{\rm SFR}$ as a function $P_{\rm grav}$. The Physics at High Angular resolution in Nearby Galaxies (PHANGS; Leroy et al. in prep.) data correspond to measurements from 28 nearby galaxies presented in \citet{Sun2020} and artificially convolved to kpc-scale by them. We also consider the spatially-resolved ULIRG data presented in \citet{Wilson2019} and The H{\sc{i}} Nearby Galaxy Survey (THINGS; \citealt{Walter2008,deBlok2008}) and DYNAMO \citep{Green2014} galactic averages presented in \citet{Fisher2019}. We caution, however, that the DYNAMO pressure data are based on ionized gas velocity dispersions and unresolved CO emission line measurements, while we are using spatially-resolved molecular gas estimates. We also show the $\sim$kpc-scale data measured for a typical star-forming galaxy at $z \sim 1.5$ \citep{Molina2019b}. 

Our data clearly fill the gap between the ULIRG the PHANGS data in terms of pressure set by self-gravity. We find that $\Sigma_{\rm SFR}$ is correlated with $P_{\rm grav}$ across nearly seven and six orders of magnitude in terms of $P_{\rm grav}$ and $\Sigma_{\rm SFR}$, respectively.

By performing a linear fit in the log-log parameter space ($\log_{10}(\Sigma_{\rm SFR}) = N \times \log_{10}(P_{\rm grav}/P_0)$) using \textsc{emcee} \citep{Foreman2013}, we find a best-fit slope $N = 0.78 \pm 0.01$ with an interceptor value $\log_{10}(P_0/k_B[{\rm cm^{-3} K}]) = 5.38\pm0.04$. This is represented by the red solid line in Fig.~\ref{fig:SFR_pressure}. In this fitting procedure, we have adopted conservative 0.2\,dex uncertainties for the PHANGS data estimates \citep{Sun2020} and we have not included the THINGS sample as we lack of measured uncertainties for those data. Nevertheless, the THINGS data are in rough agreement with the best-fit trend. We note that we have included the $z \sim 1.5$ star-forming galaxy data, but the exclusion of these data only produced a slight variation of the best-fit results ($N = 0.80 \pm 0.01$, $\log_{10}(P_0/k_B[{\rm cm^{-3} K}]) = 5.57\pm0.02$). 

Our best-fit estimates agree within 1-$\sigma$ uncertainty with the values reported for the DYNAMO \citep{Fisher2019}, and THINGS data ($N = 0.76 \pm 0.06$, $\log_{10}(P_0/k_B[{\rm cm^{-3} K}]) = 5.89\pm0.35$). However, we find that \citet{Fisher2019}'s best-fit is offset from the kpc-scale data by $\sim 0.5$\,dex, suggesting that their result was probably affected by the assumptions behind using unresolved CO data.

\citet{Sun2020} report $N = 0.84 \pm 0.01$ and $\log_{10}(P_0/k_B[{\rm cm^{-3} K}]) = 5.85 \pm 0.01$ best-fit estimates for the PHANGS data, but they caution that their best-fit uncertainties may be underestimated due to not considering systematic errors. This problem may be affecting our uncertainty estimates as well. 

To compare with their result, we estimate the r.m.s. of the best-fit residuals. We measure r.m.s.\,$\approx$\,0.34 and $\approx$\,0.36\,dex from our and \citet{Sun2020} best-fits, respectively. We note that these values are considerably increased by considering the $z \sim 1.5$ SFG data. However, these data seem to be an outlier compared to the other kpc-scale measurements and, perhaps, it is produced by an underestimated dust extinction correction applied to the observed SFR for this system. If we do not consider the $z \sim 1.5$ SFG data, we find r.m.s.\,$\approx$\,0.25\,dex from both best-fit residuals. Thus, in both cases, we obtain a good agreement between our and \citet{Sun2020} results. 

The stellar feedback regulated model predicts a $\Sigma_{\rm SFR}$ and ISM pressure correlation with the slope close to unity \citep{Ostriker2011}. In this model, stellar feedback (e.g. photoionization, radiation, supernovae, winds) heats the ISM gas while the energy and pressure losses occur via turbulent dissipation and cooling. The star formation activity gives pressure support against self-gravity (hence, $P_{\rm ISM} = P_{\rm grav}$). $\Sigma_{\rm SFR}$ and $P_{\rm ISM}$ are closely related due to a nearly constant injected feedback momentum per stellar mass formed $p_\star/m_\star$ ($\Sigma_{\rm SFR} = 4 (p_\star/m_\star)^{-1}\,P_{\rm ISM}$; e.g. \citealt{Ostriker2010,Ostriker2011,Shetty2012}). 

This is highlighted by the dot-dashed line in Fig.~\ref{fig:SFR_pressure} which represents the best-fit power-law ($N \approx 1.1$) for the hydrodynamical simulations presented in \citet{Kim2013}. This power-law overestimates the ISM pressure for the ULIRG systems. However, those systems display larger $\Sigma_{\rm SFR}$ values than that covered by the simulations ($\Sigma_{\rm SFR}^{\rm sim} \lesssim 10^{-2}$\,$M\odot$\,yr$^{-1}$\,kpc$^{-2}$; \citealt{Kim2013}).

To further test the plausibility of a $\Sigma_{\rm SFR} \propto P_{\rm grav}$ trend, we fit a linear function to our data. We obtain a best-fit interceptor value $\log_{10}(P_0/k_B[{\rm cm^{-3} K}]) = 6.74 \pm 0.01$ (we caution that the uncertainty may be underestimated due to systematic errors) and a residual r.m.s. of $\approx$0.47 and $\approx$0.31\,dex depending on whether we include the $z \sim 1.5$ SFG data or not. The r.m.s. values are slightly higher than the reported estimates from our previous best-fit. The interceptor value translates to $p_\star/m_\star \sim 4600$\,km\,s$^{-1}$, a value that is $\sim 1.5$ times higher than the estimate typically adopted for single supernovae feedback ($\sim 3000$\,km\,s$^{-1}$; e.g. \citealt{Thornton1998,Martizzi2015,Ostriker2011,Kim2015}). Enhanced feedback from clustered supernovae may be needed to explain such a $p_\star/m_\star $ value (e.g. \citealt{Sharma2014,Gentry2017,Kim2017}). However its effectiveness is still under debate \citep{Gentry2019}. 

Additional energy sources that regulate the star formation activity should also be considered. Mass transport through the galactic disc could be one of these (e.g. \citealt{Krumholz2018}). Nevertheless, this may require understanding how the gas dissipates the gravitational energy from larger scales toward smaller scales (e.g. \citealt{Bournaud2010,Combes2012}). 

Our observations suggest that, in our dusty starburst galaxies, the molecular gas properties seem to be regulated by the pressure set by self-gravity (\S~\ref{sec:vel_disp}). Therefore, a possibility is that the gravitational instabilities may induce large scale gas motions that then are dissipated through energy cascades toward small scales (e.g. \citealt{Bournaud2010,Combes2012}) until the range where stellar feedback operates and stops the local gravitational collapse through momentum/energy injection. In this picture, the observed star formation activity occurs as a response to the ISM pressure balance.

It is also worth mentioning that \citet{Sun2020} calculated $P_{\rm grav}$ by using the `dynamical equilibrium pressure' ($P_{\rm DE}$ e.g. \citealt{Elmegreen1989,Wong2002,Ostriker2010,Ostriker2011,Fisher2019}) which accounts for the gas and stellar galaxy self-gravity in the limit in which the gas disc scale height ($h_{\rm gas}$) is much smaller than the stellar disc scale height ($h_\star$; \citealt{Benincasa2016}). They neglect the gravitational pressure set by the dark matter component as this pressure source is small in galaxies analysed by them. It is also the case of our galaxies and the ULIRGs presented in \citet{Wilson2019}. We determine $P_{\rm grav}$ by using Eq.~\ref{eq:Pgrav}. This equation corresponds to a $P_{\rm DE}$ upper limit case when $h_{\rm gas} \approx h_\star$ (see Appendix A of \citealt{Benincasa2016}). Thus, a possible correction to our $P_{\rm grav}$ estimates may lead to a stepper $\Sigma_{\rm SFR} - P_{\rm grav}$ correlation.

Another possible source of uncertainty comes from the limited spatial resolution from our observations. Beam-smearing effect produces that  $\Sigma_{\rm SFR}$ and $\Sigma_{\rm H_2}$ (hence, $P_{\rm grav}$) are average $\sim$kpc-scale estimates of the patchy underlying surface density distributions. This artificial dilution effect is quantified for the PHANGS molecular gas data (see \citealt{Sun2020} for more details), however, it is unknown for the starburst galaxy population where the ISM is denser. Higher spatial resolution observations ($\sim 10-100$\,pc) may further be needed to quantify this.

\subsection{Limitations of our dynamical mass approach}
\label{sec:caveats}
One of the stronger constraints for the dynamical modelling is given by the assumption of a constant CO-to-H$_2$ conversion factor across the galactic disc. This may not be an ideal assumption due to possible $\alpha_{\rm CO}$ variation with galactocentric radius (e.g. \citealt{Sandstrom2013}). Indeed, lower $\alpha_{\rm CO}$ values are likely to be measured towards the Galactic centre \citep{Bolatto2013}. Based on the dynamical mass modelling, it is highly uncertain to determine an $\alpha_{\rm CO}$ radial variation due to the unconstrained dark matter fraction values. We need to assume a halo model. However, we can not accurately constrain the halo properties as the observed rotation curves do not extend more than $\sim$6\,kpc away from the galactic centre (Fig.~\ref{fig:1d_profiles}). A constant $\alpha_{\rm CO}$ assumption is reasonable given our data limitations.

The dynamical interpretation is mainly limited by the degeneracy between $\alpha_{\rm CO}$ and $f_{\rm DM}$ values constrained at the inner galactocentric radius considered in each galaxy ($R \approx 2.3$\,kpc, Fig.~\ref{fig:1d_profiles}). The over/under-estimation of $f_{\rm DM}$ at this radius will bias our result towards lower/higher $\alpha_{\rm CO}$ estimates. In the case that we are underestimating $f_{\rm DM}$, then the median $\alpha_{\rm CO}$ values should be considered as an upper limit of the true values, implying that both starburst galaxies might have lower molecular gas masses and $\alpha_{\rm CO}$ values even lower than that reported for ULIRGs.

On the other hand, a possible overestimation of the $f_{\rm DM}$ does not affect our results. If we assume the limit case that there is no dark matter, then the median $\alpha_{\rm CO}$ values increase by $\lesssim$14\,\% compared to these estimated values when the dark matter content is considered. This variation is smaller than the $\alpha_{\rm CO}$ PDF's 1-$\sigma$ range and it is independent of the dynamical mass model assumed (see Fig.~\ref{fig:aco_pdf}).

Another limitation comes from the adopted dynamical mass formula. We have not considered any geometrical factor that should multiply the $V_{\rm rot,CO}$ values in Eq.~\ref{eq:vcirc_sersic}. This is equivalent to the assumption of spherical geometry when calculating enclosed mass estimates from rotational motions. However, this source of uncertainty is expected to not affect our conclusions. For example, the $V_{\rm circ}$ difference between an exponential disk and the equivalent spherical mass distribution is $\lesssim 15$\% and it highly depends on the disc radius (see Fig.~2.17 in \citealt{Binney2008}). By adopting this $V_{\rm circ}$ difference value (translated into the enclosed mass difference), then the median $\alpha_{\rm CO}$ values would increase by $\sim$30\,\%. Again, this variation is smaller than the estimated $\alpha_{\rm CO}$ PDF's 1-$\sigma$ range.

\begin{figure}
\centering
\includegraphics[width=1.0\columnwidth]{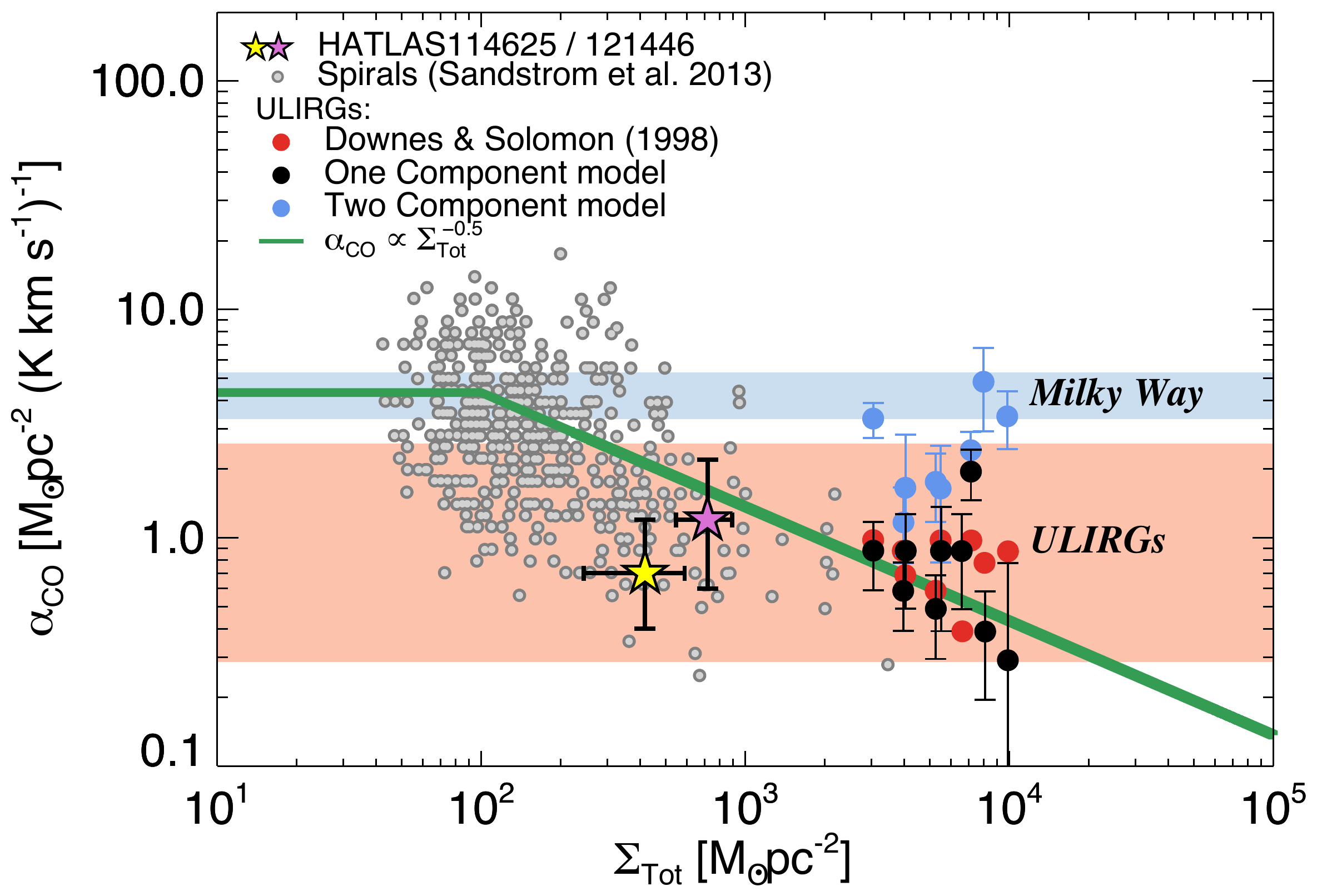}
\caption{\label{fig:aco_vs_density} Median CO-to-H$_2$ conversion factor estimates as a function of the total surface density. The grey circles correspond to the $\alpha_{\rm CO}$ values based on dust emission for nearby disc galaxies \citep{Sandstrom2013}. The ULIRG data correspond to the estimates presented in \citet{Downes1998} and \citet{Papadopoulos2012}. In the latter case, we show the $\alpha_{\rm CO}$ values reported from their `one-' and `two-component' multi-transition models. The color bands indicate the traditional CO-to-H$_2$ conversion factors and their uncertainties for Milky Way- and ULIRG-like systems \citep{Bolatto2013}. The green line corresponds to the $\alpha_{\rm CO}$ parametrization suggested by \citet{Bolatto2013} for galaxies with solar metallicity. For HATLAS114625, we find a lower  $\alpha_{\rm CO}$ estimate than the expected value from \citet{Bolatto2013}'s parametrization. In the case of HATLAS121446, we find an agreement within 1-$\sigma$ uncertainty. Figure adapted from \citet{Bolatto2013}.}
\end{figure}

The CO(1-0) emission source is also a concern. From our $\sim$kpc-scale observations, we can not separate between the CO(1-0) emission originated in GMCs and any significant molecular gas diffuse emission. This diffuse component may not be negligible in galaxies (e.g \citealt{Goldsmith2008,Schinnerer2010}) and it is enhanced in dense and high ISM pressure galactic environments \citep{Sandstrom2013}. If the diffuse molecular gas phase dominates the CO(1-0) emission, then low $\alpha_{\rm CO}$ values may underestimate the high density molecular gas mass content for both starbursts \citep{Papadopoulos2012}. This effect is highlighted by the difference in the estimated $\alpha_{\rm CO}$ values between the `one-' and `two-component' models for ULIRGs (Fig.~\ref{fig:aco_vs_density}, see \citealt{Papadopoulos2012}). 

However, it should be noted that molecular gas diffuse emission and its contribution to the estimated $\alpha_{\rm CO}$ values are uncertain. The CO-to-H$_2$ conversion factor may vary significantly for this gas component depending on local environment properties (e.g. \citealt{Liszt2012}). Additional observations tracing the dense molecular gas phase may help to determine whether our $\alpha_{\rm CO}$ estimations are biased toward low values or not.

We note that we have not considered the H{\sc i} content in our dynamical mass approach as we expect a negligible amount of H{\sc i} mass within the inner radius ($R \approx 2.3$\,kpc) at which $M_{\rm dyn}$ was calculated. We remind that the $M_{\rm dyn}$ values calculated at these radii are the ones that strongly constrain the $\alpha_{\rm CO}$ estimates in our procedure. In local spirals, the transition between an H$_2$- to H{\sc i}-dominated ISM ($\Sigma_{\rm H_2} \approx \Sigma_{\rm HI}$) occurs at $\Sigma_{\rm gas} \sim 12 \pm 6$\,$M_\odot$\,pc$^{-2}$ \citep{Leroy2008}. From the spatially-resolved CO(1-0) observations and the adopted $\alpha_{\rm CO}$ values, we estimate $\Sigma_{\rm H_2} \sim 146 \pm 93$ and $127 \pm 50$\,$M_\odot$\,pc$^{-2}$ at $R \approx 2.3$\,kpc for HATLAS114625 and HATLAS121446 galaxies, supporting our assumption.

Finally, we check if our $\alpha_{\rm CO}$ estimates are reasonable given by the theoretical expectation \citep{Bolatto2013}. From theories, it is expected that $\alpha_{\rm CO}$ mainly varies with the system metallicity and total surface density. However, before making any comparison, we note that HATLAS114625 and HATLAS121446 present supra-solar metallicity values in terms of the gas phase oxygen abundance ($12+\log({\rm O/H}) =$\,8.99 and 8.72, respectively)\footnote{We base our metallicity estimates on the \citet{Pettini2004} calibration scale and the [N{\sc ii}] and H$\alpha$ line intensities reported in Table~\ref{tab:appA}. We also adopt a solar abundance of $12+\log({\rm O/H}) = 8.69$ \citep{Asplund2009}.}. We note that above the solar metallicity, we do not expected any significant variation of $\alpha_{\rm CO}$ due to the high metal content \citep{Bolatto2013}. Thus, we simply assume a solar metallicity when comparing with \citet{Bolatto2013}'s theoretical prediction. We show this comparison in Fig.~\ref{fig:aco_vs_density}. We find that the reported median CO-to-H$_2$ values are somewhat lower than the theoretical expectation, but still consistent within the scatter. Our $\alpha_{\rm CO}$ estimates are, therefore, reasonable given the measured $12+\log({\rm O/H})$ and $\Sigma_{\rm Tot}$ values for both starbursts.

\section{Conclusions}
We present new ALMA Cycle-3 and VLT-SINFONI observations tracing the CO(1-0) and Pa$\alpha$ emission lines from three starburst galaxies at $z \sim 0.12$--0.18 taken from the VALES survey. The ALMA observations were designed to deliver spatially-resolved observations of the molecular gas content at $\sim 0\farcs5$, i.e., approximately the resolution of the seeing-limited SINFONI observations ($\sim 0\farcs4-0\farcs8$). Combining near-IR and sub-mm observations, we study the ionized and molecular gas dynamics at $\sim$kpc-scales. 

One target, HATLAS090750, presents highly asymmetric morpho-kinematics, suggesting an on-going merger. From its SINFONI observation, we also detected fainter ro-vibrational warm molecular gas transitions along with an ionized Helium line in its central galactic zone. The H$_2$ emission lines were used to determine a warm molecular gas temperature of $\sim 1700 \pm 500$\,K (Fig~\ref{fig:H2_diagram}), probably produced by supernovae remnant shocks. However, we could not discard non-thermal gas excitation sources due to the lack of H$_2$ emission line intensity measurements with different vibrational energy levels (e.g. \citealt{Davies2003}).

The other two starburst galaxies, HATLAS114625 and HATLAS121446, show disc-like morpho-kinematics, with rotation as the dominant component supporting self-gravity ($V_{\rm rot}/\sigma_v \sim 7-8$). For both systems, we model the CO and Pa$\alpha$ galactic dynamics and we aid the kinematic modelling by using $K$-band photometric models to constrain the inclination angle parameter. From those analyses, we find that the CO and Pa$\alpha$ dynamics present a good agreement in both galaxies (Fig.~\ref{fig:maps}), as suggested by similar kinematic position angles ($\Delta$PA$\sim 4$\,deg.) and emission line spatial extensions ($R_{\rm 1/2, Pa\alpha}/R_{\rm 1/2, CO} \sim 1$). Our observations suggest that the ionized and molecular gas components display roughly the same galaxy morpho-kinematics.

We estimate the total mass budget for both, HATLAS114625 and HATLAS121446 galaxies, by calculating the dynamical masses assuming the `thin-' and `thick-disc' hydrostatic equilibrium approximations \citep{Burkert2010} along with two different surface density profile models. We obtain $\alpha_{\rm CO}$ values in the range of 0.7--1.2\,$M_\odot$\,(K\,km\,s$^{-1}$\,pc$^{2}$)$^{-1}$ for both galaxies (Fig~\ref{fig:aco_pdf}), i.e., similar values to that reported for ULIRGs \citep{Downes1998}. These values do not depend strongly on the hydrostatic equilibrium and total surface density assumptions. Our conversion factor estimates are somewhat lower but still consistent with the $\alpha_{\rm CO} \propto \Sigma_{\rm Tot}^{-0.5}$ trend suggested from theoretical expectations (Fig.~\ref{fig:aco_vs_density}, \citealt{Bolatto2013}). 

By adopting the dynamically based $\alpha_{\rm CO}$ values, we obtain molecular gas fractions of the order of $\sim 0.1$ for both starbursts, far below our initial expectations and consistent with values measured in local star-forming galaxies. Therefore, the sources are not `gas-rich' as initially thought, highlighting the difficulties to estimate molecular gas masses due to the uncertainty of the CO-to-H$_2$ conversion factor.

We find that the HATLAS114625 and HATLAS121446 molecular gas velocity dispersion values are reasonably represented by a $\sigma_{v,{\rm CO}} \propto \Sigma_{\rm tot}^{0.5}$ trend, where $\Sigma_{\rm tot}$ is the total galaxy surface density (Fig.~\ref{fig:Pressure_plot}). This suggests that the molecular gas velocity dispersion values are consistent with being set by the galaxy self-gravity to maintain the vertical pressure balance.

We study the star formation activity traced at $\sim$kpc-scales in HATLAS114625 and HATLAS121446 starbursts. Both galaxies exhibit $\tau_{\rm dep}$ values in the range of $\sim$0.1--1\,Gyr (Fig.~\ref{fig:SF_plot}), i.e., values consistent with the reported estimates for ULIRGs ($\tau_{\rm dep} \sim 0.1$\,Gyr). However, both systems present $\Sigma_{\rm H_2}$ values that are comparable to those seen in local star-forming disc galaxies \citep{Leroy2013}. This suggests that the decrease of $\tau_{\rm dep}$ (or enhancement of SFE) is also produced by additional physical processes that may not only be related to high $\Sigma_{\rm H_2}$ environments. 

To further explore this, we study the correlation between $\Sigma_{\rm SFR}$ and the gravitational pressure $P_{\rm grav}$. HATLAS114625 and HATLAS121446 fill the gap between the normal galaxies and the ULIRG systems in terms of pressure set by self-gravity. We find a linear relation in the log-log space, $\log_{10}(\Sigma_{\rm SFR}) = N \times \log_{10}(P_{\rm grav}/P_0)$, where $N = 0.78 \pm 0.01$ and $\log_{10}(P_0/k_B[{\rm cm^{-3} K}]) = 5.38\pm0.04$ (Fig.~\ref{fig:SFR_pressure}). It is in agreement with the trend reported for local galaxies \citet{Sun2020}, suggesting that, in these $z \sim 0.12-0.17$ dusty starburst galaxies, the star formation activity can be a consequence of the ISM pressure balance.

\begin{acknowledgements}

We thank the anonymous referee for her/his careful reading of our manuscript as well as helpful comments and suggestions. This work was supported by the National Science Foundation of China (11421303, 11721303, 11890693, 11991052) and the National Key R\&D Program of China (2016YFA0400702, 2016YFA0400703). We acknowledge Prof. Fisher for kindly sharing his data. J.~M. and A.~E. acknowledge support from CONICYT project Basal AFB-170002 and Proyecto Regular FONDECYT (grant 1181663). E.~I. acknowledges partial support from FONDECYT through grant N$^\circ$ 1171710. N.~G. acknowledges financial support from ICM N\'ucleo Milenio de Formaci\'on Planetaria, NPF and grant support from project CONICYT-PFCHA/Doctorado Nacional/2017 folio 21170650. C.~C. is supported by the National Natural Science Foundation of China, No. 11803044, 11933003, 11673028 and by the National Key R\&D Program of China grant 2017YFA0402704. M.~J.~M.~acknowledges the support of the National Science Centre, Poland through the SONATA BIS grant 2018/30/E/ST9/00208. This work is sponsored (in part) by the Chinese Academy of Sciences (CAS), through a grant to the CAS South America Center for Astronomy (CASSACA). This paper makes use of the following ALMA data: ADS/JAO.ALMA\#2015.1.01012.S. ALMA is a partnership of ESO (representing its member states), NSF (USA) and NINS (Japan), together with NRC (Canada), MOST and ASIAA (Taiwan), and KASI (Republic of Korea), in cooperation with the Republic of Chile. The Joint ALMA Observatory is operated by ESO, AUI/NRAO and NAOJ. This work is also based on observations collected at the European Organization for Astronomical Research in the Southern Hemisphere under ESO programme ID 099.B-0479(A). 

\end{acknowledgements}

%
\bibliographystyle{aa} 
\bibliography{bibliography} 
%

\begin{appendix} 

\section{Emission line and WISE colour fluxes}
\label{sec:AppendixA}
\begin{table}[!h]
	\centering
	\tiny
	\setlength\tabcolsep{2pt}
	\renewcommand{\arraystretch}{1.5}
    	\caption[tablename=C1]{\label{tab:appA} Summary of the emission line and WISE colour fluxes used in this work and taken from the GAMA DR3 \citep{Baldry2018}. $W1$, $W2$, and $W3$ correspond to the `profile-fit photometry' WISE filter fluxes measured at 3.4, 4.6 and 12$\mu$m \citep{Cluver2014}. The uncertainties indicate 1-$\sigma$ errors. Our three galaxies are catalogued as unresolved by WISE.}
    \vspace{1mm}
	\begin{tabular}{lcccccc} 
		\hline
		\hline
		Model & HATLAS090750 & HATLAS114625 & HATLAS121446 \\
     	\hline
     	f$_{\rm H \beta}$\,($\times$\,$10^{-17}$\,erg/s/cm$^2$)  & 552$\pm$16    & 76$\pm$23    & 29$\pm$11 \\
     	f$_{\rm [O\,III]}$\,($\times$\,$10^{-17}$\,erg/s/cm$^2$)       & 402$\pm$14    & 196$\pm$34  & 11$\pm$8 \\
     	f$_{\rm H \alpha}$\,($\times$\,$10^{-17}$\,erg/s/cm$^2$) & 2680$\pm$33 & 307$\pm$18  &  588$\pm$25 \\
    		f$_{\rm [N\,II]}$\,($\times$\,$10^{-17}$\,erg/s/cm$^2$)         & 1016$\pm$17  & 444$\pm$17  & 286$\pm$17 \\
    		$W1$\,(mJy) & 1.15$\pm$0.03 & 1.57$\pm$0.03 & 0.59$\pm$0.02 \\
     	$W2$\,(mJy) & 1.04$\pm$0.03 & 2.40$\pm$0.05 & 0.66$\pm$0.02 \\
    		$W3$\,(mJy) & 16.89$\pm$0.41 & 11.99$\pm$0.34 & 8.65$\pm$0.35 \\
		\hline                                                                                        
	\end{tabular}
\end{table}

\section{Experimental observation results}
\label{sec:Jitter-results}

We performed an on-source experimental jittering pattern to optimise the S/N of the Pa$\alpha$ emission line for HATLAS090750. In this experimental observation, the pointing was kept fixed at the galaxy location. Hence, an `OOOO' jitter sequence was used. As we do not observe the sky during this observation, then the emission and telluric absorption bandpass lines need to be subtracted using sky models. We specifically used the \textsc{SkyCor} and \textsc{Molecfit} pipelines. It implies that the performance of this experimental observation is mainly limited by the accuracy of the sky emission modelling and the ability to detect the line emission (S/N$\gtrsim 15$; Godoy et al. in prep.).

Fig.~\ref{fig:1d_spectrum} clearly shows that we were able to detect the targeted emission line (Pa$\alpha$) for this galaxy (S/N\,=\,80). We were also able to observe many other fainter near-IR emission lines (Table~\ref{tab:table4}) that allowed us to study the conditions of the host molecular gas ISM (Appendix~\ref{sec:NIR-lines}). Faint tidal features were also detected for this system, demonstrating the ability of this observation to recover the galaxy spatial distribution when compared, for example, to the CO(1-0) observation (Fig.~\ref{fig:maps}). Thus, this experimental observation delivered more information than that initially expected and allowed us to study HATLAS090750 in great detail.

Even though we have not provided a reliable flux calibration for HATLAS090750, we note that this was because we were unable to accurately model the telluric standard star (HD\,56006) observation. For most of the observed telluric standards, the effective temperature agrees well in the literature with very low dispersion, allowing an appropriate calibration using a black body. However, HD\,56006 has very uncertain stellar parameters (e.g. \citealt{Niemczura2009,Lefever2010}), making a reliable calibration impossible even in the case when more sophisticated approaches were attempted. We highlight that this problem is only related to this standard star observation. If a flux standard star had been observed, then the `OOOO' observation flux calibration would have been straightforward. Thus, we conclude that our observational experiment was a success. We recommend adopting this experimental `OOOO' observational setup when the expected line emission S/N is $\sim$15 or greater.

\section{Near-IR emission lines}
\label{sec:NIR-lines}
Near-IR emission lines are useful to determine the excitation mechanism of the gaseous line-emitting ISM. Similar to the well-established optical diagnostic diagrams (e.g. BPT diagram), the ratio between star-formation line tracers (e.g. Pa$\alpha$, Pa$\beta$) and shock tracers such as ro-vibrational molecular hydrogen emission lines (e.g. H$_2$(1-0)S(1)$\lambda 2.122 \mu$m) or forbidden iron lines ([Fe{\sc ii}]$\lambda 1.64 \mu$m) can be used to determine the dominant excitation mechanism (e.g. \citealt{Larkin1998,Riffel2013,Fazeli2019}). 

\begin{table} 
	\centering
	\tiny
	\setlength\tabcolsep{4pt}
    	\caption{\label{tab:table4} Spatially-integrated Br$_\delta \lambda 1.944 \mu$m, ro-vibrational H$_2$(1-0)S(3)$\lambda 1.957 \mu$m, H$_2$(1-0)S(2)$\lambda 2.033 \mu$m, H$_2$(1-0)S(1)$\lambda 2.122 \mu$m and He{\sc i}$\lambda 2.058 \mu$m fluxes relative to the Pa$\alpha$ emission line flux. We note that the H$_2$(1-0)S(1) emission is redshifted out of the SINFONI $K$-band wavelength range for the later two galaxies. The upper emission line flux limits are calculated as r.m.s.\,$\times \, \sigma_{\rm Pa\alpha, ch}$, where the r.m.s. values are estimated within a 30 channel spectral window width centred at the expected emission line location in the spectra, and $\sigma_{\rm Pa\alpha, ch}$ is the Pa$\alpha$ emission line width.}
    	\vspace{1mm}
	\begin{tabular}{lccc} 
		\hline
		\hline
		& HATLAS090750 & HATLAS114625 & HATLAS121446 \\
		\hline
		Br$_\delta$/Pa$\alpha$ & 0.055$\pm$0.002 & <\,0.12 & <\,0.08 \\
		H$_2$(1-0)S(3)/Pa$\alpha$ & 0.063$\pm$0.005 & 0.107$\pm$0.006 & <\,0.08 \\
		H$_2$(1-0)S(2)/Pa$\alpha$ & 0.028$\pm$0.006 & <\,0.17 & <\,0.33 \\
		H$_2$(1-0)S(1)/Pa$\alpha$ & 0.072$\pm$0.016 & --- & --- \\
		He{\sc i}/Pa$\alpha$ & 0.051$\pm$0.006 & <\,0.27 & <0.29 \\	
		\hline                                                                             
	\end{tabular}
\end{table}

From the central brightest pixels from our SINFONI $K$-band observations, we detected more near-IR emission lines than only Pa$\alpha$ (Fig.~\ref{fig:1d_spectrum}). By considering a circular aperture with a radius equal to the PSF FWHM and centred at the galaxy Pa$\alpha$ luminosity peak, we measure several other emission lines fluxes (see Table~\ref{tab:table4}). These values are estimated by fitting a Gaussian function to each emission line in the spatially-collapsed spectrum and the 1-$\sigma$ uncertainties are derived by bootstrapping via Monte-Carlo simulations the flux-density errors. These ratio values are corrected by assuming a \citet{Calzetti2000} attenuation law ($A_{\rm Pa\alpha} = 0.145\,A_V$, $A_{\rm Br_\delta} = 0.132\,A_V$, $A_{\rm HeI} = 0.113\,A_V$, $A_{\rm H_2(1-0)S(3)} = 0.130\,A_V$, $A_{\rm H_2(1-0)S(2)} = 0.117\,A_V$, $A_{\rm H_2(1-0)S(1)} = 0.103\,A_V$). We note that this correction is determined by the shape of the assumed attenuation law.

We can only apply the near-IR emission line analysis to the HATLAS090750 galaxy as we lack of enough H$_2$ emission line flux measurements for the other two systems. For this galaxy, we use the Pa$\alpha$ flux intensity to differentiate between the possible excitation mechanisms. The other option, the use of the Br$_\gamma$ emission line, is impeded given that this emission line is redshifted out of the SINFONI $K$-band wavelength range. This means that the H$_2$(1-0)S(1)/Br$_\gamma \approx 0.8$ ratio that differentiates between star-forming regions and SNR shocks and compact AGN activity is translated to H$_2$(1-0)S(1)/Pa$\alpha \approx 0.07$ (assuming a Br$_\gamma$-to-Pa$\alpha$ intrinsic ratio of 12.19 (Case B recombination, \citealt{Osterbrock2006}).

HATLAS090750 shows a central H$_2$(1-0)S(1)/Pa$\alpha$ ratio consistent with the median value measured for local LIRGs ($z \lesssim 0.02$; \citealt{Colina2015}). This ratio suggests that there is no clear dominant gas excitation mechanism in this galaxy, likely with a combined effect between UV photons produced by young O-B stars, SNR shocks and weak AGN activity \citep{Colina2015}.

The detection of He{\sc i}$\lambda 2.058 \mu$m suggests that we are witnessing a young central starburst probably induced by the ongoing merger. Unfortunately, we can not constrain the ISM physical properties from this emission line as we lack of the detection of additional He{\sc i} emission lines (e.g. \citealt{Benjamin1999}).

We further characterize the HATLAS090750's ISM warm molecular gas phase by using the $\log N_{\nu,J}/g_{\nu,J} - E$ excitation diagram \citep{Davies2003}. In this diagram, $N_{\nu,J}$ is the molecular column density, $g_{\nu,J}$ is the statistical weight and $E$ is the upper energy level of the ro-vibrational transition. From the H$_2$ emission line fluxes $f_{\nu,J}$, the column densities in the upper ro-vibrational levels are computed as: 

\begin{equation}
N_{\nu,J} \equiv \left( \frac{4\pi f_{\nu,J}}{A_{ul} \Omega} \right) \left( \frac{\lambda}{hc} \right),
\end{equation}

\noindent where $\Omega$ is the spectrum extraction aperture size, $A_{ul}$ is the Einstein coefficient computed by \citet{Wolniewicz1998} and $\lambda/hc$ is the photon energy. Additionally, the $N_{\nu,J}/g_{\nu,J}$ ratios need to be normalized to a specific population distribution value $N_{\nu_0,J_0}/g_{\nu_0,J_0}$ given from a determined H$_2$ transition. We normalize them to the inferred value from the H$_2$(1-0)S(3) emission line transition (e.g. \citealt{Bedregal2009}). We note that, if the warm molecular gas phase is in LTE, then the data are well described by a simple linear function that represents a Boltzmann distribution characterized by a single excitation temperature.

\begin{figure} 
\flushleft
\includegraphics[width=1.0\columnwidth]{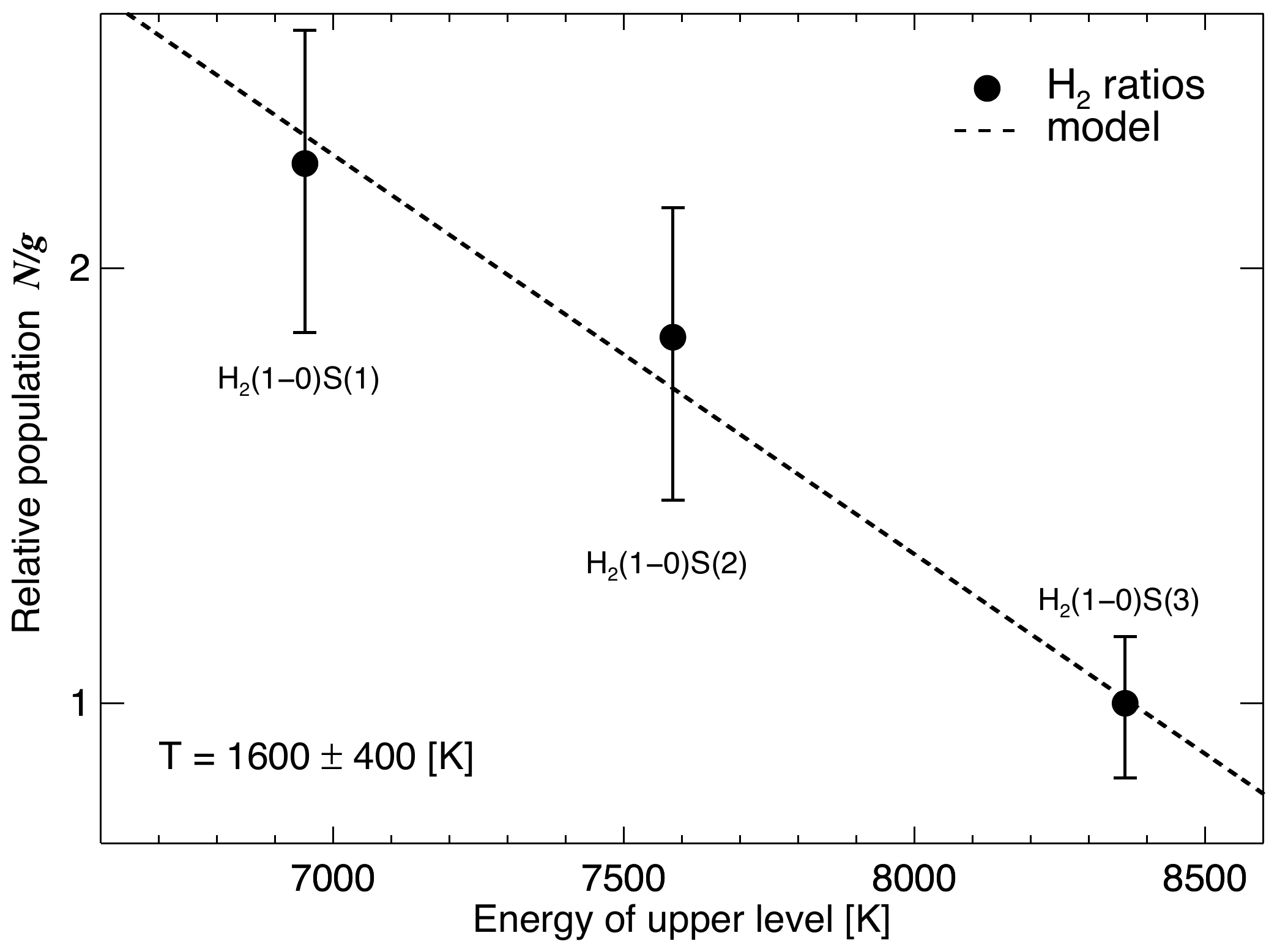}
\caption{\label{fig:H2_diagram} Gas excitation diagram for the H$_2$ emission coming from the central part of the HATLAS090750 galaxy. The $N_{\nu,J}/g_{\nu,J}$ ratios are normalized to the value inferred from the H$_2$(1-0)S(3) emission line transition. The population levels are well-characterized by a Boltzmann distribution suggesting that the warm H$_2$ gas is in LTE.}
\end{figure}

In Fig.~\ref{fig:H2_diagram} we show the excitation diagram. As we lack of H$_2$ emission line intensity measurements with different vibrational energy levels, we can only measure rotational temperature $T_{\rm rot}$ for the vibrational H$_2$($\nu = 1-0$) transition. We find that the data are well-fitted by a linear fit with $T_{\rm rot} \approx 1600 \pm 400$\,K, suggesting that the warm H$_2$ ISM phase in the central brightest zone of this galaxy is in LTE. 

This temperature value also suggests that the warm H$_2$ gas may be mainly heated by SNR shocks \citep{Brand1989,Oliva1990}, with perhaps some contribution from thermal X-ray heating from SNRs ($T_{\rm rot} \gtrsim 2000$\,K, e.g. \citealt{Draine1990}). Thermal UV heating is unlikely as models suggest lower temperature values ($T_{\rm rot} \lesssim 1000$\,K, \citealt{Sternberg1989}). However, we stress that we can not rule out non-thermal UV excitation (e.g. fluorescence; \citealt{Black1987}) as we lack of H$_2$ emission line observations with different vibrational energy levels \citep{Bedregal2009}. 

Nevertheless, our $T_{\rm rot}$ estimate is slightly higher but still consistent with the average value ($<T_{\rm rot}> \sim 1360 \pm 390$\,K) reported for local ($z < 0.08$) LIRGs \citep{Vivian2019}.

\section{Beam smearing effect on the rotation velocity values}
\label{sec:AppendixB}

Accurate rotation velocities can not be derived toward the centre of galaxies due to beam-smearing effects. Rotation velocities tend to be underestimated due to the flux-weighted nature of the data convolution with the PSF and/or synthesized beam. Thus, a beam-smearing correction needs to be applied.

Based on our best-fit two-dimensional models, we quantify the beam-smearing effect on rotation velocities by computing the beam-smeared to intrinsic \textsc{arctan} best-fit velocity model ratio and measured across the galaxy major kinematic axis. We show the estimation of this ratio in Fig.~\ref{fig:corr_factor} for the ALMA and VLT-SINFONI observations of the HATLAS114625 and HATLAS121446 galaxies. We find that a moderate beam-smearing correction ($\lesssim$10\,\%) needs to be applied to the rotation velocity values measured at a galactocentric radius longer than $1.5 \times$\,projected synthesized beam/PSF FWHM. At smaller radii, the beam-smearing effect tends to increase dramatically.

\begin{figure}[!h]
\centering
\includegraphics[width=1.0\columnwidth]{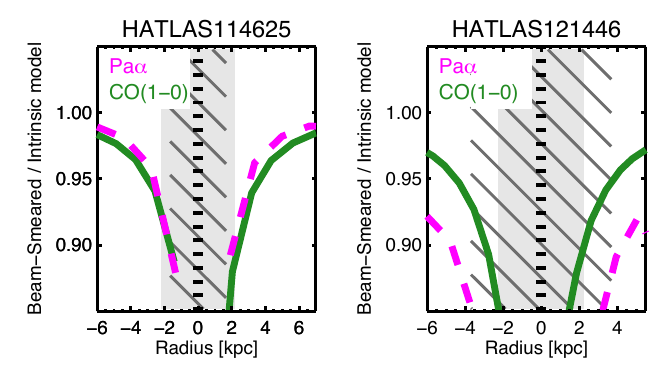}
\caption{\label{fig:corr_factor} Beam-smearing effect on the recovered rotation velocity values across the major kinematic axis for HATLAS114625 (\textit{Left}) and HATLAS121446 (\textit{Right}) galaxies. We mask the values at $R \leq 0.5 \times$\,synthesized beam/PSF FWHM due to the highly uncertain estimates and to improve visualization. We colour-code in the same way as Fig.~\ref{fig:1d_profiles}. At a radius $R \gtrsim 1.5 \times$\,synthesized beam/PSF FWHM from the dynamical centre, we estimate that a moderate correction ($\lesssim$10\,\%) needs to be applied to the observed rotation velocity values.}
\end{figure}

\section{SINFONI $K$-band continuum maps}
\label{sec:AppendixC}

In Fig~\ref{fig:SINFOcont} we show the continuum maps used to calculate the pixel-by-pixel $\Sigma_\star$ values. These maps are scaled to the respective galaxy $M_\star$ value, i.e., by assuming a constant mass-to-light ratio across the galactic disc.

\begin{figure}[!h]
\centering
\includegraphics[width=0.4\columnwidth]{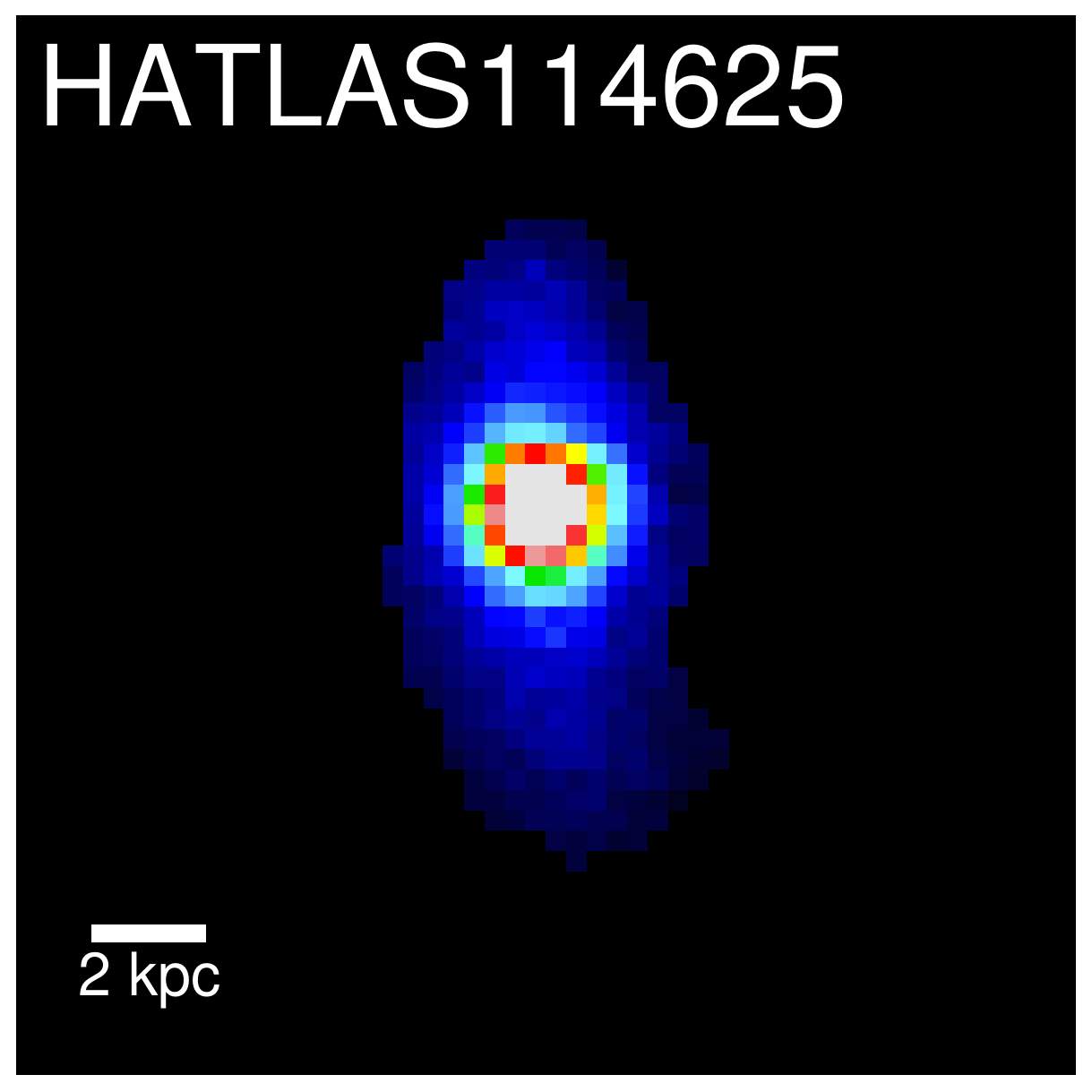}
\includegraphics[width=0.4\columnwidth]{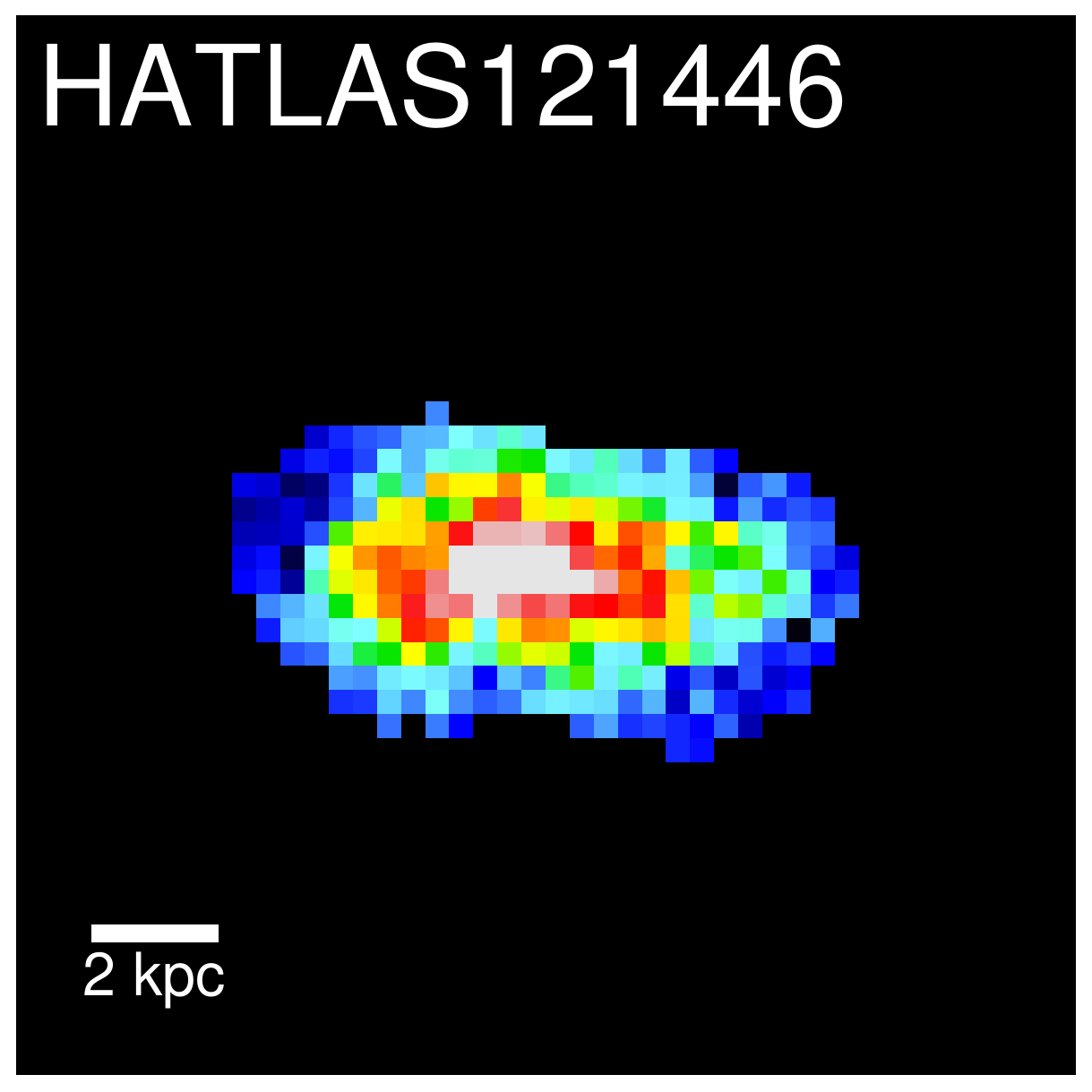}
\caption{\label{fig:SINFOcont} SINFONI $K$-band continuum maps for HATLAS114625 (\textit{Left}) and HATLAS121446 (\textit{Right}) targets.}
\end{figure}

\section{Full $\alpha_{\rm CO}$ and Dark Matter PDFs}
\label{sec:AppendixD}
Our dynamical mass approach is limited by the degeneracy between $\alpha_{\rm CO}$ and the dark matter fraction $f_{\rm DM}$ variables. Due to our assumption of a constant $\alpha_{\rm CO}$ value across the galactic disc, then the dynamical mass value estimated at the innermost galactocentric radius limits the maximum $\alpha_{\rm CO}$ value that can be obtained. At this radius, $f_{\rm DM}$ is consistent with zero, but it does not necessarily hold at longer radii. 

In Fig.~\ref{fig:corner}, we show the $\alpha_{\rm CO}$ and $f_{\rm DM}(R)$ PDFs derived from the `thick-disc plus S\'ersic' dynamical mass model, i.e. the model that determines the adopted median $\alpha_{\rm CO}$ value in our work. There is no major difference between the $\alpha_{\rm CO}$ PDF obtained from this model and those derived from other dynamical models as the broad $\alpha_{\rm CO}$ PDF shapes probe (Fig.~\ref{fig:aco_pdf}).

HATLAS114625 galaxy shows a higher increase of $f_{\rm DM}$ as a function of galactocentric radius compared to the HATLAS121446, suggesting that baryonic matter is distributed more compactly in this system. It is consistent with the reported $K$-band S\'ersic index value (Table~\ref{tab:table2}) and the steeper velocity gradient that is seen in HATLAS114625 (Fig.~\ref{fig:1d_profiles}).

\begin{figure*}[!b]
\centering
\includegraphics[width=1.3\columnwidth]{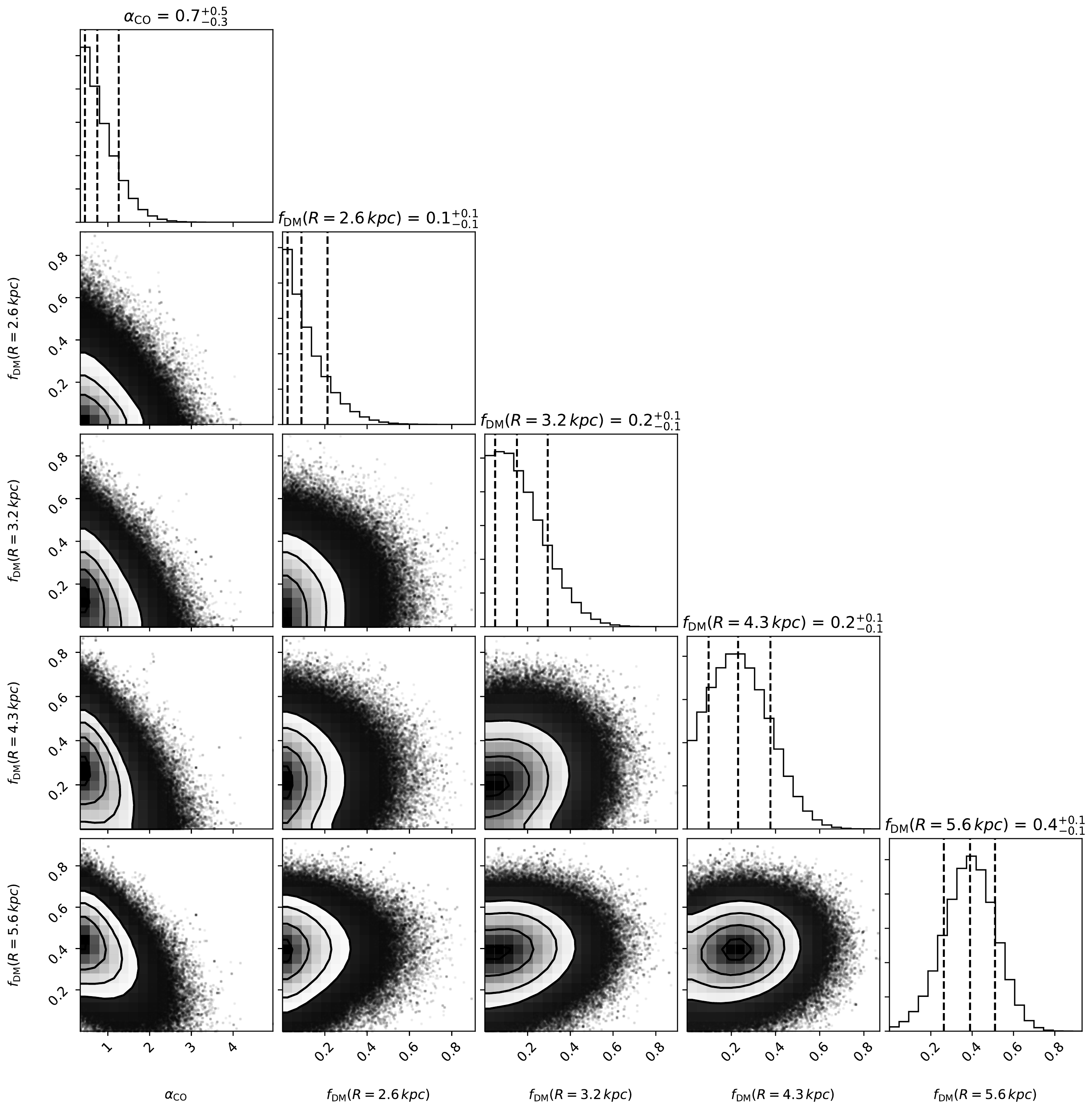}\\
\includegraphics[width=1.3\columnwidth]{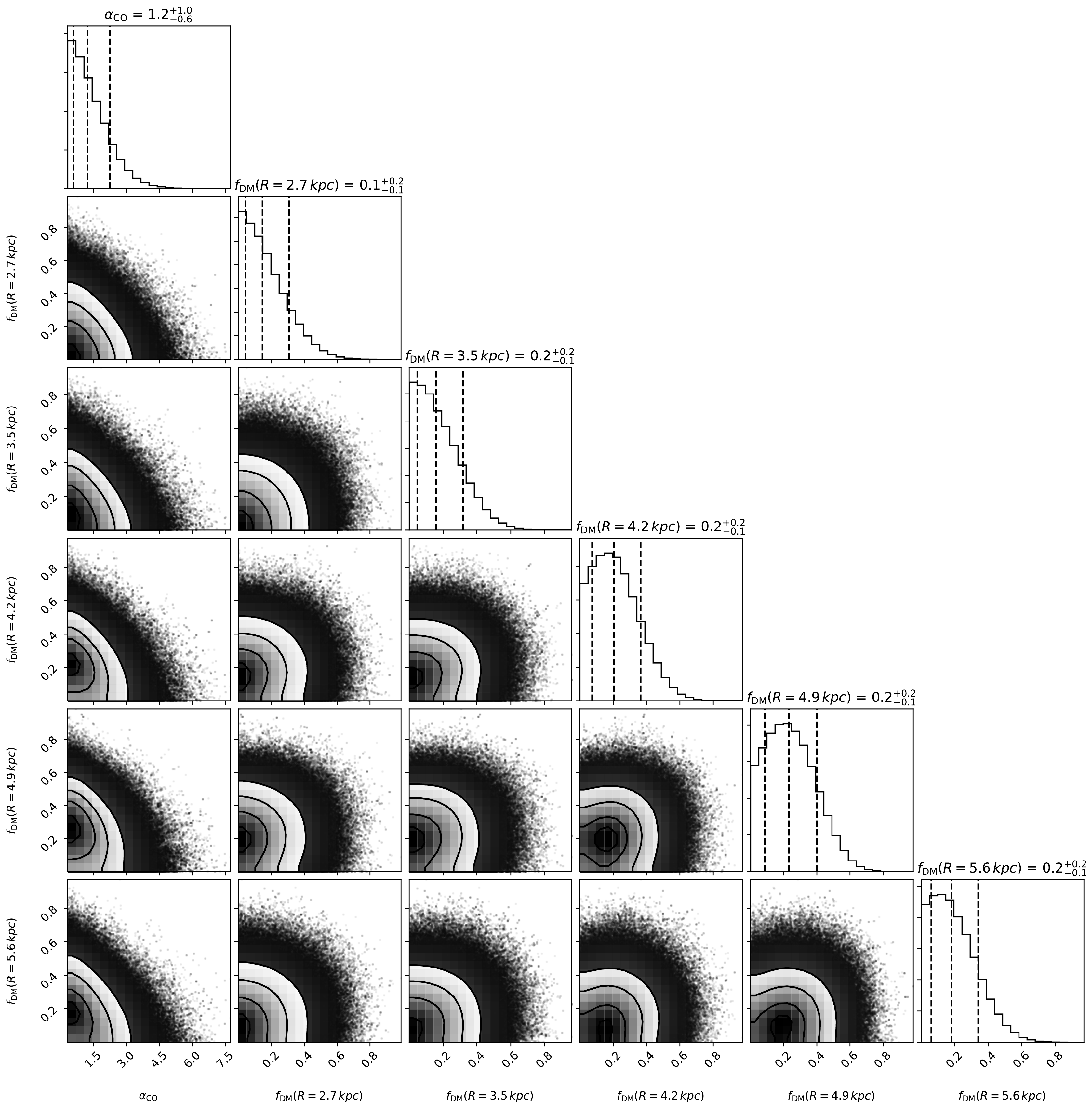}
\caption{\label{fig:corner} Corner plot for $\alpha_{\rm CO}$ and $f_{\rm DM}(R)$ variables for HATLAS114625 (\textit{Top}) and HATLAS121446 (\textit{Bottom}) starbursts.}
\end{figure*}

\end{appendix}

\end{document}